\documentclass[showpacs,aps,prd,amsmath,amsfonts,amssymb,eqsecnum,nofootinbib,
  floatfix,secnumarabic,preprintnumbers,superscriptaddress]{revtex4}
\usepackage{bbold,epsfig,undertilde, color}

\begin{document} 
\renewcommand{\baselinestretch}{1.3}
\newcommand{\nc}{\newcommand}
\nc{\postscript}[2]{\setlength{\epsfxsize}{#2\hsize}\centerline{\epsfbox{#1}}}
\nc{\be}{\begin{equation}}   \nc{\ee}{\end{equation}}
\nc{\beq}{\begin{equation}}   \nc{\eeq}{\end{equation}}
\nc{\bea}{\begin{eqnarray}}   \nc{\eea}{\end{eqnarray}}
\nc{\baa}{\begin{array}}      \nc{\eaa}{\end{array}}
\nc{\bit}{\begin{itemize}}    \nc{\eit}{\end{itemize}}
\nc{\ben}{\begin{enumerate}}  \nc{\een}{\end{enumerate}}
\nc{\bce}{\begin{center}}     \nc{\ece}{\end{center}}
\nc{\non}{\nonumber}
\nc{\sh}{\hat s}
\nc\bpm{\begin{pmatrix}}      \nc\epm{\end{pmatrix}} 
\def\bsp#1\esp{\begin{split}#1\end{split}}

\newcommand{\etc}{{\it etc.}}
\newcommand{\ie}{{\it i.e.}}
\newcommand{\eg}{{\it e.g.}}

\nc\lag{{\cal L}}
\nc\e{\varepsilon} 
\nc\Nbar{{\bar N}}
\nc\lbar{{\bar l}}
\nc\qbar{{\bar q}}
\nc\xibar{{\bar \xi}}
\nc\psibar{{\bar \psi}}
\nc\Psibar{{\bar \Psi}}
\nc\Lbar{{\bar L}}
\def\d{\mathrm{d}}

\preprint{CERN-PH-TH/2013-157}
\preprint{CUMQ/HEP 178}

\title{Doubly-charged particles at the Large Hadron Collider}

\author{Adam Alloul}
\email{adam.alloul@iphc.cnrs.fr}
\affiliation{Institut Pluridisciplinaire Hubert Curien/D\'epartement Recherches
 Subatomiques, Universit\'e de Strasbourg/CNRS-IN2P3, 23 Rue du Loess,
 F-67037 Strasbourg, France}
\author{Mariana Frank}
\email{mfrank@alcor.concordia.ca}
\affiliation{Department of Physics, Concordia University, 7141 Sherbrooke St. 
West, Montreal, Quebec, Canada H4B 1R6}
\author{Benjamin Fuks}
\email{fuks@cern.ch}
\affiliation{Institut Pluridisciplinaire Hubert Curien/D\'epartement Recherches
 Subatomiques, Universit\'e de Strasbourg/CNRS-IN2P3, 23 Rue du Loess,
 F-67037 Strasbourg, France}
\affiliation{Theory Division, Physics Department, CERN, CH-1211 Geneva 23, Switzerland}
\author{Michel Rausch de Traubenberg}
\email{michel.rausch@iphc.cnrs.fr}
\affiliation{Institut Pluridisciplinaire Hubert Curien/D\'epartement Recherches
 Subatomiques, Universit\'e de Strasbourg/CNRS-IN2P3, 23 Rue du Loess,
 F-67037 Strasbourg, France}

\date{\today}
\begin{abstract}
  In this work we investigate the production and signatures of doubly-charged particles
  at the Large Hadron Collider. We start with the Standard Model particle content and
  representations and add generic doubly-charged exotic particles.  We classify these
  doubly-charged states according to their spin, considering scalar, fermionic and
  vectorial fields, and according to their $SU(2)_L$ representation, being chosen to be
  either trivial, fundamental, or adjoint. We write the most general interactions between
  them and the Standard Model sector and study their production modes and possible decay
  channels. We then probe how they can most likely be observed and how particles with
  different spin and $SU(2)_L$ representations could be possibly distinguished.
\end{abstract}


\maketitle

\section{Introduction}
The search for new particles has been given a boost with the discovery of a Higgs boson
at the Large Hadron Collider (LHC) at CERN \cite{ATLAS:2012ae,Chatrchyan:2012tx}.
While we are still awaiting confirmation of whether this is the long-awaited Higgs boson of the Standard Model (SM) or, as it appears at present, a Higgs boson with possibly
tantalizing signs of new physics, phenomenologists are trying to predict
benchmarks for beyond the Standard Model
(BSM) physics that could follow. 
Some adhere to long-hyped scenarios such as weak-scale supersymmetry (either in its minimal reincarnation, or
in a non-minimal form),
extended gauge structures, extra-dimensional models, or Higgs composite models,
and base their analyses  on most likely and telling signatures of the models.
Yet there has been also some interest in testing more generic collider features, that would have clear signatures and may be common to more models. The advantage is the fact that several models might yield the same collider signals, so that a simple but general model with minimal additions (particles and interactions) over the SM might be useful for obtaining  results easy to compare with data.
A small set of model parameters is usually involved, such as the masses of the new particles and the coupling strength of their interactions.
In this setup, one could think of the SM as a limiting case when the new physics sector decouples.

Adopting this bottom-up approach for new physics phenomenology,
the cleanest and clearest results are most likely obtained for exotic particles, \ie, particles with quantum numbers unlike those from  the SM particle content. In this work, we
propose to investigate the possibility for supplementing the Standard Model with a very
simple kind of exotic states, which we denote by $X^{++}$ and which are
either scalar, fermion or vector fields with two units of
electric charge. We keep our analysis as general as possible by allowing the new
particles to belong to different
$SU(2)_L$ representations, but further restrict ourselves to the simplest ones, namely the singlet, doublet and triplet cases.
This goes along the same lines as several recent works investigating doubly-charged particles
in more or less generic
situations \cite{Cuypers:1996ia,DelNobile:2009st,Rentala:2011mr,Meirose:2011cs,Hisano:2013sn,delAguila:2013yaa}, our work
being however the only analysis studying in detail the collider implications
of various spin and weak isospin quantum numbers.

From a top-down point of view, it must be noted that such
doubly-charged particles appear in many BSM scenarios,
and thus are of particular interest to model builders.
As examples, doubly-charged scalar states, often dubbed doubly-charged
Higgs fields, 
appear in left-right symmetric models \cite{Pati:1974yy, Mohapatra:1974hk,
Mohapatra:1974gc,Senjanovic:1975rk, Mohapatra:1977mj, Senjanovic:1978ev,Mohapatra:1979ia} or
in see-saw models for neutrino masses with Higgs triplets \cite{Cheng:1980qt,
Gelmini:1980re, Zee:1980ai, Han:2005nk, Lee:2005kd, Picek:2009is, Majee:2010ar,Aoki:2011yk,
Chen:2011de,Kumericki:2011hf,
Aoki:2011pz,Chiang:2012dk,Kumericki:2012bh,Sugiyama:2012yw,Picek:2012ei,
Kanemura:2013vxa}.  Doubly-charged fermions can appear
in extra-dimensional models including custodian taus \cite{Csaki:2008qq,
Chen:2009gy, Kadosh:2010rm, delAguila:2010vg, delAguila:2010es,Delgado:2011iz}, in
new physics models inspired by string theories \cite{Cvetic:2011iq} or as
the supersymmetric partners for
the doubly-charged scalar fields in supersymmetric extensions of left-right symmetric models
\cite{Demir:2008wt, Frank:2007nv,Babu:2013ega,Franceschini:2013aha}. Finally, BSM
theories with an extended gauge group often include doubly-charged vector
bosons \cite{Frampton:1989fu, Pal:1990xw, Pisano:1991ee, Frampton:1991wf,
Frampton:1992wt} although it is also possible to consider vector states with a double electric charge
independently of any gauge-group structure, as in models with a non-commutative geometry
or in composite or technicolor theories
\cite{Farhi:1980xs, Harari:1982xy, Cabibbo:1983bk, Eichten:1984eu,
Pancheri:1984sm, Buchmuller:1985nn, Stephan:2005uj, Gudnason:2006ug, Biondini:2012ny}.

Subsequently, same-sign dilepton and/or doubly-charged Higgs bosons resonances have been the
topic of many accelerator analyses in the past. Usually assumed to be produced either singly
or in pairs, no events have been observed by experiments around the Large Electron Positron
collider (LEP) \cite{OPAL, OPAL_single, L3, DELPHI}, the Hadron Electron Ring Accelerator
(HERA) \cite{H1,H1bis} and the Tevatron \cite{D0, D0bis, CDF, CDFFV}.
The most up-to-date bounds have however been more recently
derived by the LHC experimental collaborations.
Both ATLAS and CMS have searched for long-lived doubly-charged states
\cite{Aad:2013pqd,Chatrchyan:2013oca}, basing
their analyses either on
identification of the new particles by using their longer time-of-flight to the outer subdetectors or
on their anomalous energy loss along their tracks.
Assuming a Drell-Yan-like pair production, the long-lived doubly-charged state masses have been
constrained to lie above 685 GeV after analyzing
5 fb$^{-1}$ of LHC collisions at a center-of-mass energy of
$\sqrt{S_h}=7$ TeV and 18.8 fb$^{-1}$ of collisions at $\sqrt{S_h}$ = 8 TeV \cite{Chatrchyan:2013oca}.
Nevertheless, these limits do not hold for promptly-decaying doubly-charged particles.
In this case, dedicated studies only exist for
doubly-charged Higgs bosons. Being pair-produced, they are then assumed to
decay into a pair of leptons with the same electric charge
through Majorana-type interactions \cite{Chatrchyan:2012ya,ATLAS:2012hi}.
Assuming a branching fraction of 100\% decays into leptons, \ie,
neglecting the possible decays into a $W$-boson pair,
the doubly-charged Higgs mass has been constrained to be larger than about 450 GeV.
All these existing mass bounds can however be easily
evaded by relaxing the rather constraining
new physics assumptions. We hence follow the approach of most model-builders and
consequently assume the new particle mass to be a free parameter.

Our work is organized as follows. In Section \ref{sec:themodel} we describe in detail our
model for particles and interactions, differentiating in the discussion into spin and weak-isospin
fields. We also compute analytical cross sections for the production of the new states
at hadron colliders and for their decays into Standard Model particles. We then dedicate
Section \ref{sec:numerics} to a detailed numerical analysis of doubly-charged particle signals
at the LHC and briefly discuss, in Section~\ref{sec:MC}, different kinematical variables
that would allow for distinguishing the spin and/or $SU(2)_L$ quantum numbers of the
new states.  Finally, we summarize our results in Section~\ref{sec:conclusion}.

\section{Simplified models for exotic doubly-charged states}
\label{sec:themodel}
Following the approach of the LHC New Physics Working Group \cite{Alves:2011wf},
specific final state topologies are described by means of dedicated simplified models.
They consist of minimal extensions of the Standard Model
where the number of new states and operators is maximally reduced. Moreover,
the model parameters are translated in terms of relevant products of cross sections and
branching ratios so that LHC data can be
easily reinterpreted in terms of constraints on these quantities.
In this work, we construct a set of simplified models describing all the mechanisms
yielding the production of doubly-charged particles, followed by their subsequent
decays into pairs of charged leptons (possibly together with additional neutral
states when relevant) with the same electric charges. Only final
state signatures with three leptons or more will be considered, as the associated
Standard Model background is known to be under good control.

We start with the Standard Model field content and the $SU(3)_C \times
SU(2)_L \times U(1)_Y$ gauge group and then add an exotic doubly-charged,
non-colored, state lying in a specific representation of the Lorentz group and
$SU(2)_L$. Motivated by the most common existing new physics
theories, we restrict ourselves to scalar, spin $1/2$ and vector
states which we assume to lie either in the trivial, fundamental or adjoint representation of 
$SU(2)_L$. However, higher spin states or higher-dimensional representations of $SU(2)_L$
could be considered, such as in Ref.~\cite{Biondini:2012ny} where the phenomenology of
excited leptons in the $\utilde{\bf 4}$ representation of $SU(2)_L$ is investigated.
In addition, the hypercharge quantum numbers of the new multiplet are
chosen so that the doubly-charged component is always the state with the
highest electric charge. Finally, any interaction allowed by model symmetries but
irrelevant for our study is omitted from the Lagrangians presented in this section.

In the rest of this section, we construct simplified models following a classification of 
the doubly-charged states by their spin. We focus on their signals
at the
LHC and analytically compute cross sections and decay widths relevant for the production
of final states with
three leptons or more, the corresponding numerical analysis being performed in Section
\ref{sec:numerics}. This will guide us to a choice of benchmark scenarios to be
considered for a more advanced study based on Monte Carlo
simulations, as in Section \ref{sec:MC}.

\subsection{Spin $0$ doubly-charged particles}\label{sec:modsc}
In the following, we focus on simplified models describing the dynamics of $SU(2)_L$
multiplets containing, as a component with the highest electric charge, a
doubly-charged state. We denote by $\phi$, $\Phi$ and $\mathbf{\Phi}$ three complex
scalar fields lying in the $\utilde{\bf 1}$, $\utilde{\bf 2}$ and $\utilde{\bf
3}$ representations of $SU(2)_L$, respectively, with hypercharge quantum numbers set to
$Y_\phi = 2$, $Y_\Phi = 3/2$,  $Y_{\mathbf\Phi} = 1$, 
\be\label{eq:scalarfields}
  \phi \equiv \phi^{++} \ , \qquad
  \Phi^i \equiv \bpm \Phi^{++} \\  \Phi^+ \epm \qquad\text{and}\qquad
   \mathbf \Phi^i{}_j \equiv
      \bpm \frac{\mathbf \Phi^+}{\sqrt{2}} & \mathbf \Phi^{++}\\
           \mathbf \Phi^0 & -\frac{\mathbf \Phi^+}{\sqrt{2}}
  \epm \ .
\ee 
In the last expression, we have employed the matrix representation for triplet fields
defined by
\be\label{sec:adjfundeq}
 \mathbf \Phi^i{}_j = \frac{1}{\sqrt{2}} (\sigma_a)^i{}_j\ \mathbf\Phi^a \ ,
\ee
the matrices $\sigma^a$ being the Pauli matrices and $a=1,2,3$ a 
$SU(2)_L$ adjoint gauge index\footnote{In our notations, we always employ Latin letters
of the middle of the alphabet for fundamental indices and Latin letters
of the beginning of the alphabet for adjoint indices.}. Diagonalizing the third
generator of $SU(2)_L$ in the adjoint representation, the gauge eigenstates
$\mathbf\Phi^a$ can be linked to the physical mass-eigenstates $\mathbf\Phi^0$,
$\mathbf \Phi^+$ and $\mathbf \Phi^{++}$ by means of 
\be
  \mathbf\Phi^1 = \frac1{\sqrt{2}} \Big[\mathbf\Phi^0 + \mathbf\Phi^{++}\Big]  
    \ , \quad
  \mathbf\Phi^2 = \frac1{\sqrt{2} i} \Big[\mathbf\Phi^0 - \mathbf\Phi^{++}\Big]
    \quad\text{and}\quad
  \mathbf\Phi^3 = \mathbf\Phi^+ \ .
\ee

Kinetic and gauge interaction terms for the three fields of Eq.\
\eqref{eq:scalarfields} are fixed by gauge invariance, 
\be
  \lag_{\rm kin} = D_\mu \phi^\dag D^\mu \phi + 
    D_\mu \Phi_i^\dag D^\mu \Phi^i + \
    D_\mu \mathbf\Phi_a^\dag D^\mu \mathbf\Phi^a +\ldots \ , 
\label{eq:Lsck}\ee
the covariant derivatives being given by
\be\bsp
  D_\mu\phi =&\ \partial_\mu\phi - 2 i g' B_\mu\phi \ , \\
  D_\mu\Phi^i =&\ \partial_\mu\Phi^i - \frac{3}{2} i g' B_\mu\Phi^i - i g\
    \frac{(\sigma_a)^i{}_j}{2}\ \Phi^j\ W_\mu^a \ ,  \\
  D_\mu\mathbf\Phi^a =&\ \partial_\mu\mathbf\Phi^a - i g' B_\mu\mathbf\Phi^a + g\
    \e_{bc}{}^a\ \mathbf\Phi^c\ W_\mu^b \ ,
\esp\label{eq:covder}\ee
and the dots standing for mass terms.
In the expressions above, $g$ and $g'$ are the weak and hypercharge coupling constants,
respectively, and we have normalized the structure
constants of $SU(2)$ as $\e_{12}{}^3=1$. We have also introduced the electroweak gauge
bosons denoted by $B_\mu$ and $W_\mu^a$.
The Lagrangian above also allows the components of fields lying in a
non-trivial representation of $SU(2)_L$ to decay into each other,
together with an accompanying gauge boson, if kinematically allowed.
However, we choose to focus in this work on the low mass region of the
parameter space so that splittings, induced, \eg, by electroweak
symmetry breaking, are assumed smaller than the weak boson masses.
In order to allow for the extra fields $\phi$, $\Phi$ and $\mathbf\Phi$ to decay,
new Yukawa interactions are hence required,
\be
  \lag_{\rm yuk} = 
    \frac12 y^{(1)}\ \phi\ \lbar^c_R l_R +
    \frac{y^{(2)}}{\Lambda}\ \Phi^i\ \Lbar^c_i \gamma_\mu D^\mu l_R +
    \frac12 y^{(3)}\ \mathbf \Phi^i{}_j\ \Lbar^c_i L^j +
    {\rm h.c.} \ ,
\label{eq:Lscy}\ee
where the three quantities $y^{(1)}$, $y^{(2)}$ and $y^{(3)}$ are $3 \times 3$
matrices in generation space and where flavor indices have been omitted for clarity.
Moreover, we have explicitly indicated the chirality of the
lepton fields, or equivalently their $SU(2)_L$ representations so that there is no
confusion about the action of the gauge-covariant derivative $D_\mu$. The
(four-component) spinorial field $l_R$ stands hence for the right-handed charged
lepton singlet of $SU(2)_L$ and the objects 
\be\label{eq:deflep}
  L^i = \bpm \nu_L\\ l_L \epm \qquad\text{and}\qquad
  L_i = \e_{ij}\ L^j = \bpm l_L \\ - \nu_L \epm
\ee
are two representations of the weak doublet comprised of
the left-handed neutrino $\nu_L$ and charged lepton $l_L$ fields. The terms included
in the Lagrangian of Eq.\ \eqref{eq:Lscy} are consistent with gauge and 
Lorentz invariance, which leads to the appearance of the charge conjugation
operator denoted by the superscript $^c$.
Moreover, care is taken so that each component of the new scalar fields is allowed to decay
into Standard Model particles. In particular, this implies the 
use of a higher-dimensional operator suppressed by an
effective scale $\Lambda$ for the coupling of the weak doublet
$\Phi$ to the lepton fields since the four-component spinor 
product $\xibar^c_R \lambda_L=0$ for any fermionic fields.

The Lagrangian of Eq.\ \eqref{eq:Lsck} includes couplings of the new
fields to the electroweak gauge bosons so that the former can then be produced at hadron
colliders either from quark-antiquark scattering or through vector boson fusion.
The only processes giving rise to a signature with three charged leptons or more 
consist of the pair production of two doubly-charged fields or of the associated production
of one singly-charged and one doubly-charged state. Considering first the neutral current
channels, $q\bar q \to \phi^{++} \phi^{--}$, $q\bar q\to \Phi^{++} \Phi^{--}$ and 
$q\bar q \to \mathbf{\Phi}^{++} \mathbf{\Phi}^{--}$,
the relevant partonic cross sections read, as a function of the partonic
center-of-mass energy $\sh$,
\be\bsp
 \hat\sigma^{NC}_1 = &\ 
    \frac{4 \pi \alpha^2 \sh}{9} \big[ 1 - 4x^2_{\phi^{++}}\big]^{\frac32} \bigg[
    \frac{e^2_q}{\sh^2} -
    \frac{ e_q (L_q+R_q) (\sh-M_Z^2)}{2 c_W^2 \sh |\sh_Z|^2} +
    \frac{L_q^2 + R_q^2}{8 c_W^4 |\sh_Z|^2}
  \bigg]  \ , \\
 \hat\sigma^{NC}_2 = &\ 
    \frac{4 \pi \alpha^2 \sh}{9} \big[ 1 - 4x^2_{\Phi^{++}}\big]^{\frac32} \bigg[  
    \frac{e^2_q}{\sh^2} +
    \frac{ e_q (1-4s_W^2) (L_q+R_q) (\sh-M_Z^2)}{8 c_W^2 s_W^2 \sh |\sh_Z|^2}+
    \frac{(1-4s_W^2)^2 (L_q^2 + R_q^2)}{128 c_W^4 s_W^4 |\sh_Z|^2}
  \bigg]  \ , \\
  \hat\sigma^{NC}_3 = &\
    \frac{4 \pi \alpha^2 \sh}{9} \big[ 1 - 4x^2_{\mathbf{\Phi}^{++}}\big]^{\frac32} \bigg[  
    \frac{e^2_q}{\sh^2} +
    \frac{ e_q (1-2s_W^2) (L_q+R_q) (\sh-M_Z^2)}{4 c_W^2 s_W^2 \sh |\sh_Z|^2} +
    \frac{(1-2s_W^2)^2 (L_q^2 + R_q^2)}{32 c_W^4 s_W^4 |\sh_Z|^2}
  \bigg]  \ , 
\esp\label{eq:xsecscNC}\ee
where $\hat\sigma_i$ is associated with the pair production of doubly-charged components
of multiplets in the $\utilde{\mathbf{i}}$ representation of $SU(2)_L$.
In those expressions, we have introduced the sine and cosine of the weak mixing angle
$s_W$ and $c_W$, the quark electric charges $e_q$, their weak isospin quantum numbers
$T_{3 q}$ and their $Z$-boson coupling strengths $L_q=2 (T_{3q}-e_q s_W^2)$ and 
$R_q=-2 e_q s_W^2$. Moreover, we are employing the reduced kinematical variables
$x_\phi^2 = \frac{M^2_\phi}{\sh}$ and 
$\sh_Z = \sh - M_Z^2 + i \Gamma_Z M_Z$, where $M_Z$ and $\Gamma_Z$ are the $Z$-boson mass and
width, respectively.

For a new $SU(2)_L$ doublet (triplet) of scalar fields, trilepton signatures can also
rise from the charged current production of a doubly-charged
state in association with a singly-charged state,
$u_i \bar d_j \to \Phi^{++} \Phi^-$ ($u_i \bar d_j \to \mathbf{\Phi}^{++} \mathbf{\Phi}^-$).
The corresponding cross sections read
\be\bsp
 \hat\sigma^{CC}_2 = &\
   \frac{\pi \alpha^2 \sh}{72 s_W^4 |\sh_W|^2}  |V_{ij}^{\rm CKM}|^2\ \lambda^{\frac32}(1,x^2_{\Phi^{++}},x^2_{\Phi^+}) \ , \\
 \hat\sigma^{CC}_3 = &\
   \frac{\pi \alpha^2 \sh}{36 s_W^4 |\sh_W|^2}  |V_{ij}^{\rm CKM}|^2\ \lambda^{\frac32}(1,x^2_{\Phi^{++}},x^2_{\Phi^+}) \ ,
\esp\label{eq:xsecscCC}\ee
where as for the neutral currents, we have employed the reduced propagator
$\sh_W =\sh - M_W^2 + i
\Gamma_W M_W$, $M_W$ and $\Gamma_W$ being the $W$-boson mass and width, in addition to
the K\"allen function $\lambda(a,b,c) = a^2+b^2+c^2-2ab -2bc -2 ca$ and to the CKM matrix
$V^{\rm CKM}$.

After electroweak symmetry breaking, the neutral component of the $\mathbf\Phi$ multiplet can
develop a vacuum expectation value (vev) $v_{\mathbf\Phi}$,
\be\label{eq:vev}
  \mathbf\Phi^0 \to \frac{1}{\sqrt{2}} \Big[v_{\mathbf\Phi} + \mathbf H^0
    + i \mathbf A^0 \Big] \ ,
\ee
where we are now distinguishing the scalar $\mathbf H^0$ and pseudoscalar $\mathbf A^0$
degrees of freedom of the neutral field. In
principle, $v_{\mathbf\Phi}$ is constrained to be small by
the electroweak $\rho$-parameter which only slightly deviates from unity
\cite{Gunion:1989in, Csaki:2002qg, Chen:2003fm, Csaki:2003si}, in addition to
strong constraints arising from the neutrino sector \cite{Gelmini:1980re, Zee:1980ai, Han:2005nk,
Lee:2005kd}. We have however decided to consider
both scenarios with a very small $v_{\mathbf\Phi}$ value and those with a larger value,
as in left-right symmetric models where the vacuum expectation values of the neutral fields
are less constrained. In this case, the Lagrangian of Eq.~\eqref{eq:Lsck} allows for
the production of a single doubly-charged or singly-charged new state through
vector boson fusion \cite{Chiang:2012dk, Rommerskirchen:2007jv} or in
association with a weak gauge boson \cite{Akeroyd:2005gt}. Since vector boson fusion processes do not
yield final states with more than two charged leptons, we restrict ourselves to the study
of the $q \bar q' \to \mathbf{\Phi}^+ Z$ and $q \bar q' \to \mathbf{\Phi}^{++} W$ channels, the
corresponding partonic cross sections being given by
\be\bsp
 \hat\sigma^{W\mathbf\Phi^{++}} = &\ 
   \frac{\pi^2 \alpha^3 v^2_{\mathbf\Phi}}{18 s_W^6 \sh_W^2}  |V_{ij}^{\rm CKM}|^2\ \lambda^\frac12 (1,x^2_{\Phi^{++}},x^2_W)
      \bigg[ \frac{\lambda(1,x^2_{\Phi^{++}},x^2_W)}{x_W^2} + 12 \bigg]\ , \\ 
 \hat\sigma^{Z\mathbf\Phi^+} = &\
   \frac{\pi^2 \alpha^3 v^2_{\mathbf\Phi}(1+s_W^2)^2}{72 s_W^6 c_W^2  \sh_W^2}  |V_{ij}^{\rm CKM}|^2\ \lambda^\frac12 (1,x^2_{\Phi^{++}},x^2_Z)
      \bigg[ \frac{\lambda(1,x^2_{\Phi^{++}},x^2_Z)}{x_Z^2} + 12 \bigg]\ .
\esp\label{eq:xsecscvev}\ee
In addition, the neutral ${\bf A^0}$ and ${\bf H^0}$ fields can
also decay into multileptonic final states so that their single- and pair-production
should be considered.
However, a correct treatment of these fields implies to
also add their mixings with the components of the Standard Model Higgs doublet. This renders
the approach rather non-minimal so that we neglect them from the present analysis. Let us note
that including them would not change our conclusions in the following sections and even
increase any possible new physics signals.

As mentioned above, the Lagrangian of Eq.\ \eqref{eq:Lscy} allows for the new states to decay. We
assume that these states are all close enough in mass so that they cannot decay into each other and
compute below the relevant partial widths.
Considering the decays of the doubly-charged states into a
same-sign lepton pair, we obtain, assuming lepton flavor conservation,
\be\bsp
  \Gamma_{1,\ell}^{++} = &\ \frac{M_{\phi^{++}}|y^{(1)}|^2}{32 \pi} \Big[1-2 x_\ell^2\Big] \sqrt{1-4 x_\ell^2}\ , \\
  \Gamma_{2,\ell}^{++} = &\ \frac{M_{\Phi^{++}} M_\ell^2 |y^{(2)}|^2}{8 \pi \Lambda^2} \Big[ 1-2 x_\ell^2\Big] \sqrt{1-4 x_\ell^2} \ , \\
  \Gamma_{3,\ell}^{++} = &\ \frac{M_{\mathbf{\Phi}^{++}} |y^{(3)}|^2}{32 \pi} \Big[1 - 2 x_\ell^2\Big] \sqrt{1 -4 x_\ell^2}\ ,
\esp\label{eq:BRsc++1}\ee
using again different subscripts to distinguish the scalar field representations.
Whereas flavor-violating effects could have also been considered,
we nevertheless
neglect them as they are constrained to be small due to lepton
decay processes such as $\mu \to 3 e$, $\mu\to e\gamma$,
\etc, which are bound to be extremely rare~\cite{Swartz:1989qz,
Rizzo:1981xx,Gunion:1989in,Kosmas:1993ch,Mohapatra:1992uu}.
In addition, the triplet field can also decay into a pair of $W^+$-bosons,
the associated width being given by
\be\label{eq:BRsc++2}
  \Gamma_{3,WW}^{++} =  
    \frac{M^3_{\mathbf{\Phi}^{++}} \alpha^2 \pi v_{\mathbf{\Phi}}^2}{4 M_W^4 s_W^4} \sqrt{1 - 4 x_W^2} \Big[1 - 4 x_W^2 + 12 x_W^4\Big] \ .
\ee
Turning to the singly charged components of the new multiplets, the partial widths for the
leptonic decays $\Phi^+ \to \ell^+ \bar\nu_l$ and $\mathbf{\Phi}^+ \to \ell^+ \bar\nu_l$ are
computed as
\be\bsp
  \Gamma_{2,\ell}^+ = &\ \frac{M_{\Phi^+} M_\ell^2 |y^{(2)}|^2}{16 \pi \Lambda ^2} \Big[1-x_\ell^2\Big]^2\ , \\
  \Gamma_{3,\ell}^+ = &\ \frac{M_{\mathbf{\Phi}^+} |y^{(3)}|^2}{32 \pi} \Big[1 - x_\ell^2\Big]^2\ , 
\esp\label{eq:BRsc+1} \ee
while the one for the decay of a triplet field into a pair of weak gauge bosons,
$\mathbf{\Phi}^+ \to W^+ Z$, is found to be
\be\label{eq:BRsc+2}\bsp
  \Gamma_{3,WZ}^+ = &\ \frac{ M^3_{\mathbf{\Phi}^+} \alpha^2\pi v_{\mathbf{\Phi}}^2 (1 + s_W^2)^2}{8  c_W^2 s_W^4 M_Z^2 M_W^2 }
    \Big[\lambda(1,x^2_W,x^2_Z) + 12 x_Z^2 x_W^2\Big]  \sqrt{\lambda(1,x^2_W,x^2_Z)} \ .
\esp \ee

\subsection{Spin $1/2$ doubly-charged particles}\label{sec:femod}
\subsubsection{Simplified models with three generations of leptons}\label{sec:femodA}
We now turn to the building of Standard Model extensions containing one extra
fermionic multiplet and 
like in Section \ref{sec:modsc}, we restrict ourselves to states with an
electric charge not higher than two. We therefore consider three fermionic fields
$\psi$, $\Psi$ and $\mathbf\Psi$ lying in the
singlet, doublet and adjoint representation of $SU(2)_L$, respectively, and we fix their
hypercharge quantum numbers to
$Y_\psi = 2$, $Y_\Psi = 3/2$ and $Y_{\mathbf\Psi} = 1$,
\be\label{eq:ferfields}
  \psi \equiv \psi^{++} \ , \qquad
  \Psi^i \equiv \bpm \Psi^{++} \\  \Psi^+ \epm \qquad\text{and}\qquad
   \mathbf \Psi^i{}_j \equiv
      \bpm \frac{\mathbf \Psi^+}{\sqrt{2}} & \mathbf \Psi^{++}\\
           \mathbf \Psi^0 & -\frac{\mathbf \Psi^+}{\sqrt{2}}
  \epm \ . 
\ee
The associated kinetic and gauge interaction terms are standard and induced by gauge covariant derivatives
that  can be derived from Eq.\ \eqref{eq:covder},
\be
  \lag_{\rm kin} = i \psibar \gamma^\mu D_\mu \psi + 
     i \Psibar_i \gamma^\mu D_\mu \Psi^i + 
     i \mathbf{\Psibar}_a \gamma^\mu D_\mu \mathbf\Psi^a + \ldots\ ,
\label{eq:Lfek}\ee
where mass terms are included in the dots.
As in Section~\ref{sec:modsc}, we assume the new states to be
almost mass-degenerate so that they cannot decay into each other.
In order to allow for the decays of the $\psi$, $\Psi$ and
$\mathbf\Psi$ states, it is thus
necessary to introduce at least
one additional fermionic particle $N$, which we choose to be gauge singlet
as in massive neutrino models \cite{King:2003jb, Altarelli:2004za,
Asaka:2005an}. Avoiding the introduction of more new states to our simplified model,
three-body decays of the $\psi$, $\Psi$ and
$\mathbf \Psi$ fermions into a lepton pair and a $N$ particle
are permitted by means of non-renormalizable four-fermion interactions,
\bea
  \lag_{\rm F} &=&
  \frac{G^{(1,1)}}{2\Lambda^2} \big[\lbar^c_R l_R\big] \big[\Nbar P_L \psi\big] +
  \frac{G^{(1,2)}}{2\Lambda^2} \big[\lbar^c_R l_R\big] \big[\Nbar P_R \psi\big] 
+
  \frac{G^{(2,1)}}{\Lambda^2} \big[\lbar^c_R \Psi^i\big] \big[\Nbar L_i\big] +
  \frac{G^{(2,2)}}{\Lambda^2} \big[\lbar^c_R N\big] \big[\Lbar^{ic} \Psi_i\big]
\non \\ & +&
  \frac{G^{(3,1)}}{2\Lambda^2} \big[\Lbar^c_i L^j\big] \big[\Nbar P_L
    \mathbf\Psi^i{}_j\big] +
  \frac{G^{(3,2)}}{2\Lambda^2} \big[\Lbar^c_i L^j\big] \big[\Nbar P_R
    \mathbf\Psi^i{}_j\big]
 + {\rm h.c.} \ .
\label{eq:LfeF}\eea 
In this Lagrangian, we have only considered a set of independent effective operators
and omitted all generation indices for clarity. In addition, $P_L$ and $P_R$ are
chirality
projectors acting on spin space, the left-handed and right-handed fermionic components
of the Standard Model fields have been defined in Section \ref{sec:modsc} and the
interaction strengths suppressed by a new physics scale $\Lambda$ are
$3\times 3$ matrices in flavor space $G^{(1,1)}$, $G^{(1,2)}$, $G^{(2,1)}$,
$G^{(2,2)}$,  $G^{(3,1)}$ and $G^{(3,2)}$.

The gauge interactions included in the Lagrangian of Eq.~\eqref{eq:Lfek} imply the
possible hadroproduction of pairs of  $\psi$, $\Psi$ and $\mathbf \Psi$
states from quark-antiquark
scattering.  Focusing on final state signatures with at least three
leptons, the partonic cross sections associated with the relevant neutral current
processes $q \bar q \to \psi^{++} \psi^{--}$, $q \bar q \to \Psi^{++}
\Psi^{--}$ and $q \bar q \to \mathbf{\Psi}^{++} \mathbf{\Psi}^{--}$ are
\be\bsp
 \hat\sigma^{NC}_1 = &\ 
    \frac{16 \pi \alpha^2 \sh}{9} \big[1 + 2 x^2_{\psi^{++}}\big] 
    \sqrt{1 - 4x^2_{\psi^{++}}} \bigg[
    \frac{e^2_q}{\sh^2} -
    \frac{ e_q (L_q+R_q) (\sh-M_Z^2)}{2 c_W^2 \sh |\sh_Z|^2} +
    \frac{L_q^2 + R_q^2}{8 c_W^4 |\sh_Z|^2}
  \bigg]  \ , \\
 \hat\sigma^{NC}_2 = &\ 
    \frac{16 \pi \alpha^2 \sh}{9} \big[1 + 2 x^2_{\Psi^{++}}\big] 
    \sqrt{1 - 4x^2_{\Psi^{++}}} \bigg[  
    \frac{e^2_q}{\sh^2} +
    \frac{ e_q (1-4s_W^2) (L_q+R_q) (\sh-M_Z^2)}{8 c_W^2 s_W^2 \sh |\sh_Z|^2} +
    \frac{(1-4s_W^2)^2 (L_q^2 + R_q^2)}{128 c_W^4 s_W^4 |\sh_Z|^2}
  \bigg]  \ , \\
 \hat\sigma^{NC}_3 = &\ 
    \frac{16 \pi \alpha^2 \sh}{9} \big[1 + 2 x^2_{\mathbf \Psi^{++}}\big] \sqrt{1 - 
    4x^2_{\mathbf \Psi^{++}}} \bigg[
    \frac{e^2_q}{\sh^2} +
    \frac{ e_q (1-2s_W^2) (L_q+R_q) (\sh-M_Z^2)}{4 c_W^2 s_W^2 \sh |\sh_Z|^2} +
    \frac{(1-2s_W^2)^2 (L_q^2 + R_q^2)}{32 c_W^4 s_W^4 |\sh_Z|^2}
  \bigg]  \ , 
\esp\label{eq:xsecfeNC}\ee
whereas those related to the charged current processes $u_i\bar d_j\to
\Psi^{++}\Psi^-$ and $u_i\bar d_j\to \mathbf{\Psi}^{++}\mathbf{\Psi}^-$ are
\be\bsp
 \hat\sigma^{CC}_2 = &\ 
   \frac{\pi \alpha^2 \sh}{36 s_W^4 |\sh_W|^2}  |V_{ij}^{\rm CKM}|^2\ \sqrt{\lambda(1,x^2_{\Psi^{++}},x^2_{\Psi^+})} 
     \Big[ 1-(x_{\Psi^{++}}-x_{\Psi^+})^2\Big]   \Big[ 2+(x_{\Psi^{++}}+x_{\Psi^+})^2\Big] \ , \\
 \hat\sigma^{CC}_3 = &\ 
   \frac{\pi \alpha^2 \sh}{18 s_W^4 |\sh_W|^2}  |V_{ij}^{\rm CKM}|^2\ \sqrt{\lambda(1,x^2_{\Psi^{++}},x^2_{\Psi^+})} 
     \Big[ 1-(x_{\Psi^{++}}-x_{\Psi^+})^2\Big]   \Big[ 2+(x_{\Psi^{++}}+x_{\Psi^+})^2\Big] \ .
\esp\label{eq:xsecfeCC}\ee
We recall that our conventions for the notations have been introduced in
Section~\ref{sec:modsc}. Since there
are no closed formulas for the widths of the new
particles that can be calculated from Eq.\ \eqref{eq:LfeF}, we refer to
Section~\ref{sec:numerics} for the corresponding numerical analysis.
As in the scalar case, flavor violating effects are again neglected
since they are constrained from rare leptonic decays, as illustrated
in the left-right
supersymmetric case in Ref.~\cite{Frank:2000dw}.

\subsubsection{Simplified models with four charged lepton species}\label{sec:femodB}
Since the singly-charged component of the $\Psi$
and $\mathbf\Psi$ multiplets have the same quantum numbers as the charged leptons,
they could mix, as in $R$-parity
violating supersymmetric theories \cite{Akeroyd:1997iq} or in custodian tau models
\cite{delAguila:2010es}. This mixing is however highly constrained by LEP data,
by measurements of the muon anomalous magnetic
moment as well as by limits on leptonic flavor violating processes and on conversions in nuclei.
We therefore restrict the most general case to a $2\times 2$ mixing with
tau leptons,
\be
  \bpm e' \\ \mu' \\ \tau' \\ E' \epm = 
  \bpm e \\ \mu \\ \tau' \\ E' \epm = 
  \bpm 1 &0&0&0\\ 0&1&0&0\\ 0&0&c_\tau&s_\tau\\ 0&0&-s_\tau&c_\tau\epm
  \bpm e \\ \mu \\ \tau \\ \Psi^+ \epm
  \qquad\text{and}\qquad
  \bpm e'' \\ \mu'' \\ \tau'' \\ E'' \epm =
  \bpm e \\ \mu \\ \tau'' \\ E'' \epm = 
  \bpm 1 &0&0&0\\ 0&1&0&0\\ 0&0&c_\tau&s_\tau\\ 0&0&-s_\tau&c_\tau\epm
  \bpm e \\ \mu \\ \tau \\ \mathbf\Psi^+ \epm\  ,
\ee
where $c_\tau$ ($s_\tau$) denotes the cosine (sine) of the associated mixing angle.
In those notations, the $E'$ and $\tau'$ fields are related to lepton mixing with the singly-charged
component of the $SU(2)_L$ doublet $\Psi$ while the $E''$ and $\tau''$ fields to mixing with
the $SU(2)_L$ triplet ${\mathbf \Psi}$.
In contrast to the model constructed in Section \ref{sec:femodA}, the decays
of the new fields are possible via this mixing, without requiring the addition
of any extra particle to the theory.

Assuming mass-degenerate $\Psi$ states, the doubly-charged
component of the $SU(2)_L$ doublet always decays into a 
$W$-boson and a $\tau$ lepton, which further yields a final state with zero, one or two charged leptons. The associated
partial width can be computed from the Lagrangian of Eq.~\eqref{eq:Lfek} and is
\be
  \Gamma^{++}_{2, \tau W} = \frac{M_{\Psi^{++}}^3 \alpha s_\tau^2}{8 M_W^2 s_w^2} 
    \sqrt{\lambda(1,x^2_W,x^2_\tau)} \bigg[ \lambda(1,x^2_W,x^2_\tau) - 3 x_W^2 \Big(x_W^2 - (1-x_\tau)^2\Big)\bigg] \ .
\label{eq:decfemix1}\ee
Similarly, the $E'$ lepton can decay either to a $\tau Z$ or to a $\nu_\tau W$ final state, the corresponding widths being
\be\bsp
  \Gamma^+_{2, \tau Z} =&\ \frac{M_{E'}^3 \alpha s_\tau^2 c_\tau^2}{32 M_Z^2 c_W^2 s_W^2} 
    \sqrt{\lambda(1,x_Z^2,x_\tau^2)} 
     \bigg[ 5\lambda(1,x_Z^2,x_\tau^2) + 3 x_Z^2 \Big(5 (1 + x_\tau^2 - x_Z^2) - 8 x_\tau\Big) \bigg] \ ,  \\
  \Gamma^+_{2,\nu_\tau W} =&\ \frac{M_{E'}^3 \alpha s^2_\tau}{16 M_W^2 s_W^2} \Big[1-x_W^2\Big]^2 \Big[1+2x_W^2\Big]\ .
\esp\label{eq:decfemix2}\ee
After accounting for the decays of the Standard Model particles, other production mechanisms,
in addition to the one of a pair of doubly charged states (whose cross section is given by Eq.~\eqref{eq:xsecfeNC}),
can induce signatures with at
least three leptons. The two processes $q \bar q \to E'^+ E'^-$
and $q \bar q \to E'^{\pm} \tau^\mp$ hence possibly lead to final states containing up to six leptons,
the corresponding partonic cross sections being respectively given by
\be\bsp
 \hat\sigma_2^{E'E'} =&\
    \frac{4 \pi \alpha^2 \sh}{9} \sqrt{1 - 4x^2_{E'}} \bigg[ \big[1 + 2 x^2_{E'}\big] \frac{e^2_q}{\sh^2} -
    \frac{e_q (L_q+R_q)}{8c_W^2 s_W^2} \  \frac{\sh-M_Z^2}{ \sh |\sh_Z|^2}\ \big[1 + 2 x^2_{E'}\big] \big[2 + 4 s_W^2 - 3 s^2_\tau\big]\
      \\ &\ +
    \frac{(L_q^2+R_q^2)}{64 c_W^4 s_W^4} \  \frac{1}{ |\sh_Z|^2}\ \Big( \big[1 + 2 x^2_{E'}\big] \big[ 2(1+2 s_W^2)^2
      - 6 s_\tau^2 (1+2 s_W^2) \big] + s_\tau^4 (5+7x^2_{E'})\Big) \bigg]\ ,\\
  \hat\sigma_2^{E'\tau} = &\
    \frac{\pi \alpha^2 (L_q^2+R_q^2) s_\tau^2 c_\tau^2}{288 c_W^4 s_W^4} \frac{\sh}{|\sh_Z|^2}\ \sqrt{\lambda(1,x_\tau^2,x_{E'}^2)} 
    \Big[ 5\big( 3 - 3 x_{E'}^2 -3 x_\tau^2- \lambda(1,x_\tau^2,x_{E'}^2) \big) + 24 x_{E'} x_\tau\Big] \ ,
\esp\ee
whilst the production channels of an associated pair comprised of a singly and a doubly-charged particle, $u_i \bar d_j \to \Psi^{++} E'^-$
and $u_i \bar d_j \to \Psi^{++} \tau^-$, can yield up to five charged leptons.
In this case, the production cross sections are
\be\bsp
 \hat\sigma_2^{\Psi^{++}E'} = &\ 
   \frac{\pi \alpha^2 \sh c_\tau^2}{36 s_W^4 |\sh_W|^2}  |V_{ij}^{\rm CKM}|^2\ \sqrt{\lambda(1,x^2_{\Psi^{++}},x^2_{E'})} 
     \Big[ 1-(x_{\Psi^{++}}-x_{E'})^2\Big]   \Big[ 2+(x_{\Psi^{++}}+x_{E'})^2\Big] \ , \\
 \hat\sigma_2^{\Psi^{++} \tau} = &\ 
   \frac{\pi \alpha^2 \sh s_\tau^2}{36 s_W^4 |\sh_W|^2}  |V_{ij}^{\rm CKM}|^2\ \sqrt{\lambda(1,x^2_{\Psi^{++}},x^2_\tau)} 
     \Big[ 1-(x_{\Psi^{++}}-x_\tau)^2\Big]   \Big[ 2+(x_{\Psi^{++}}+x_\tau)^2\Big] \ .
\esp\ee

In the same way, all the components of the $\mathbf\Psi$ multiplet can decay to final states containing up to three leptons, after
accounting for subsequent decays of the gauge bosons and tau leptons. The corresponding partial widths are
\be\bsp
  \Gamma^{++}_{3,\tau W} = &\ 
    \frac{M_{\mathbf{\psi}^{++}}^3 \alpha s_\tau^2}{4 M_W^2 s_W^2}\sqrt{\lambda(1,x^2_W, x_\tau^2)}
      \Big[ \lambda(1,x^2_W, x_\tau^2) + 3 x_W^2 \big( (1-x_\tau)^2-x_W^2\big)\Big] \ ,  \\
  \Gamma^+_{3,\tau Z} = &\ 
    \frac{ M_{E''}^3 \alpha s_\tau^2 c_\tau^2}{32 M_Z^2 c_W^2 s_W^2} \sqrt{\lambda(1,x^2_Z, x_\tau^2)}
      \Big[ \lambda(1,x^2_Z, x_\tau^2) + 3 x_Z^2 ( 1+x_\tau^2-x_Z^2) \Big] \ ,  \\
  \Gamma^+_{3,\nu_\tau W} =&\ \frac{M_{E''}^3 \alpha s_\tau^2}{16  M_W^2 s_W^2} \Big[1-x_W^2\Big]^2 \Big[1 + 2 x_W^2\Big] \  , \\
  \Gamma^0_{3, \tau W} =&\  \frac{ M_{\mathbf{\psi}^0}^3 \alpha s_\tau^2}{4 M_W^2 s_W^2} 
    \sqrt{\lambda(1,x^2_W, x_\tau^2)}  \Big[ \lambda(1,x^2_W, x_\tau^2) + 3 x_W^2 \big( (1-x_\tau)^2-x_W^2\big)\Big] \ .
\esp\label{eq:decfemix3}\ee
Therefore, the production of any pair of components of the $\mathbf\Psi$ field can lead to signatures with three or more charged
leptons. While the partonic cross section related to the production of two doubly-charged particles is given
in Eq.~\eqref{eq:xsecfeNC}, all the other relevant cross sections, associated with the $u_i \bar d_j \to \mathbf{\Psi}^{++} E''^-$, 
$u_i \bar d_j \to \mathbf{\Psi}^{++} \tau^-$, $q\bar q\to E''^+ E''^-$, $q\bar q\to E''^\pm \tau^\mp$,
$u_i \bar d_j \to \mathbf{\Psi}^0 E''^+$, $u_i \bar d_j \to \mathbf{\Psi}^0 \tau^+$ and $q \bar q\to \mathbf{\Psi}^0 \mathbf{\Psi}^0$
modes, are respectively computed as
\be\bsp
 \hat\sigma_3^{\mathbf{\Psi}^{++}E''}  = &\ 
   \frac{\pi \alpha^2 \sh c_\tau^2}{18 s_W^4 |\sh_W|^2}  |V_{ij}^{\rm CKM}|^2\ \sqrt{\lambda(1,x^2_{\Psi^{++}},x^2_{E''})}
     \Big[ 1-(x_{\Psi^{++}}-x_{E''})^2\Big]   \Big[ 2+(x_{\Psi^{++}}+x_{E''})^2\Big] \ ,\\
  \hat\sigma_3^{\mathbf{\Psi}^{++}\tau} = &\ 
   \frac{\pi \alpha^2 \sh s_\tau^2}{18 s_W^4 |\sh_W|^2}  |V_{ij}^{\rm CKM}|^2\ \sqrt{\lambda(1,x^2_{\Psi^{++}},x^2_\tau)} 
     \Big[ 1-(x_{\Psi^{++}}-x_\tau)^2\Big]   \Big[ 2+(x_{\Psi^{++}}+x_\tau)^2\Big] \ ,\\
 \hat\sigma_3^{E'' E''} =&\ 
    \frac{4 \pi \alpha^2 \sh}{9} \sqrt{1 - 4x^2_{E''}} \bigg[ \big[1 + 2 x^2_{E''}\big] \frac{e^2_q}{\sh^2} -
    \frac{e_q (L_q+R_q)}{8c_W^2 s_W^2} \  \frac{\sh-M_Z^2}{ \sh |\sh_Z|^2}\ \big[1 + 2 x^2_{E''}\big] \big[4 s_W^2 - s_\tau^2\big]\
      \\& \ +
    \frac{(L_q^2+R_q^2)}{64 c_W^4 s_W^4} \  \frac{1}{ |\sh_Z|^2}\ \Big( \big[1 + 2 x^2_{E''}\big] \big[ 8 s_W^4
      - 4 s_\tau^2 s_W^2 \big] + s^4_\tau (1-x^2_{E''})\Big) \bigg]\ ,  \\
  \hat\sigma_3^{E''\tau} = & \
    \frac{\pi \alpha^2 (L_q^2+R_q^2)s_\tau^2 c_\tau^2}{288 c_W^4 s_W^4} \frac{\sh}{|\sh_Z|^2}\ \sqrt{\lambda(1,x_\tau^2,x_{E''}^2)} 
    \Big[ 3\big( 1 - x_{E''}^2 - x_\tau^2\big) - \lambda(1,x_\tau^2,x_{E''}^2) \Big] \ , \\
 \hat\sigma_3^{E'' \Psi^0} =&\ 
    \frac{\pi \alpha^2 \sh c_\tau^2}{18 s_W^4 |\sh_W|^2}  |V_{ij}^{\rm CKM}|^2\ \sqrt{\lambda(1,x^2_{\Psi^0},x^2_{E''})}
       \Big[ 1-(x_{\Psi^0}-x_{E''})^2\Big]   \Big[ 2+(x_{\Psi^0}+x_{E''})^2\Big] \ ,\\
 \hat\sigma_3^{\tau\Psi^0}  =&\ 
    \frac{\pi \alpha^2 \sh s_\tau^2}{18 s_W^4 |\sh_W|^2}  |V_{ij}^{\rm CKM}|^2\ \sqrt{\lambda(1,x^2_{\Psi^0},x^2_\tau)}
       \Big[ 1-(x_{\Psi^0}-x_\tau)^2\Big]   \Big[ 2+(x_{\Psi^0}+x_\tau)^2\Big] \ ,\\
 \hat\sigma_3^{\Psi^0\Psi^0} =&\ 
    \frac{\pi \alpha^2 (L_q^2+R_q^2)}{18 c_W^4 s_W^4} \frac{\sh}{|\sh_Z|^2}\ \sqrt{1-4x_{\mathbf{\Psi}^0}^2} 
    \Big[ 1 + 2 x_{\mathbf{\Psi}^0}^2 \Big] \ .
\esp\ee

\subsection{Spin $1$ doubly-charged particles}\label{sec:modve}
In this section, we adjoin to the Standard Model particle content additional complex vectorial fields $V$, $\cal V$ and $\mathbf V$ lying respectively in 
the singlet, fundamental and adjoint representation of $SU(2)_L$. The hypercharge quantum numbers are again set to $Y_V=2$, $Y_{\cal V} = 3/2$ and
$Y_{\mathbf{V}} = 1$, so that the component fields read
\be\label{eq:lkvec}
  V_\mu \equiv V_\mu^{++} \ , \qquad
  {\cal V}_\mu^i \equiv \bpm {\cal V}_\mu^{++} \\  {\cal V}_\mu^+ \epm \qquad\text{and}\qquad
   \mathbf V^i{}_j \equiv
      \bpm \frac{\mathbf V_\mu^+}{\sqrt{2}} & \mathbf V_\mu^{++}\\
           \mathbf V_\mu^0 & -\frac{\mathbf V_\mu^+}{\sqrt{2}}
  \epm \ . 
\ee
Gauge interactions and kinetic terms are described by means of a gauge-covariant version of the Proca Lagrangian
\be\bsp
  \lag_{\rm kin}= &\ -\frac12 \Big[ D_\mu V_\nu^\dag - D_\nu V_\mu^\dag \Big] \Big[ D^\mu V^\nu - D^\nu V^\mu \Big]
   - \frac12 \Big[ D_\mu {\cal V}_\nu^\dag - D_\nu {\cal V}_\mu^\dag \Big] \Big[ D^\mu {\cal V}^\nu - D^\nu {\cal V}^\mu \Big]
\\ &\ 
   - \frac12 \Big[ D_\mu {\mathbf V}_\nu^\dag - D_\nu {\mathbf V}_\mu^\dag \Big] \Big[ D^\mu {\mathbf V}^\nu - D^\nu {\mathbf V}^\mu \Big] + \ldots \ ,
\esp\ee
the dots being identified with omitted mass terms and the covariant derivatives being derived from Eq.~\eqref{eq:covder}. As in the
previous sections, we again forbid the new states to decay into each
other and model
the decays of the new vector fields into pairs of charged leptons via the Lagrangian
\be
  {\cal L}_{\rm dec} = \frac{\tilde g^{(1)}}{\Lambda} V_\mu\ \bar l^c_R \sigma^{\mu \nu} D_\nu l_R 
    + \tilde g^{(2)} {\cal V}_\mu^i\  \bar{L}^c_i \gamma^\mu  l_R
    + \frac{\tilde g^{(3)}}{\Lambda} (\mathbf V_\mu)^i{}_j \ \bar L^c_i \sigma^{\mu \nu} D_\nu L^j \ , 
\label{eq:ldvec}\ee
where the coupling strengths are denoted by $3\times 3$ matrices in flavor space $\tilde g^{(1)}$,  $\tilde g^{(2)}$
and $\tilde g^{(3)}$
and where $\sigma^{\mu \nu}=\frac{i}{4}[\gamma^\mu, \gamma^\nu]$. Gauge invariance makes use of the
standard gauge-covariant derivatives $D_\mu L$ and $D_\mu l_R$ as well as of the dual object $L_i$ defined in Eq.~\eqref{eq:deflep}.
In addition,
dimension-four operators such as $\bar l_R^c\gamma^\mu l_R$ identically vanish so that the use of higher-dimensional operators
suppressed by a new physics scale $\Lambda$ is required.

The Lagrangian of Eq.~\eqref{eq:lkvec} allows to pair produce the new doubly charged fields via quark-antiquark scattering
processes, $q\bar q\to V^{++} V^{--}$,  $q\bar q\to {\cal V}^{++} {\cal V}^{--}$ and  $q\bar q\to {\mathbf V}^{++} {\mathbf V}^{--}$,
which leads to a four leptons signature after accounting for the decays permitted by the interactions described
in Eq.~\eqref{eq:ldvec}. The associated partonic cross sections are calculated as
\be\bsp
 \hat\sigma^{NC}_1 = &\ \frac{4 \pi \alpha^2 \sh}{9} \sqrt{1 - 4x^2_{V^{++}}}\ \frac{1-x^2_{V^{++}}-12x^4_{V^{++}}}{x^2_{V^{++}}}\bigg[
    \frac{e^2_q}{\sh^2}  -
    \frac{ e_q (L_q+R_q) (\sh-M_Z^2)}{2 c_W^2 \sh |\sh_Z|^2} +
    \frac{L_q^2 + R_q^2}{8 c_W^4 |\sh_Z|^2} \bigg] \ , \\
 \hat\sigma^{NC}_2= &\ \frac{4 \pi \alpha^2 \sh}{9} \sqrt{1 - 4x^2_{{\cal V}^{++}}}\
    \frac{1-x^2_{{\cal V}^{++}}-12x^4_{{\cal V}^{++}}}{x^2_{{\cal V}^{++}}}\\ &\quad \times
    \bigg[\frac{e^2_q}{\sh^2}  +
    \frac{ e_q (1-4s_W^2)(L_q+R_q) (\sh-M_Z^2)}{8 s_W^2 c_W^2 \sh |\sh_Z|^2} +
    \frac{(1-4 s_W^2)^2(L_q^2 + R_q^2)}{128 s_W^4 c_W^4 |\sh_Z|^2} \bigg] \ , \\
 \hat\sigma^{NC}_3 = &\ \frac{4 \pi \alpha^2 \sh}{9} \sqrt{1 - 4x^2_{{\mathbf V}^{++}}}\
    \frac{1-x^2_{{\mathbf V}^{++}}-12x^4_{{\mathbf V}^{++}}}{x^2_{{\mathbf V}^{++}}}\\ &\quad \times
    \bigg[\frac{e^2_q}{\sh^2}  +
    \frac{ e_q (1-2s_W^2)(L_q+R_q) (\sh-M_Z^2)}{4 s_W^2 c_W^2 \sh |\sh_Z|^2} +
    \frac{(1-2 s_W^2)^2(L_q^2 + R_q^2)}{32 s_W^4 c_W^4 |\sh_Z|^2} \bigg] \ , 
\esp\ee
respectively. In the case of fields lying in the fundamental or adjoint representation of $SU(2)_L$, three
lepton final states also result from the production of an associated pair comprised of one
doubly-charged state and one singly charged state, $u_i \bar d_j \to {\cal V}^{++} {\cal V}^-$
and  $u_i \bar d_j \to {\mathbf V}^{++} {\mathbf V}^-$ whose cross sections respectively are
\be\bsp
 \hat\sigma^{CC}_2 = &\ 
   \frac{\pi \alpha^2 \sh}{288 s_W^4 |\sh_W|^2}  |V_{ij}^{\rm CKM}|^2\ \sqrt{\lambda(1,x^2_{{\cal V}^{++}},x^2_{{\cal V}^+})} 
     \frac{\big[ 1-(x_{{\cal V}^{++}}-x_{{\cal V}^+})^2\big]\big[ 1-(x_{{\cal V}^{++}}+x_{{\cal V}^+})^2\big]}{x_{{\cal V}^{++}}^2 x^2_{{\cal V}^+}} 
\\ &\quad
   \times \Big[ \lambda(1,x^2_{{\cal V}^{++}},x^2_{{\cal V}^+}) - 1 +4 (x_{{\cal V}^+}^2 + x_{{\cal V}^{++}}^2 + 3 x_{{\cal V}^+}^2 x_{{\cal V}^{++}}^2)\Big]
   \ , \\ 
 \hat\sigma^{CC}_3 = &\ 
   \frac{\pi \alpha^2 \sh}{144 s_W^4 |\sh_W|^2}  |V_{ij}^{\rm CKM}|^2\ \sqrt{\lambda(1,x^2_{{\mathbf V}^{++}},x^2_{{\mathbf V}^+})} 
     \frac{\big[ 1-(x_{{\mathbf V}^{++}}-x_{{\mathbf V}^+})^2\big]\big[ 1-(x_{{\mathbf V}^{++}}+x_{{\mathbf V}^+})^2\big]}{x_{{\mathbf V}^{++}}^2 x^2_{{\mathbf V}^+}} 
\\ &\quad
   \times \Big[ \lambda(1,x^2_{{\mathbf V}^{++}},x^2_{{\mathbf V}^+}) - 1 +4 (x_{{\mathbf V}^+}^2 + x_{{\mathbf V}^{++}}^2 
      + 3 x_{{\mathbf V}^+}^2 x_{{\mathbf V}^{++}}^2)\Big] \ .
 \esp\ee
The partial widths associated with the decays of the states above into charged leptons are finally given by
\be\bsp
  \Gamma_{1,\ell}^{++} = &\ \frac{M_{V^{++}} M_\ell^2 (\tilde g^{(1)})^2}{96 \pi \Lambda^2} 
    \Big[ 1-4 x_\ell^2\Big]^{3/2}\ , \\
  \Gamma_{2,\ell}^{++} = &\ \frac{M_{{\cal V}^{++}} (\tilde g^{(2)})^2}{24 \pi} \Big[1 -4 x_\ell^2\Big]^{3/2}\ , \\
  \Gamma_{3,\ell}^{++} = &\ \frac{M_{\mathbf{V}^{++}} M_\ell^2 (\tilde g^{(3)})^2}{96 \pi \Lambda^2} 
    \Big[ 1-4 x_\ell^2\Big]^{3/2}\ , \\
  \Gamma_{2,\ell}^+ = &\ \frac{M_{{\cal V}^+} (\tilde g^{(2)})^2}{48 \pi}\Big[2 - 3 x_\ell^2 + x_\ell^6\Big] \ , \\
  \Gamma_{3,\ell}^+ = &\ \frac{M_{{\mathbf V}^+} M_\ell^2 (\tilde g^{(3)})^2}{384 \pi \Lambda^2}\Big[1-x_\ell^2\Big]^2 \Big[2 +  x_\ell^2\Big] \ .
\esp\label{eq:vecwidths}\ee
As in the previous sections, flavor violating decays have not been
considered as the underlying interactions could lead to visible
signals in rare leptonic decay experiments.

\section{Multilepton production at the LHC}\label{sec:numerics}

In this section, we present numerical predictions of total cross
sections for the production of multileptonic final states originating from
the pair production and decay of the various new particles introduced in Section \ref{sec:themodel}.
For the sake of simplicity, we assume that all new physics
masses are equal. We focus on $pp$ collisions such as produced
at the CERN LHC collider running at a center-of-mass energy $\sqrt{S_h}=8$ TeV. Thanks
to the QCD factorization theorem, hadronic cross sections are calculated
by convolving the partonic cross sections
$\hat\sigma$ computed in Section\ \ref{sec:themodel} with the universal parton
densities $f_a$ and $f_b$ of partons $a$ and $b$ in the proton,
which depend on the longitudinal momentum fractions of the two partons
$x_{a,b} = \sqrt{\tau}e^{\pm y}$ and on the unphysical factorization scale
$\mu_F$,
\be
 \sigma =
 \int_{4\tilde M^2/S_h}^1\!\d\tau
 \int_{-1/2\ln\tau}^{1/2\ln\tau}\d y\
 f_a(x_a,\mu_F^2) \ f_b(x_b,\mu_F^2) \ \hat\sigma(x_a x_b S_h)  \ .
\ee
We employ the leading order set L1 of the CTEQ6 global
parton density fit \cite{Pumplin:2002vw}, which includes five light
quark flavors and the gluon
and we identify the factorization
scale to be the average mass of the produced final state
particles $\tilde M$.

For the masses and widths of the electroweak gauge bosons, we use the
current values given in the Particle Data Group review \cite{Beringer:2012zz},
\ie, $M_Z=91.1876$ GeV, $M_W=80.385$
GeV, $\Gamma_Z=2.4952$ GeV and $\Gamma_W=2.085$ GeV, and we evaluate the
CKM-matrix elements by using the Wolfenstein parameterization.
The corresponding four free parameters are set to
$\lambda=0.22535$, $A=0.811$, $\bar\rho=0.131$ and $\bar\eta=0.345$.
The squared sine of the electroweak mixing angle $\sin^2\theta_W=1-
M_W^2/M_Z^2$ and the electromagnetic fine structure constant $\alpha=
\sqrt{2} G_F M_W^2\sin^2\theta_W/\pi$ are calculated in the improved
Born approximation using a value of $G_F=1.16638\cdot
10^{-5}$ GeV$^{-2}$ for the Fermi coupling constant, which is derived from muon
lifetime measurements.

\begin{figure}[!t]
  \centering
  \includegraphics[width=.32\columnwidth]{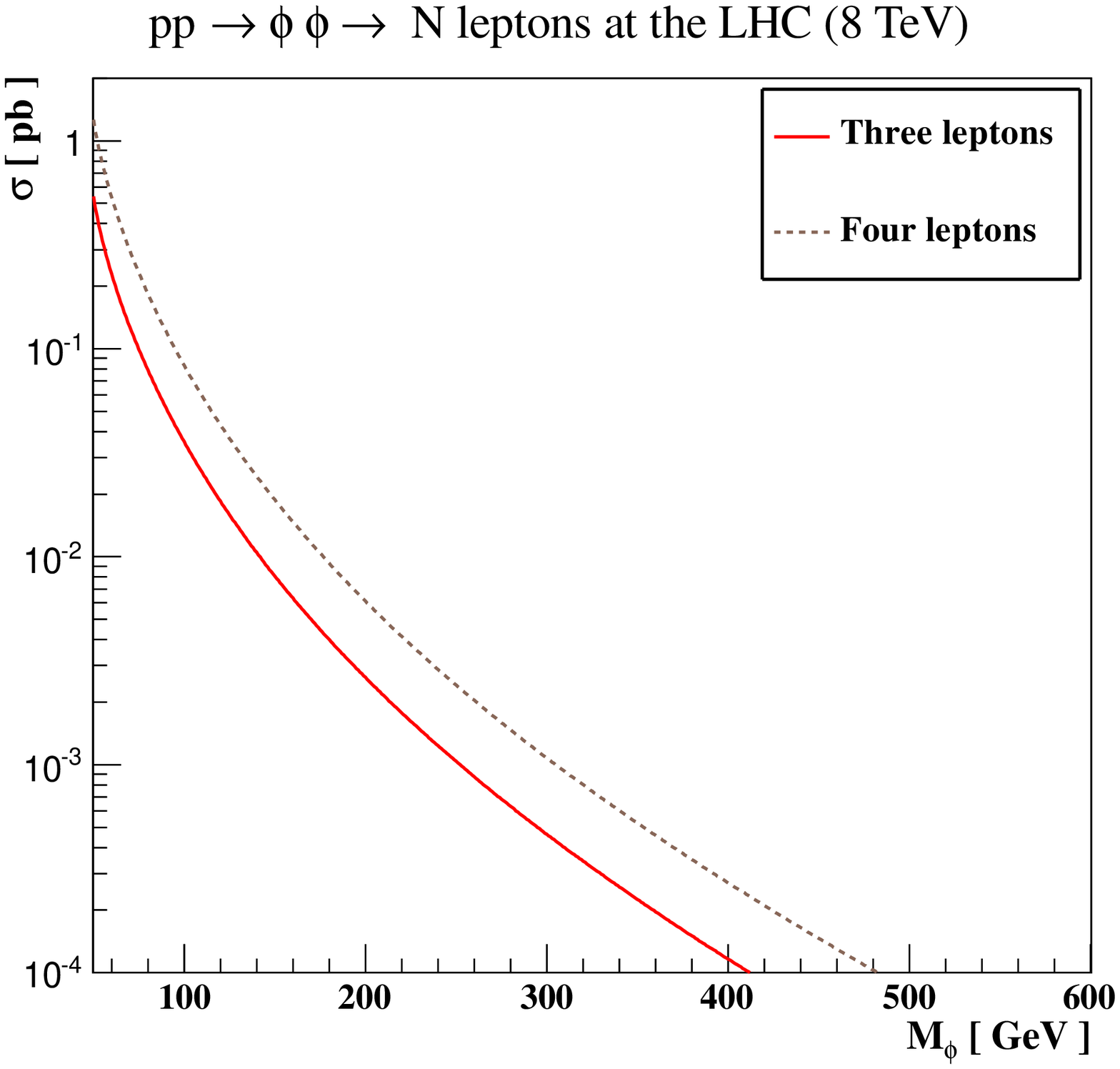}
  \includegraphics[width=.32\columnwidth]{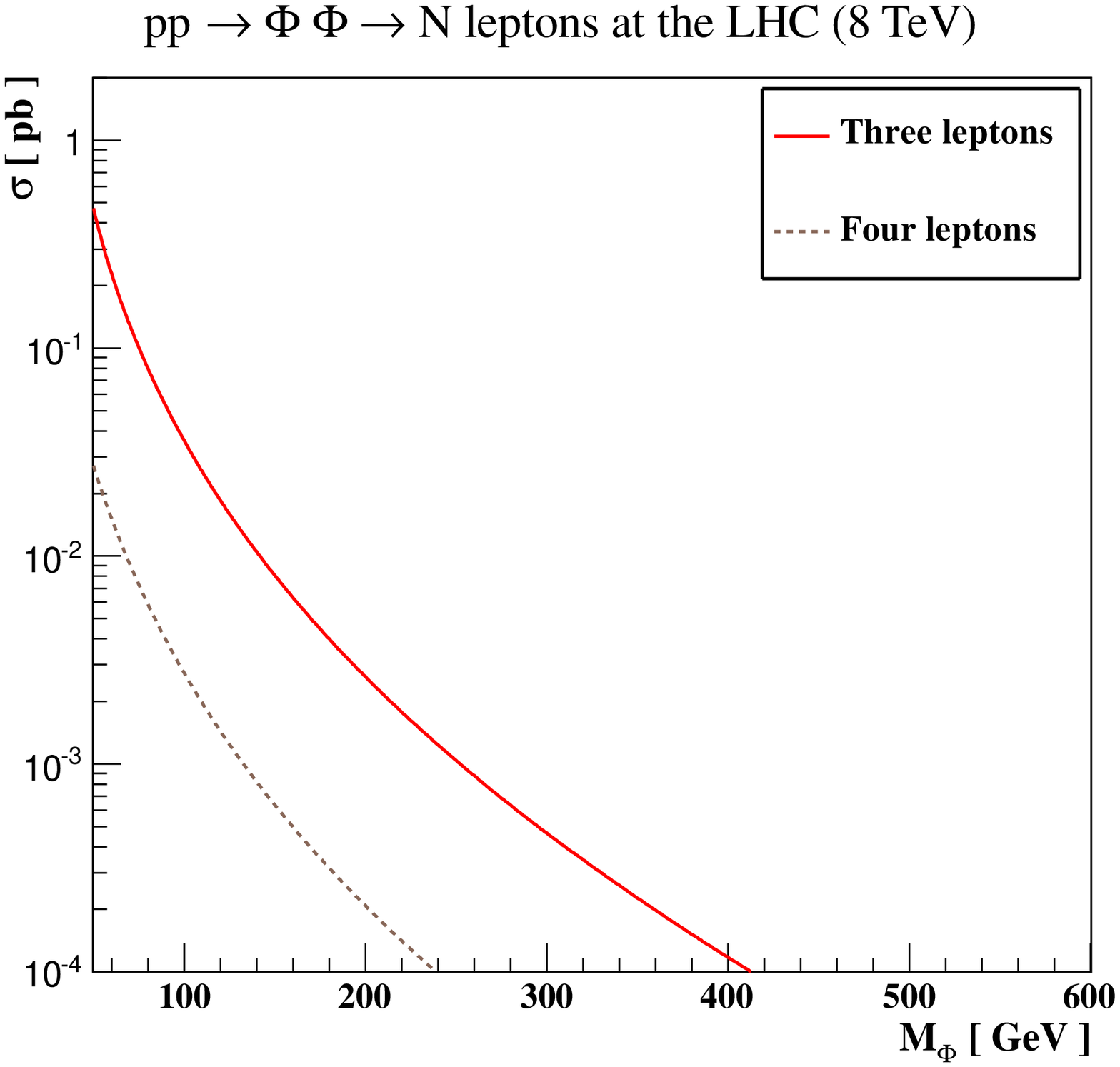}
  \includegraphics[width=.32\columnwidth]{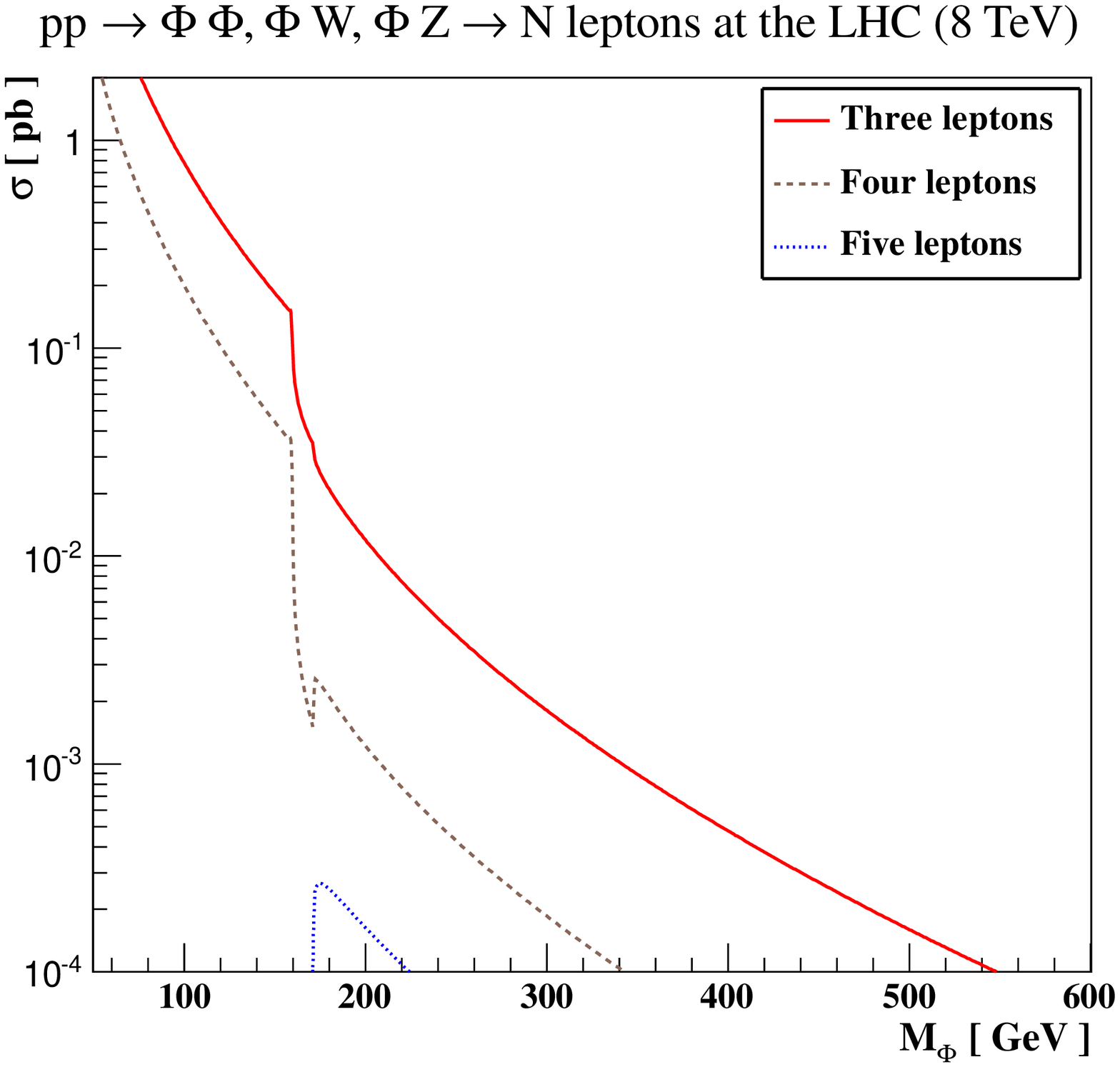}
  \caption{\label{fig:scalar}Doubly-charged particle contributions to the
    production rate of final states containing more than three charged
    leptons at the LHC, running at a center-of-mass energy of 8 TeV. We consider non-standard fields
    lying in the singlet (left),
    doublet (center) and triplet (right) representations of $SU(2)_L$ and impose that its component
    with the highest electric charge to be doubly-charged.
    }
\end{figure}

We start by focusing on mechanisms yielding
the production of pairs of components of the $SU(2)_L$ multiplets
$\phi$, $\Phi$ and $\mathbf\Phi$ introduced in Section \ref{sec:modsc}.
In the first step, we evaluate the different branching ratios of their doubly-charged, singly-charged and neutral
states to a given number of electrons, muons and taus by using
Eq~\eqref{eq:BRsc++1}, Eq.~\eqref{eq:BRsc++2},
Eq.~\eqref{eq:BRsc+1} and Eq.~\eqref{eq:BRsc+2}. We then derive,
in Figure~\ref{fig:scalar}, BSM contributions to the
production rates of final states containing $N_\ell=3$,
$N_\ell = 4$ and
$N_\ell = 5$ charged leptons\footnote{From now on, we denote
by the terminology \textit{charged leptons}, electrons and muons.}
originating from the production and decay of the new scalar particles after
accounting for subsequent decays of tau leptons, $W$-bosons and $Z$-bosons
where relevant. For our numerical analysis, we choose to couple
the $\phi$, $\Phi$ and $\mathbf\Phi$ fields to charged leptons in a
flavor-conserving way, as already mentioned
in the previous section of this paper,
and we fix the parameters of the Lagrangian of Eq.~\eqref{eq:Lscy} to
\be
  y^{(1)} =  y^{(2)} =   y^{(3)} = 0.1 \cdot \mathbb{1}\ .
\ee
Moreover, we consider the effective interactions
of the doublet fields to be suppressed by an energy scale  $\Lambda = 1$~TeV and allow the neutral
component of the triplet field $\mathbf\Phi^0$
to acquire a non-vanishing vacuum expectation value
equal to $v_{\mathbf\Phi} = 100$~GeV. Finally, all new particles are assumed mass-degenerate.

On the left panel of the figure,
we present results for a field lying in the singlet representation
of $SU(2)_L$. The unique component of such a field, \ie,
the doubly charged particle $\phi^{++}$, always decays into
a charged lepton pair or a tau pair. Four lepton signatures are
therefore expected to be copiously
produced, although the possible hadronic decays of the tau
allow for contributions to final states containing less than four electrons and muons.
Consequently, doubly-charged particles lying in the singlet
representation of $SU(2)_L$ contribute, for moderate masses
$M_\phi \lesssim 330$~GeV, to the production of final states with a
leptonic multiplicity $N_\ell \geq 3$, with a cross section larger than 1~fb.

\begin{figure}
\centering
  \includegraphics[width=.32\columnwidth]{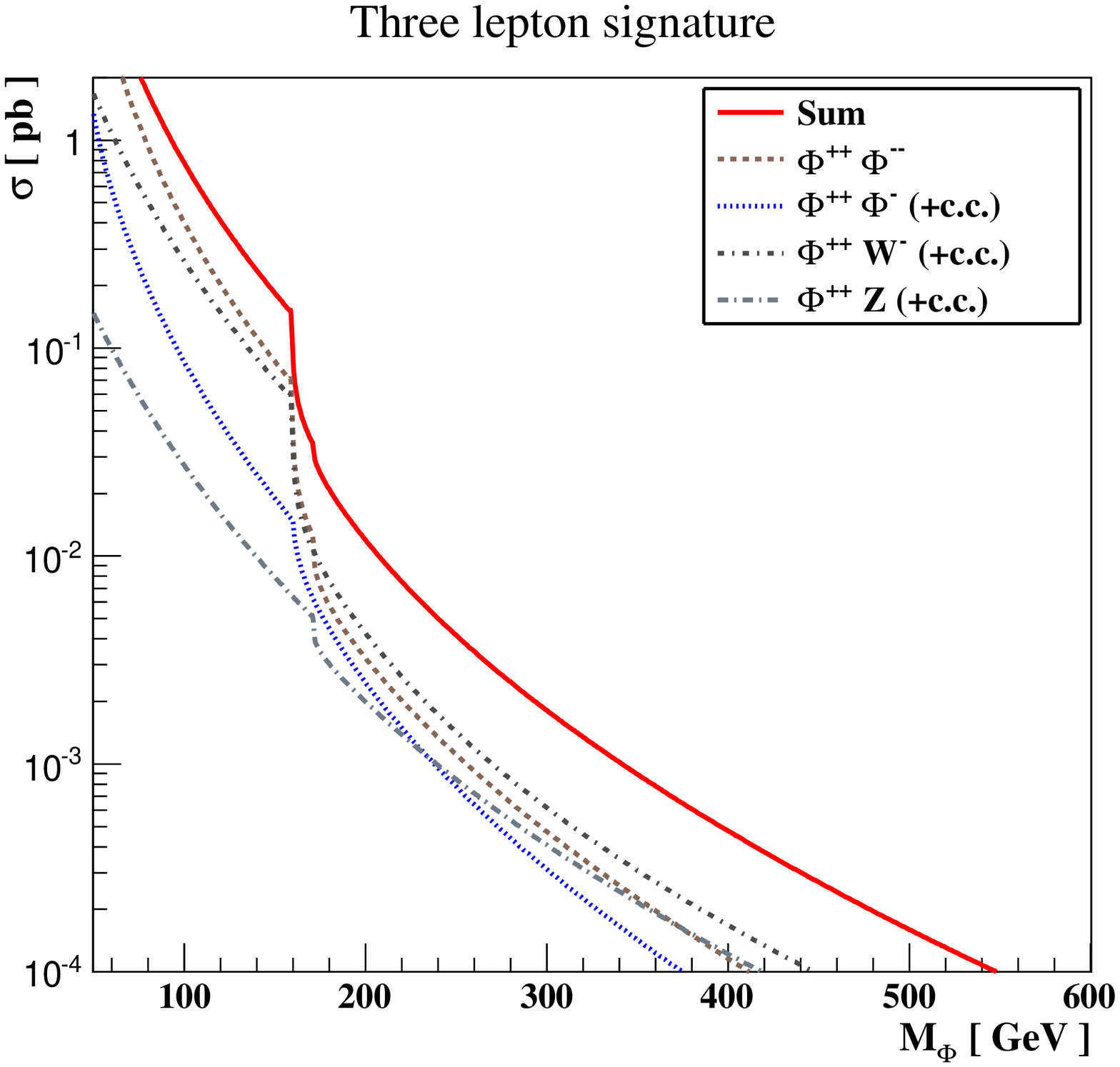}
  \includegraphics[width=.32\columnwidth]{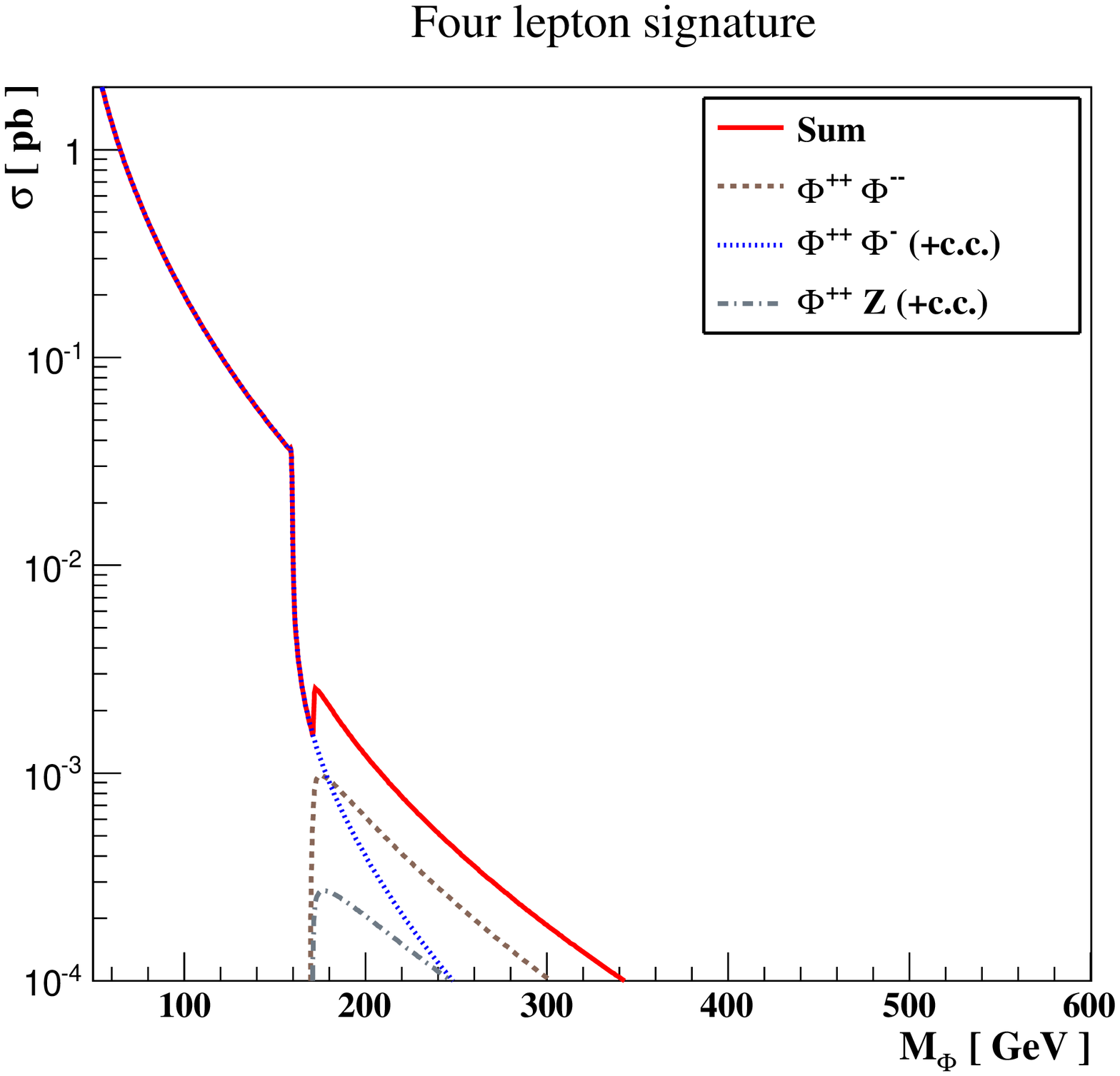}
  \caption{\label{fig:NLscalartri}
    Different new physics contributions to the production of three (left) and four
    (right) leptons at the LHC,
    running at a center-of-mass energy of 8 TeV, in the context of a simplified model containing
    one additional $SU(2)_L$ triplet of scalar fields whose highest component is a doubly-charged state.
    }
\end{figure}
The situation changes when the Standard Model is extended by fields lying
in a non-trivial representation of $SU(2)_L$ due to the possibility of production of associated pairs of
different new states or of a single new state together with a Standard Model gauge boson.
This is illustrated on the central and right panels of Figure~\ref{fig:scalar} for the doublet
and triplet cases, respectively. For fields lying in the fundamental representation of $SU(2)_L$,
the BSM contributions to the production cross section
of final states with $N_\ell \geq 3$ charged leptons
are found to be
larger than 1~fb only for a rather low mass scale $M_\Phi \lesssim 250$~GeV.
For fields in the adjoint representation, the
predictions however highly depend on the size of the vev $v_{\mathbf\Phi}$.
For very small $v_{\mathbf \Phi}$ values, the
cross sections are sizable and a large mass reach can be foreseen, as extrapolated from the results
presented in the region of the right panel of Figure \ref{fig:scalar} located
below the dibosonic thresholds, where the value of
the vev is irrelevant.
However, as soon as the dibosonic decay modes of the $\mathbf\Phi^+$ and
$\mathbf\Phi^{++}$ fields are open, important vev values render them dominant
so that the production rate of final states with more than three charged leptons drops,
the Standard Model weak bosons preferably decaying into quarks. In our example, where
the vev has a value close to the weak scale, the new physics masses have to be below 350~GeV
in order to contribute to the production cross sections of multileptonic final
states (with $N_\ell\geq 3$) with at least $\sigma = 1$~fb.
We show in detail the effects of a large $v_{\mathbf\Phi}$ in Figure~\ref{fig:NLscalartri},
where we split the different production channels contributing to final state signatures with
$N_\ell=3$ (left panel) and
$N_\ell=4$ (right panel) charged leptons and present the respective contribution of each subprocess.

In addition,
it should be noted that final states with five and six leptons can also be produced, since both
the doubly-charged and singly-charged fields can decay into three leptons after accounting
for leptonic $W$-boson and $Z$-boson decays. However, the related rates
are very low and render the possible observation of five-lepton
or six-lepton events rather unlikely when one takes into account the available integrated luminosity
of 20 fb$^{-1}$ recorded at the LHC in 2012.

\begin{figure}[!t]
  \centering
  \includegraphics[width=.32\columnwidth]{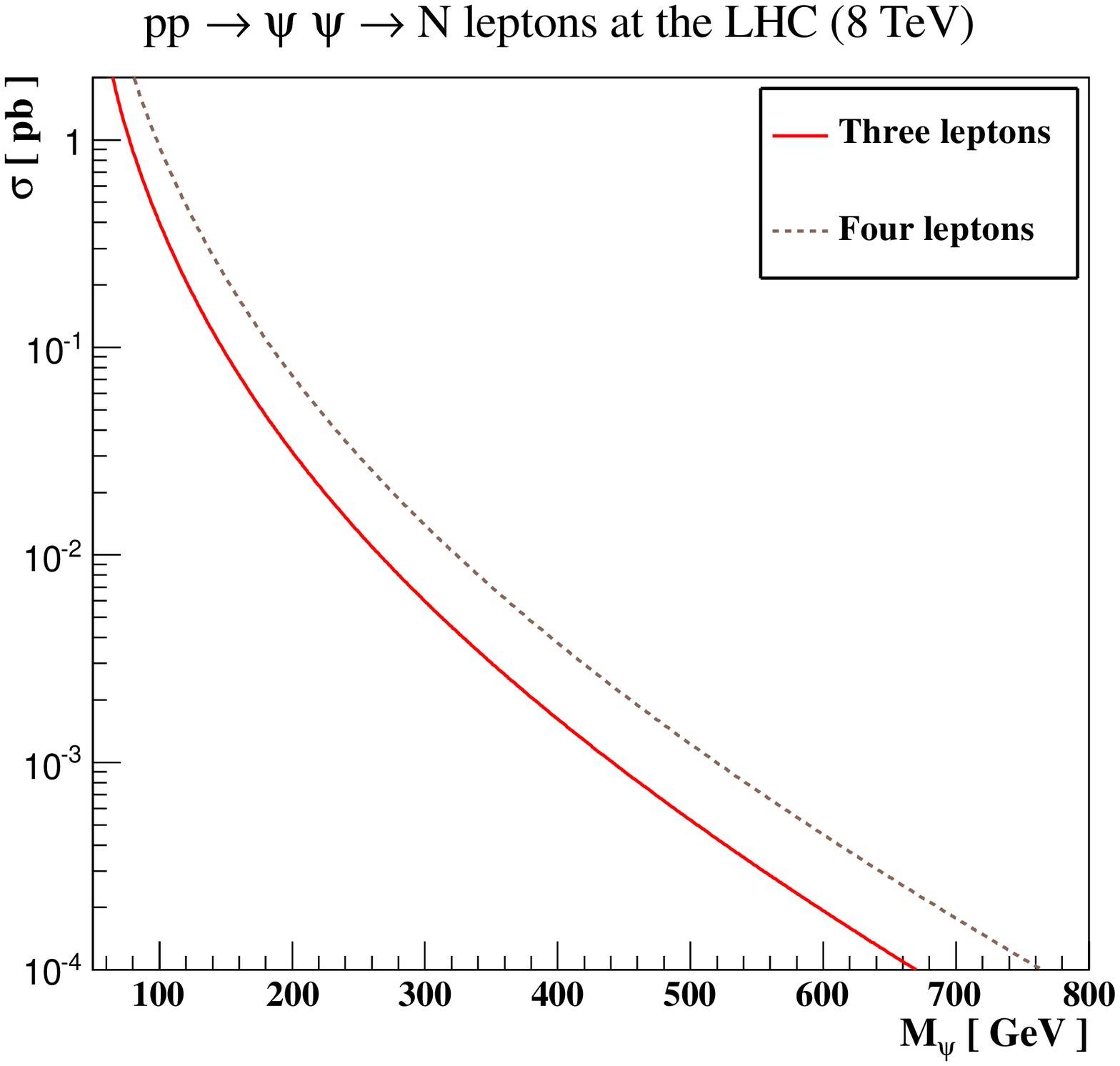}
  \includegraphics[width=.32\columnwidth]{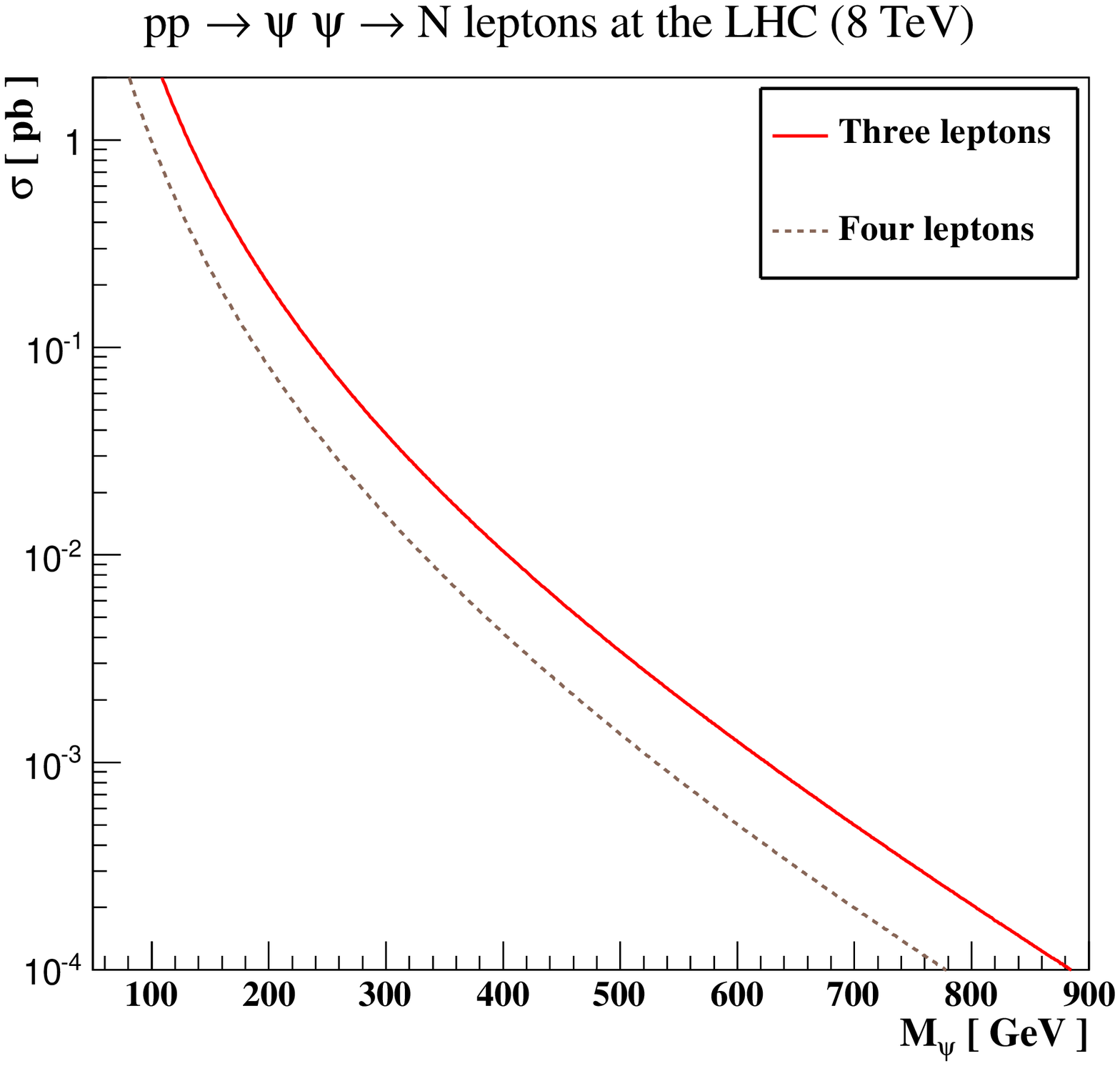}
  \includegraphics[width=.32\columnwidth]{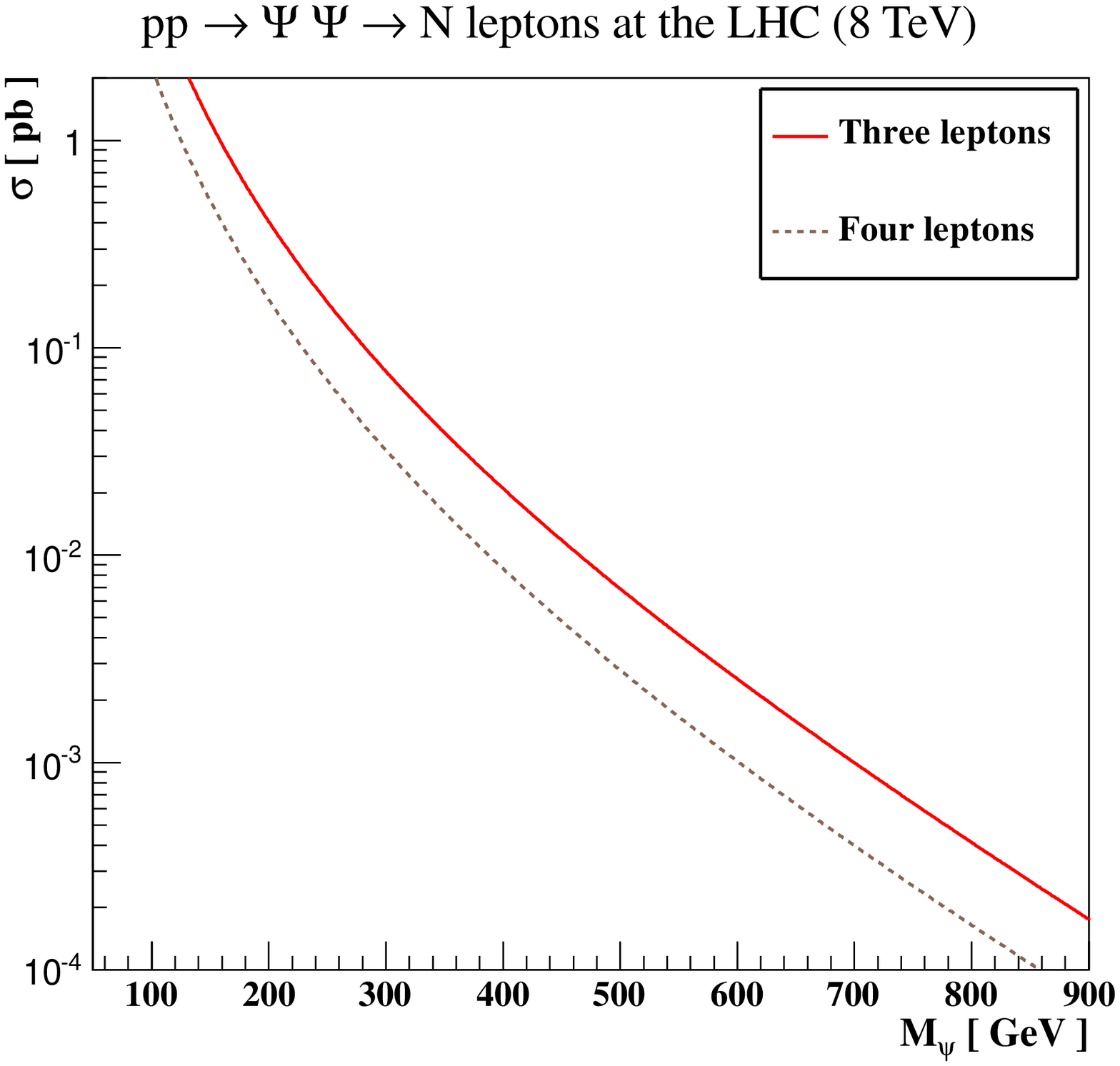}
  \caption{\label{fig:fermionA}Doubly-charged particle production rate of final states containing more than three
   charged
    leptons at the LHC, running at a center-of-mass energy of 8 TeV. We consider
    extensions of the Standard Model with an extra fermionic field lying in the singlet (left),
    doublet (center) or triplet (right) representation of $SU(2)_L$ and impose
    that its component with the highest electric charge to be doubly-charged.
    In addition, its singly-charged component is prevented from mixing with the SM sector.
    }
\end{figure}

We now turn to the pair-production of the components of the fermionic fields $\psi$,
$\Psi$ and $\mathbf\Psi$ introduced in Section \ref{sec:femod} and calculate their effect
on the production of final states containing three leptons or more at the LHC.
We investigate first scenarios with three generations of fermions as presented in Section \ref{sec:femodA}.
In this case, the new states decay through a four-fermion interaction into a pair of leptons with
the same electric charge and a fermionic field $N$ similar to a right-handed neutrino,
as described by Eq.~\eqref{eq:LfeF}. In order to compute the related partial widths,
we implement the Lagrangian of
Eq.~\eqref{eq:LfeF} into the {\sc FeynRules}\ package \cite{Christensen:2008py,%
Christensen:2009jx,Christensen:2010wz,Duhr:2011se,Fuks:2012im,Alloul:2013fw}, then
export the Feynman rules to a UFO module \cite{Degrande:2011ua} that is subsequently linked to
{\sc MadGraph}~5 \cite{Alwall:2011uj}. For our numerical analysis, we set the effective scale
$\Lambda = 1$~TeV and impose the coupling strengths to be flavor-diagonal,
\be
 G^{(1,1)} = G^{(1,2)} = G^{(2,1)} = G^{(2,2)} = G^{(3,1)} = G^{(3,2)} = 0.1 \cdot \mathbb{1} \ .
\ee
We also set the mass of the $N-$field to 50 GeV, in order to be compatible
with the current limits on the existence of heavy stable neutral leptons \cite{Beringer:2012zz} and
assume, for the sake of simplicity, that all components of the new fermionic multiplet
containing the doubly-charged state have the same mass.
In Figure~\ref{fig:fermionA}, we derive the associated contributions to the production of
final states with at least three charged leptons in the singlet (left panel), doublet (central panel)
and triplet (right panel) cases. In contrast to scalar extra fields, we observe that
a much larger mass range is expected to give rise to cross sections larger than 1~fb, which
is guaranteed for $M_\psi \lesssim 550$~GeV, $M_\Psi\lesssim 650$~GeV and $M_{\mathbf\Psi}\lesssim 725$~GeV
for new fermionic fields lying in
the singlet, doublet and triplet representations of $SU(2)_L$, respectively.

\begin{figure}[!t]
  \centering
  \includegraphics[width=.32\columnwidth]{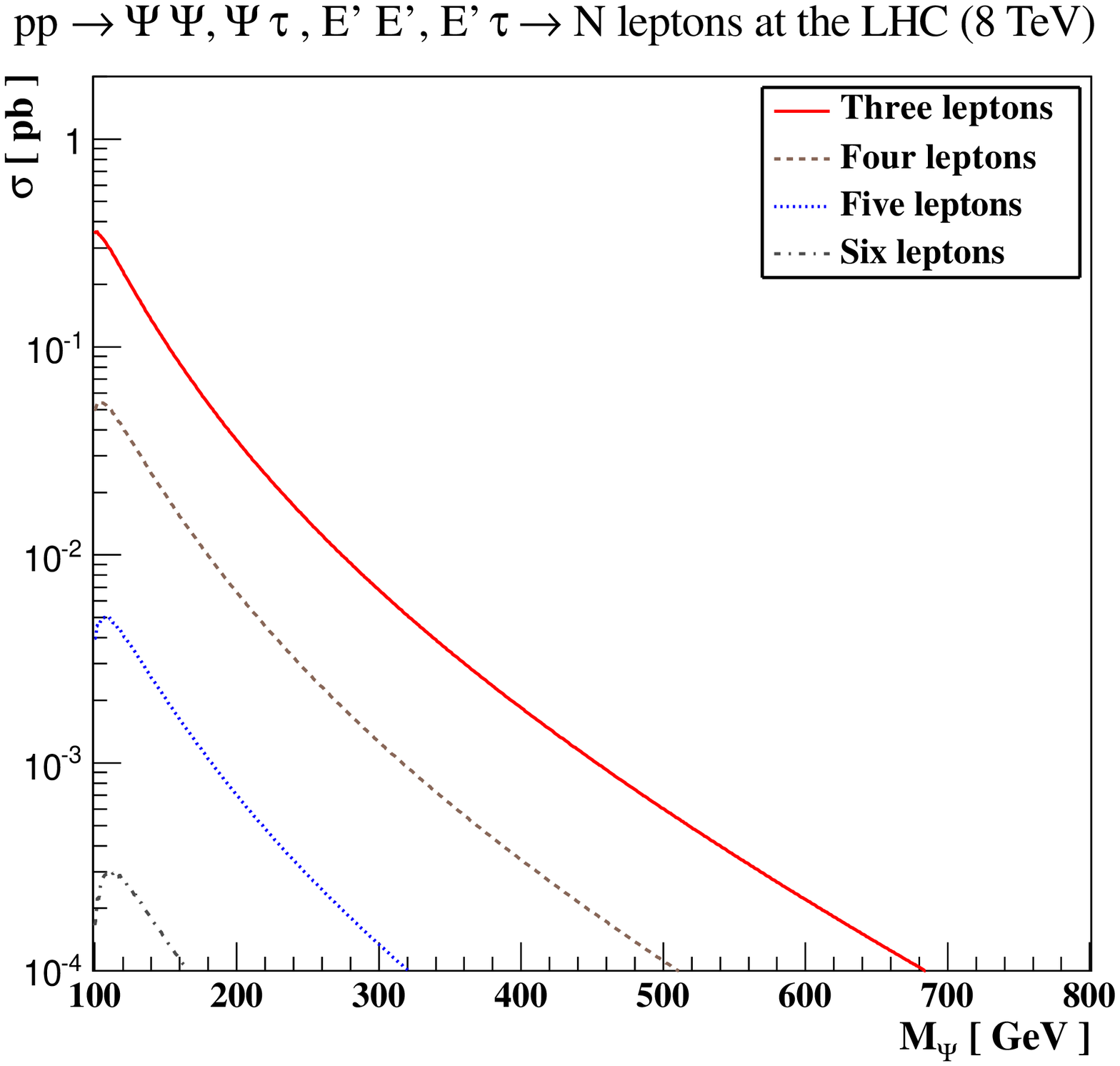}
  \includegraphics[width=.32\columnwidth]{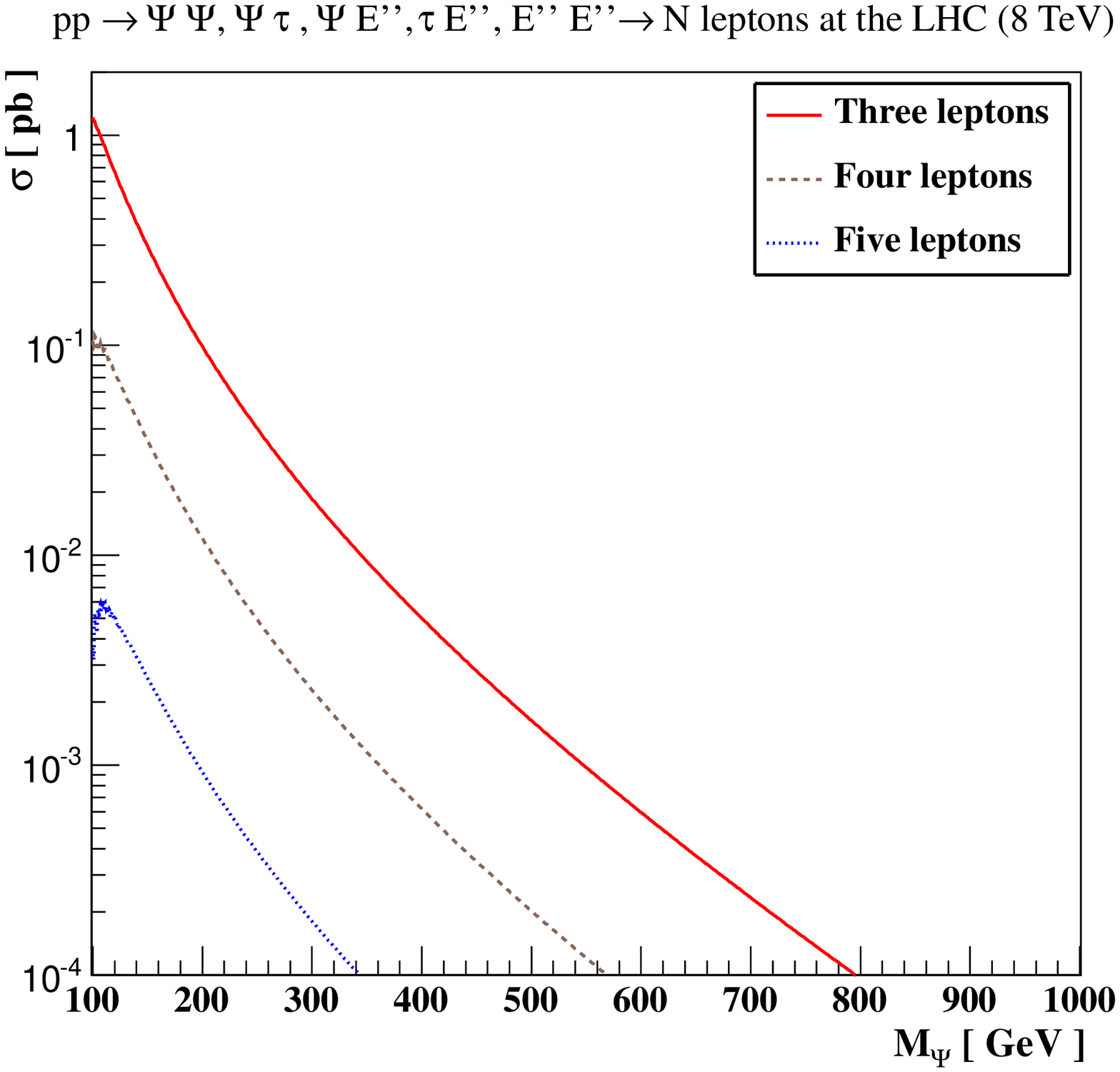}
  \caption{\label{fig:fermionB}Doubly-charged particle production rate of final states containing more than three
     charged
     leptons at the LHC, running at a center-of-mass energy of 8 TeV. We consider
     extensions of the Standard Model with an extra fermionic field lying in the singlet (left),
     doublet (center) or triplet (right) representation of $SU(2)_L$ and impose
     its component with the highest electric charge ti be doubly-charged.
     In addition, its singly-charged component mixes with the SM tau lepton.
  }
\end{figure}

In the series of fermionic scenarios presented in
Section~\ref{sec:femodB}, the Standard Model is only supplemented by a fermionic
field lying either in the fundamental or in the adjoint representation of $SU(2)_L$. In this case,
the singly-charged component mixes with the Standard Model tau lepton, which allows for the decays
of the new states to the Standard Model sector
and open new production mechanisms giving rise to signatures with 
$N_\ell \geq 3$ charged leptons. In order not to challenge the very precisely measured properties of the tau
lepton inferred by its coupling to the $Z$-boson, we fix the mixing angle
\be
  \sin\theta_\tau = 0.01 \ .
\ee
Computing the different branching ratios using the formulas of Eq.~\eqref{eq:decfemix1},
Eq.~\eqref{eq:decfemix2} and Eq.~\eqref{eq:decfemix3} we then show in Figure~\ref{fig:fermionB}
the new physics contributions to the production of final states with at least three leptons.
It is found that for masses $M_\Psi \lesssim 470$~GeV and $M_{\mathbf\Psi} \lesssim 550$~GeV
in the doublet and triplet cases, respectively, the corresponding cross sections are higher than 1~fb.

\begin{figure}[!t]
  \centering
  \includegraphics[width=.32\columnwidth]{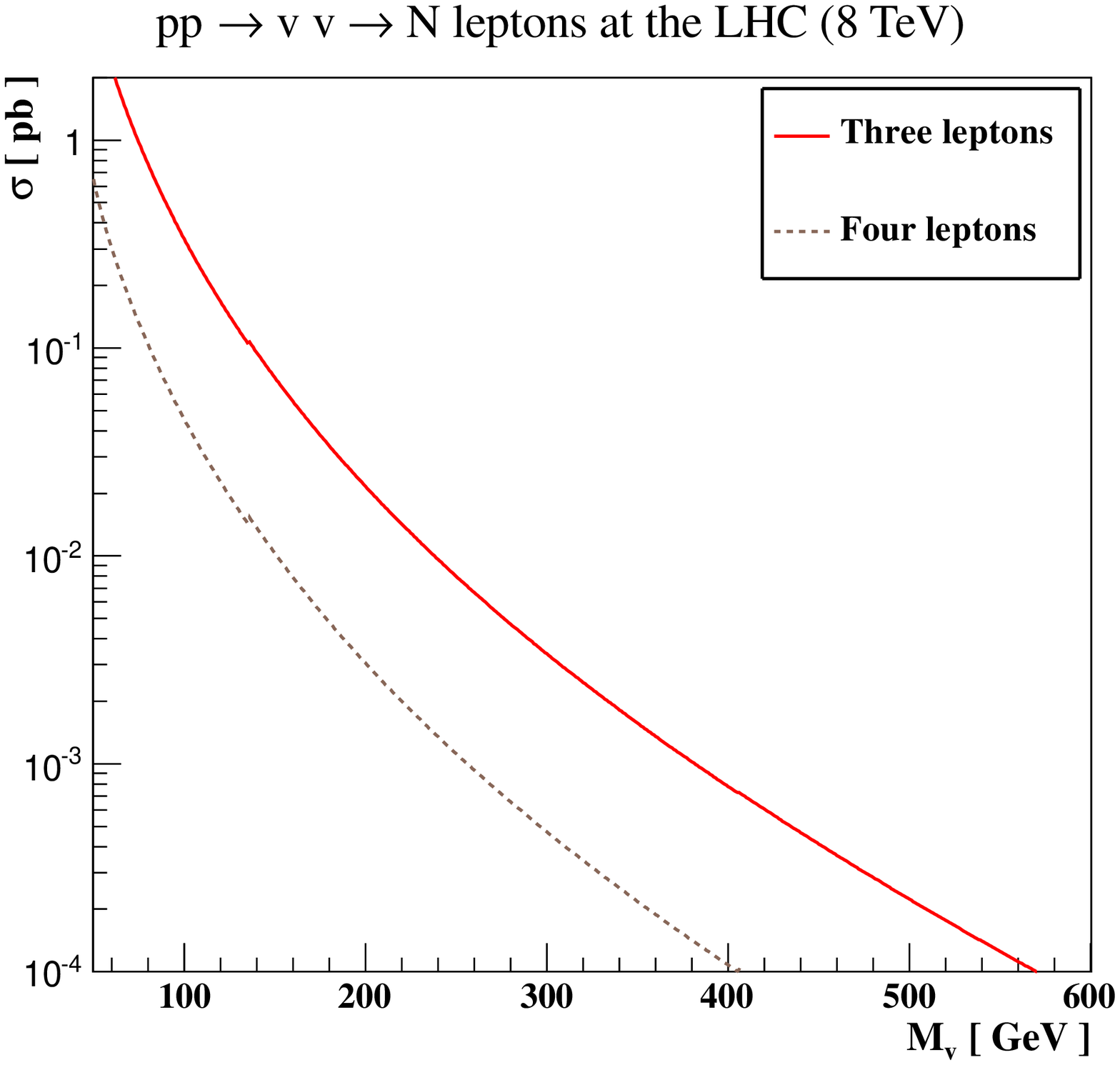}
  \includegraphics[width=.32\columnwidth]{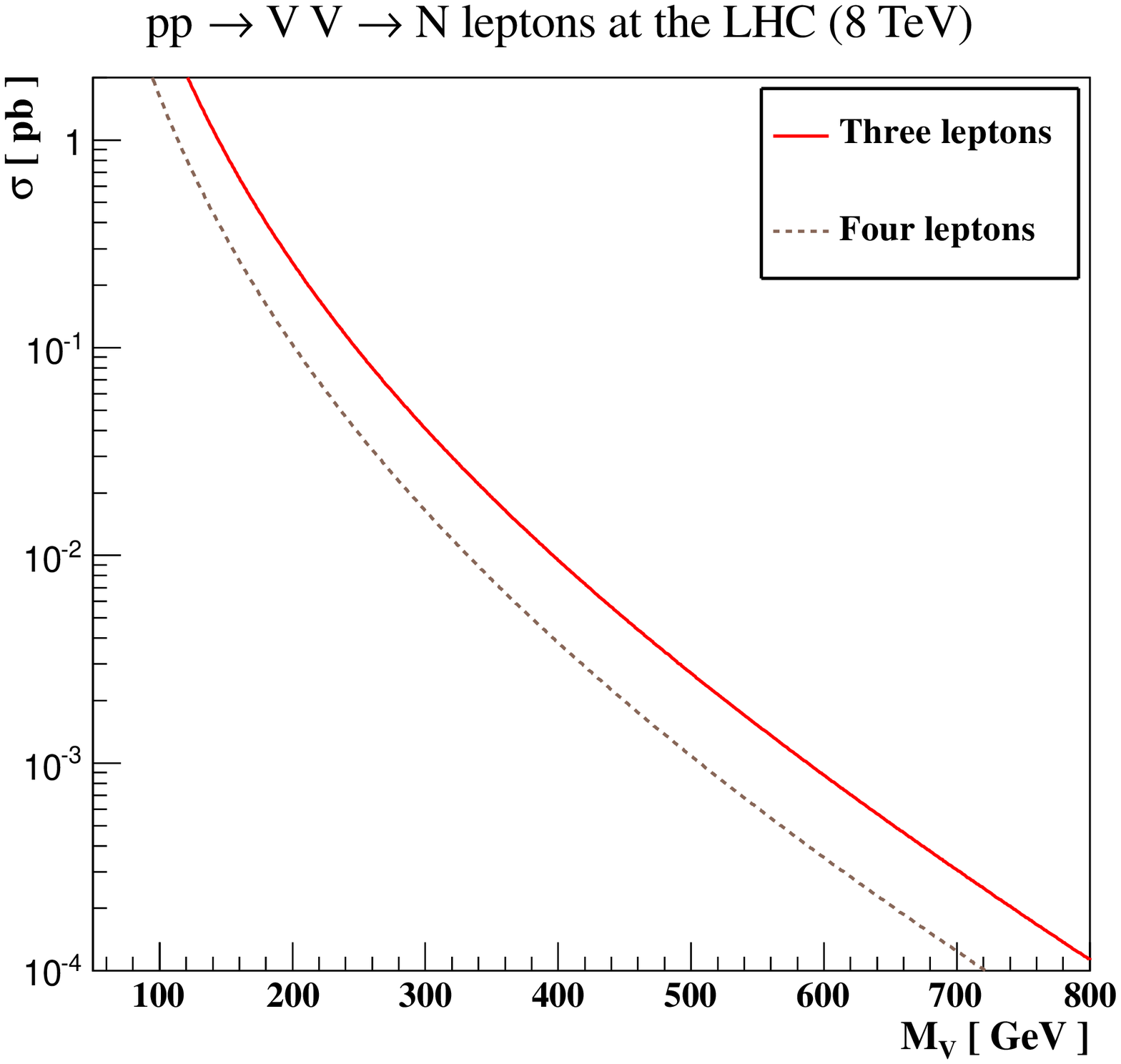}
  \includegraphics[width=.32\columnwidth]{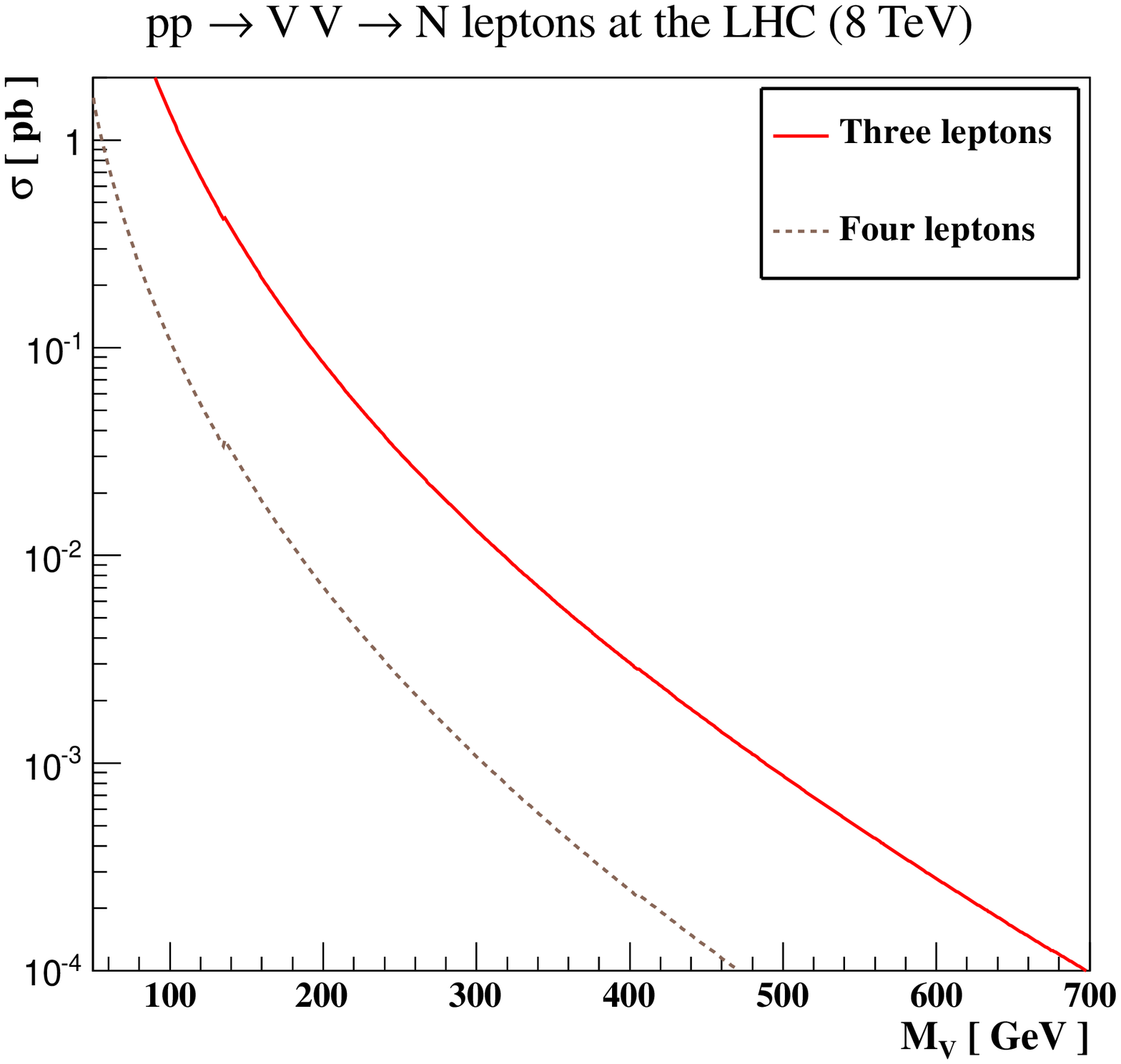}
  \caption{\label{fig:vector}Doubly-charged particle production rate of final states containing more than three
    charged leptons at the LHC, running at a center-of-mass energy of 8 TeV. We consider
     extensions of the Standard Model with an extra vectorial field lying in the singlet (left),
     doublet (center) or triplet (right) representation of $SU(2)_L$ and impose that
     its component with the highest electric charge to be doubly-charged.
     }
\end{figure}

We finally address the production of the vectorial multiplets
$V$, $\cal V$ and $\mathbf V$ defined in Section~\ref{sec:modve}. These fields lie
in the trivial, fundamental and adjoint representation of $SU(2)_L$, respectively, and
their component with the highest electric charge is doubly-charged.
In order to derive the associated contributions to the production of signatures with three leptons
and more, we start by computing the different partial widths of the new fields according to the
results of Eq.~\eqref{eq:vecwidths}. As for the other types of models, we choose the coupling
strengths of the new field to the Standard Model leptons and neutrinos, presented in Eq.\eqref{eq:ldvec},
to be flavor-conserving,
\be
  \tilde g^{(1)} = \tilde g^{(1)} = \tilde g^{(1)} = 0.1 \cdot \mathbb{1} \ ,
\ee
and suppressed, if relevant, by an effective scale fixed to $\Lambda = 1$~TeV. Moreover, all
new states are once again assumed mass-degenerate.
We show in  Figure \ref{fig:vector} the hadronic cross sections related to
trilepton and tetralepton production induced by the pair-production
of the component of the $V$, $\cal V$ and $\mathbf V$ fields. We find that these cross sections
are larger than 1~fb, which possibly implies the observation of some events during the 2012 LHC run,
for new physics masses satisfying
$M_V \lesssim 400$~GeV, $M_{\cal V} \lesssim 600$~GeV and $M_{\mathbf V} \lesssim 
500$~GeV in the singlet, doublet
and triple cases, respectively.

\begin{table}
\begin{center}
\caption{\label{tab:summary}Upper bound on doubly-charged particle mass scale so that the sum of all
  BSM contributions to the production rate, at the LHC running at a center-of-mass energy of 8 TeV,
  of multileptonic final states (with $N_\ell \geq 3$) is larger than 1 fb.}
\begin{tabular}{l||c|c|c}
& $~~~SU(2)_L$ singlet$~~~$ & $~~~SU(2)_L$ doublet$~~~$ & $~~~SU(2)_L$ triplet$~~~$ \\
\hline
Scalar fields                            & 330 GeV & 257 GeV & 350 GeV\\
Fermionic fields (no mixing with the SM) & 555 GeV & 661 GeV & 738 GeV\\
Fermionic fields (mixing with the SM)    & -       & 471 GeV & 560 GeV\\
Vector fields                            & 392 GeV & 619 GeV & 495 GeV\\
\end{tabular}
\end{center}
\end{table}

In Table \ref{tab:summary}, we summarize the different mass ranges expected to give rise to new physics
contributions to multilepton production at the LHC collider, running at a center-of-mass energy
of 8~TeV,
larger than 1~fb. This motivates us to select three benchmark scenarios
for a more careful analysis, based on Monte Carlo simulations, of the effects associated with
the presence of fields containing a doubly-charged component, with the aim of
defining some ways to distinguish their spin and $SU(2)_L$ representation. We hence choose first
a series of scenarios where the common mass is fixed to a rather optimistic value of 100~GeV, recalling
that no LHC constraints has been derived for promptly decaying fields and/or when 
non-leptonic decay channels are open. We then define two other classes of scenarios where the new physics
mass scale lies well above the dibosonic thresholds. We fix it to 250 GeV and 350 GeV, respectively.

\section{Probing spin and $SU(2)_L$ representations with Monte Carlo
simulations} \label{sec:MC}
In the previous section, we have shown that the mass range for possibly observing doubly-charged
particles at the LHC depends on their spin and $SU(2)_L$ representations. In this section, we focus
on the study of various kinematical distributions that should allow to probe the nature
of a doubly-charged particle, if one assumes that it is responsible for
the observation of excesses in multilepton final states
with $N_\ell \geq 3$ charged leptons. Towards this goal, we implement all the models presented in
Section~\ref{sec:themodel} in {\sc MadGraph}~5  \cite{Alwall:2011uj} via {\sc FeynRules}
\cite{Christensen:2008py,Christensen:2009jx,Christensen:2010wz,Duhr:2011se,Fuks:2012im,%
Alloul:2013fw}. We then present, within the {\sc MadAnalysis}~5 framework \cite{Conte:2012fm}, results
that are based on a hadron-level simulation of the signal
describing 20~fb$^{-1}$ of collisions at a center-of-mass energy of
				8 TeV. For this, the parton-level events as generated by {\sc Madgraph}~5 
have been showered and hadronized by means of the {\sc Pythia}~6 package~\cite{Sjostrand:2006za},
tau lepton decays being handled by using the {\sc Tauola} program \cite{Davidson:2010rw}.

We start our analysis by preselecting events after imposing
a set of basic selections ensuring that most (non-simulated) background contributions
are well under control. 
\begin{itemize}
\item We start by removing from the event final states all charged leptons not having
a transverse momentum $p_T~\ge~10$~GeV and a pseudorapidity satisfying $|\eta| \leq 2.5$. 
\item Jets are
reconstructed by means of the anti-$k_{t}$ algorithm \cite{Cacciari:2008gp}
 as implemented in
{\sc FastJet}~\cite{Cacciari:2005hq,Cacciari:2011ma} after we set the radius parameter to $R=0.4$.
We only consider jet candidates with 
$p_T \geq 20$~GeV  and $|\eta| \leq 2.5$ which are not too close to an electron, \ie, which lie outside a cone of
radius $R=0.1$ centered around the electron.
\item Lepton isolation is then enforced by rejecting all leptons lying in a cone of radius $R=0.4$
centered on any of the remaining jets.
\item We require the presence in the final state of at least
three isolated charged leptons.
\end{itemize}
While a complete simulation of the Standard Model background goes beyond
the scope of this work, we refer to an existing phenomenological study
of leptonic final states to demonstrate that the background
remaining after the selection above\footnote{Although an
additional veto on events containing identified $b$-jets is applied, this
does not affect our purely leptonic signal.} is under
control~\cite{Alloul:2013fra}. This analysis shows that we can
indeed expect about 5500
background events, originating in 99.5\% of the cases from diboson
production processes, so that already 230 signal events can induce
a $3\sigma$ deviation from the Standard Model expectation.

\begin{figure}[!t]
  \centering
  \begin{picture}(400,2)
\put(80,2){$p p \to N_\ell$ charged leptons at the LHC (8 TeV), with $N_\ell \geq 3$.}
  \end{picture}
  \hspace*{-0.6cm}
  \includegraphics[width=.33\columnwidth]{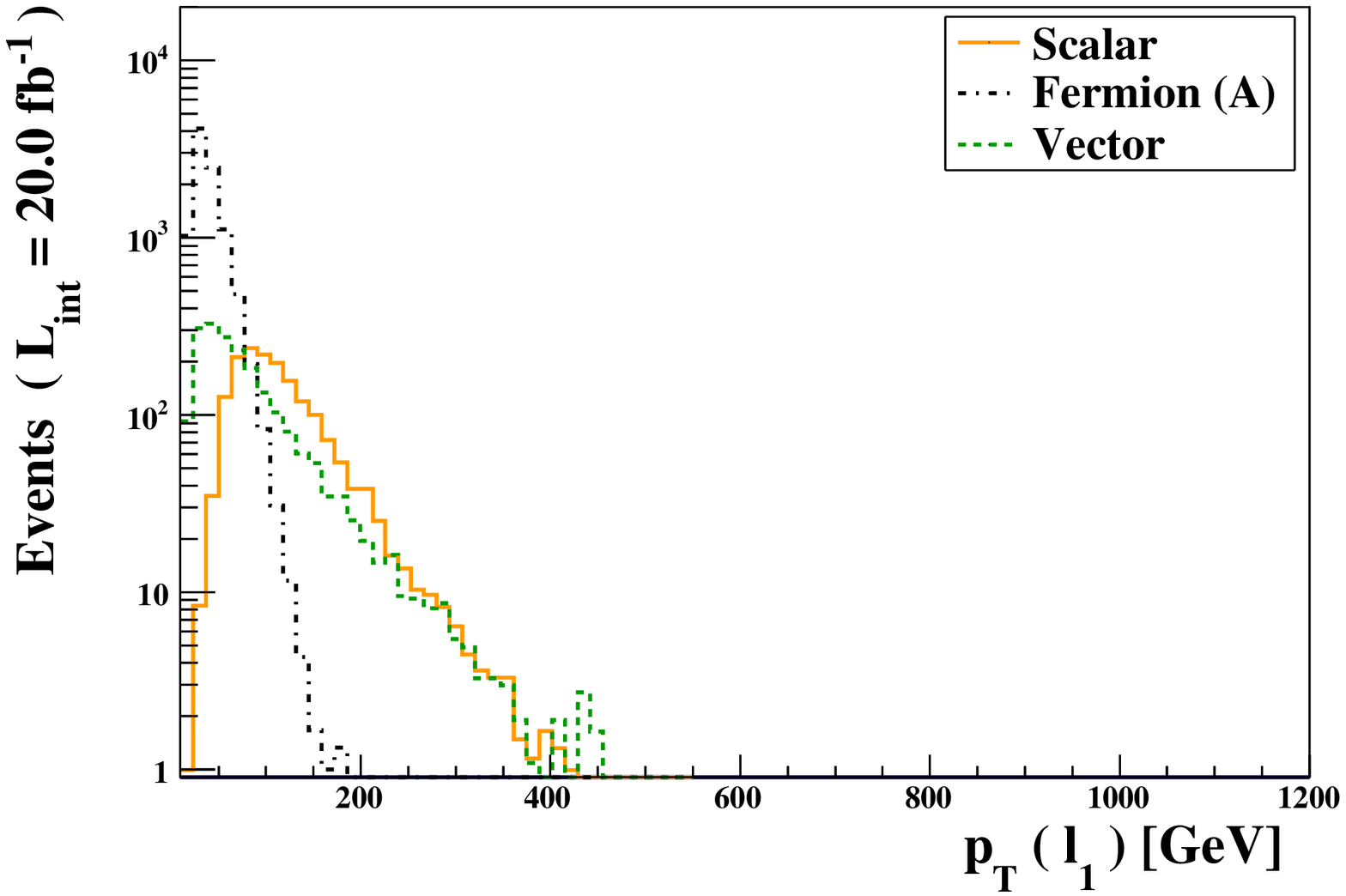}
  \includegraphics[width=.33\columnwidth]{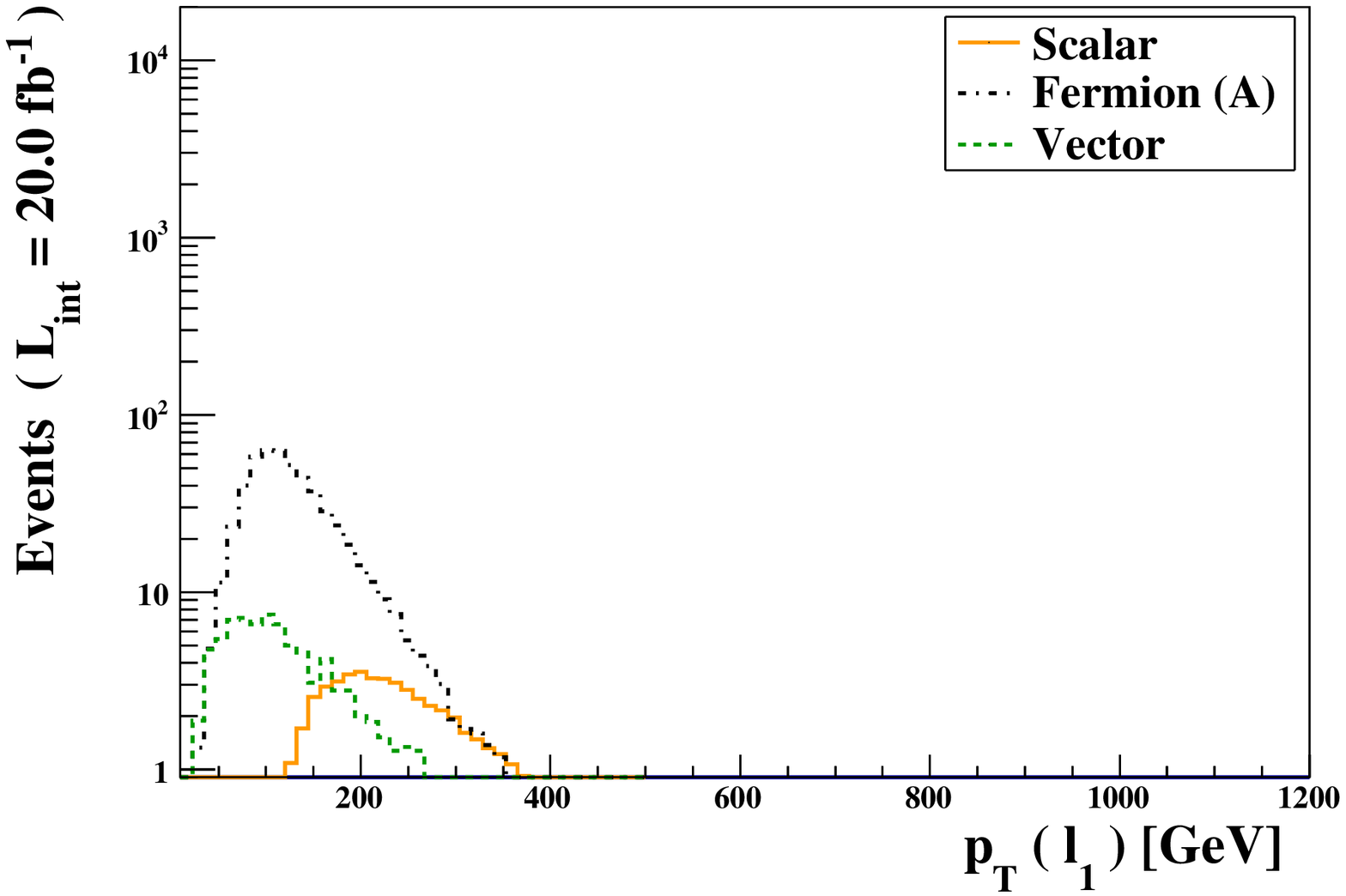}
  \includegraphics[width=.33\columnwidth]{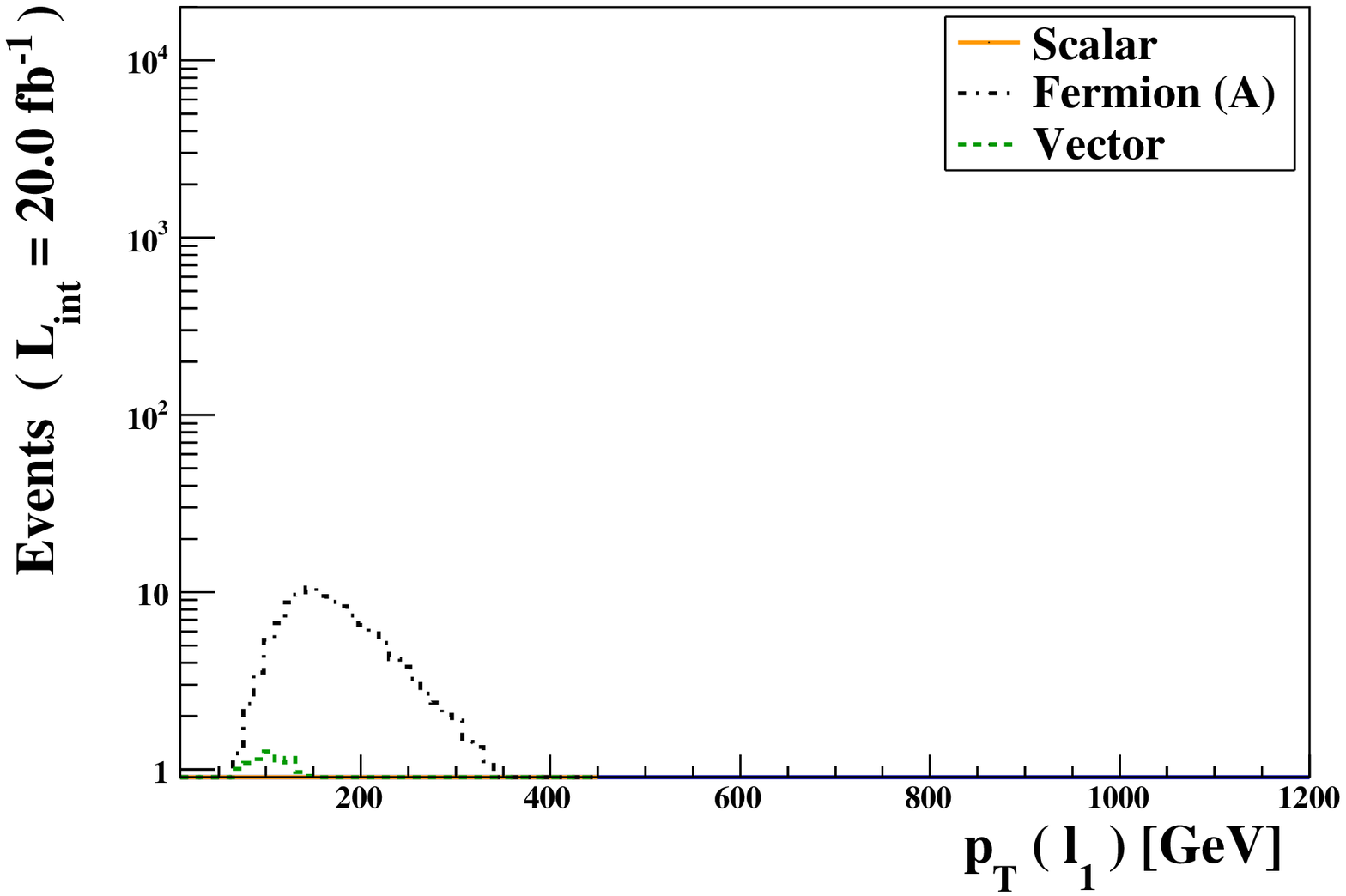}\\
  \hspace*{-0.6cm}
  \includegraphics[width=.33\columnwidth]{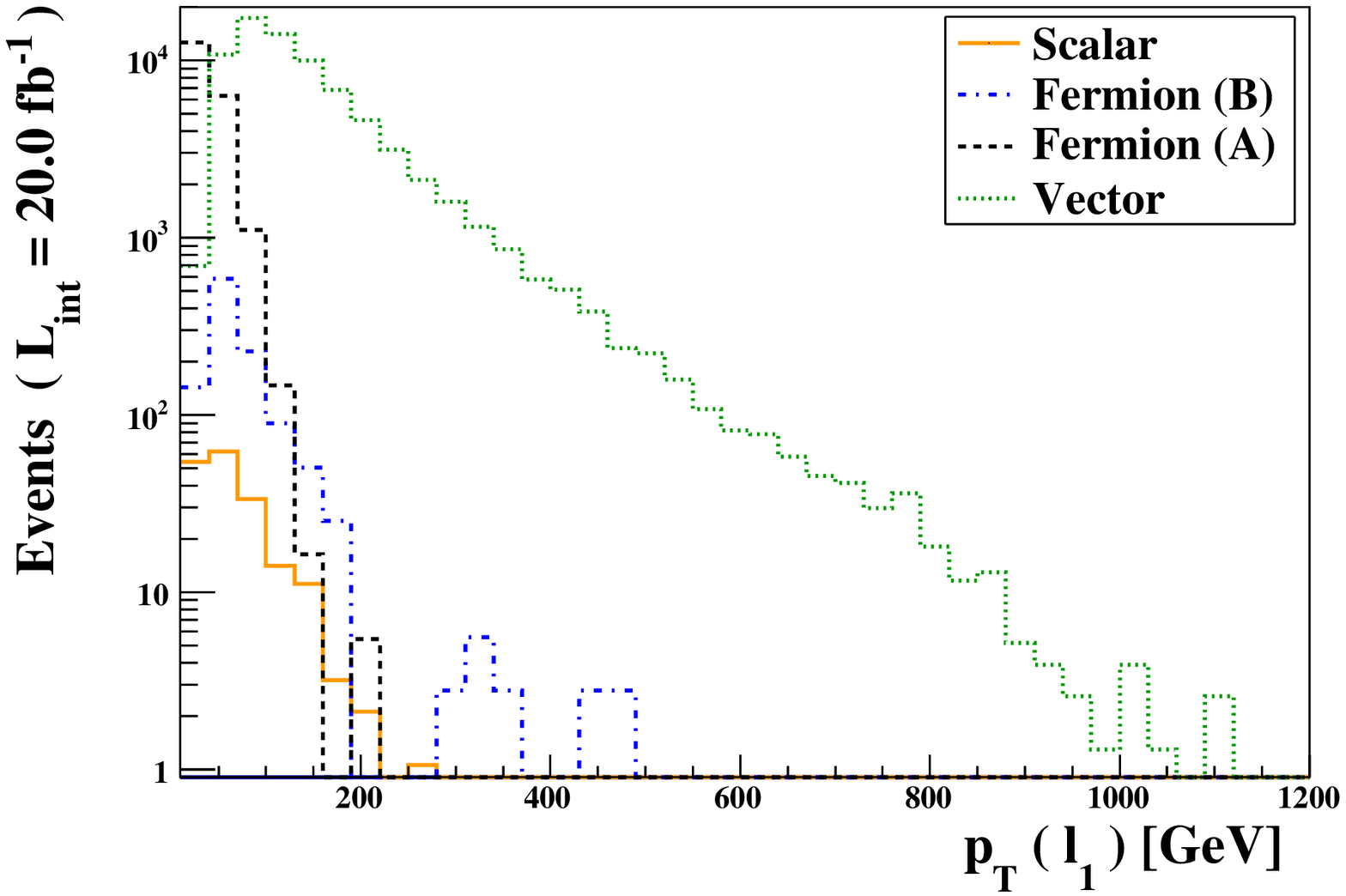}
  \includegraphics[width=.33\columnwidth]{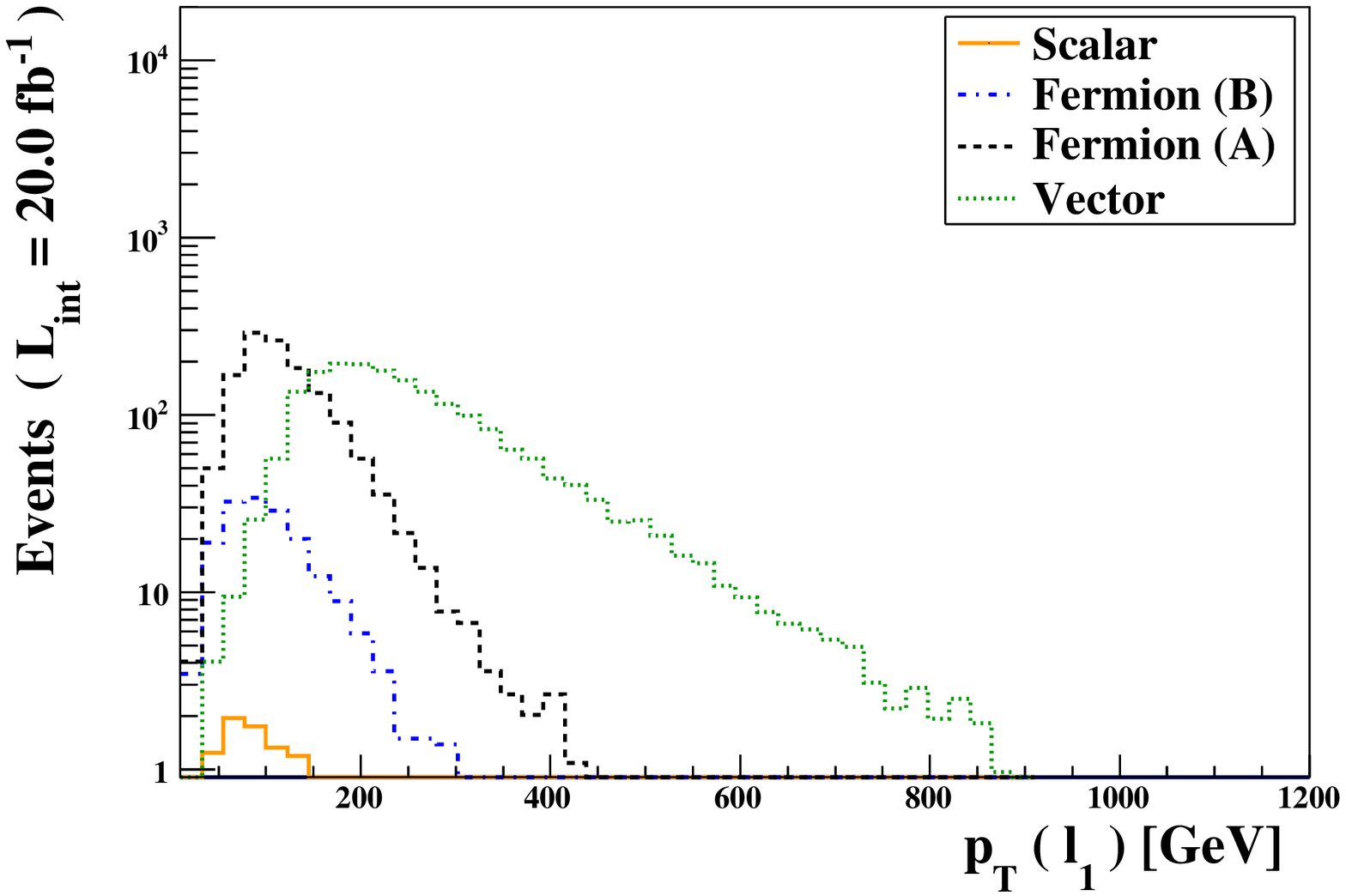}
  \includegraphics[width=.33\columnwidth]{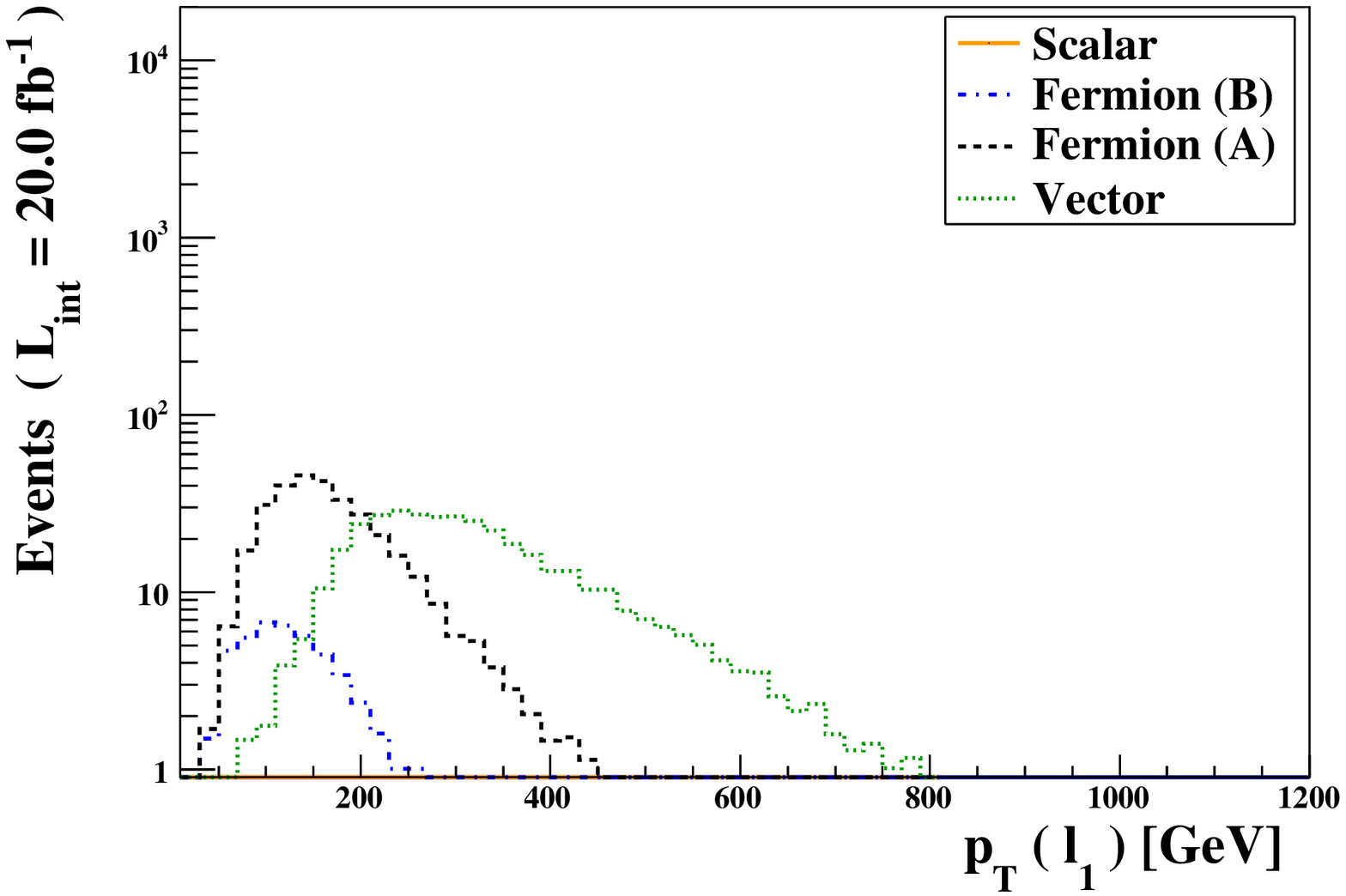}\\
   \hspace*{-0.6cm}
  \includegraphics[width=.33\columnwidth]{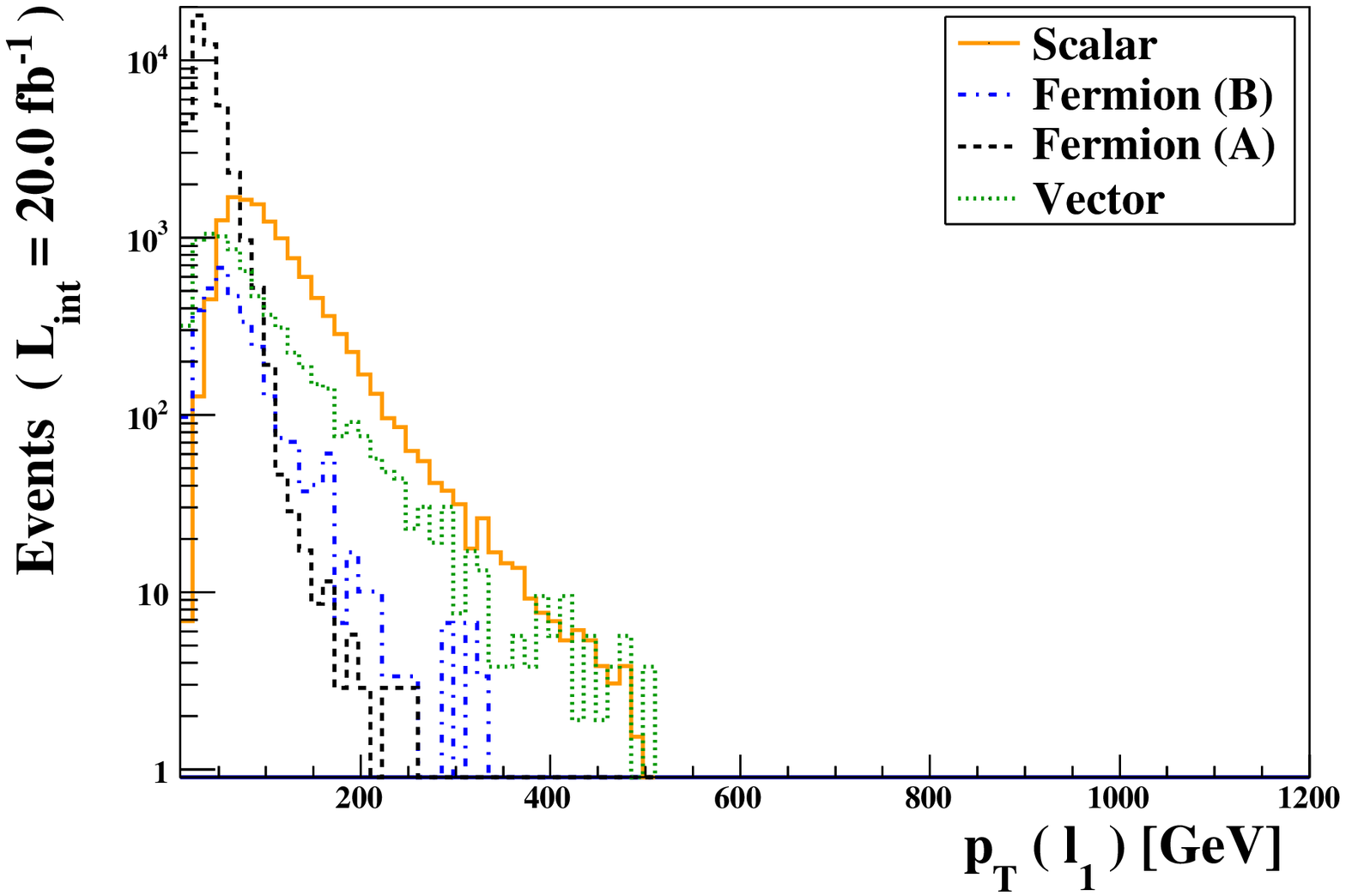}
  \includegraphics[width=.33\columnwidth]{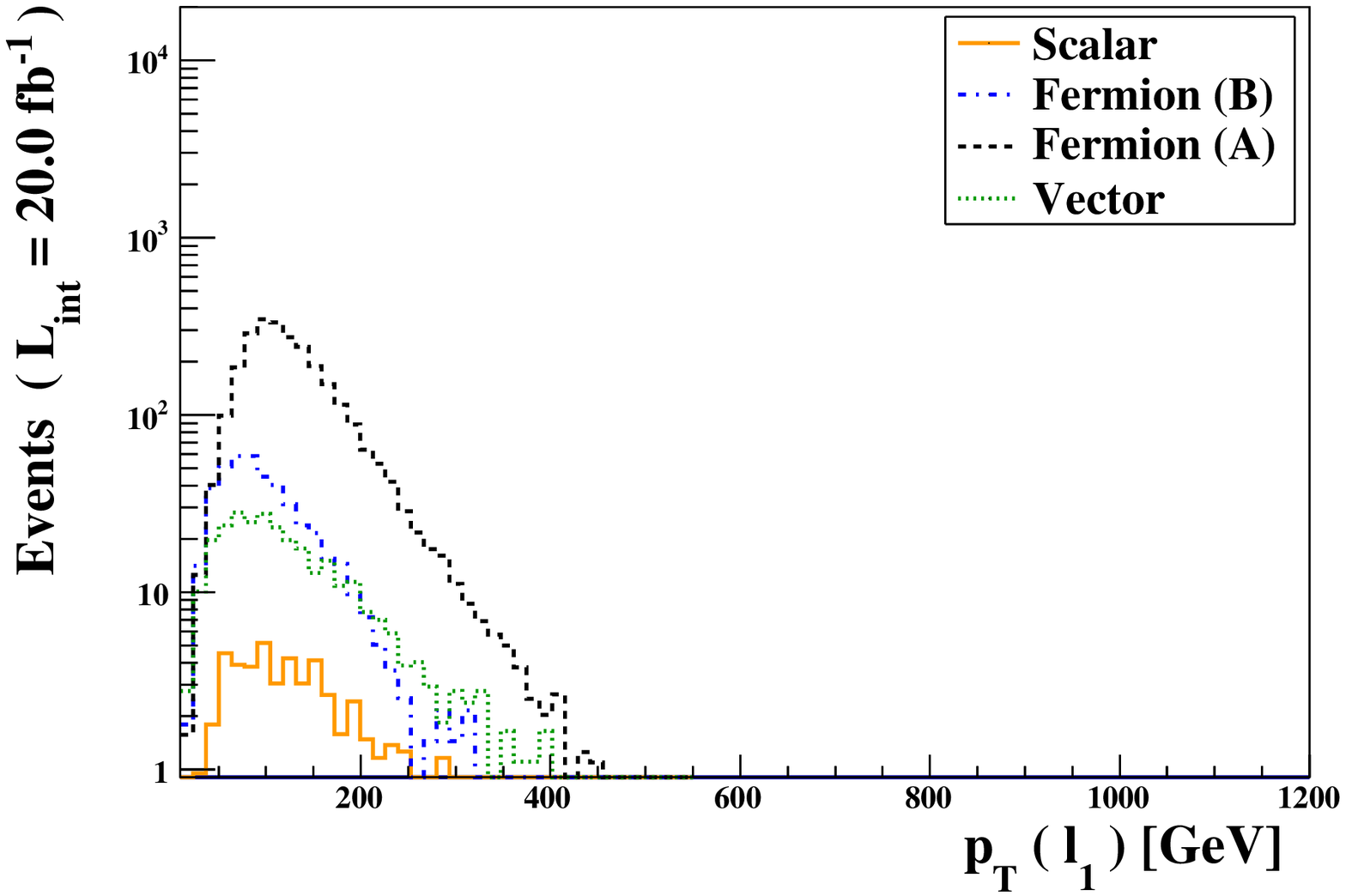}
  \includegraphics[width=.33\columnwidth]{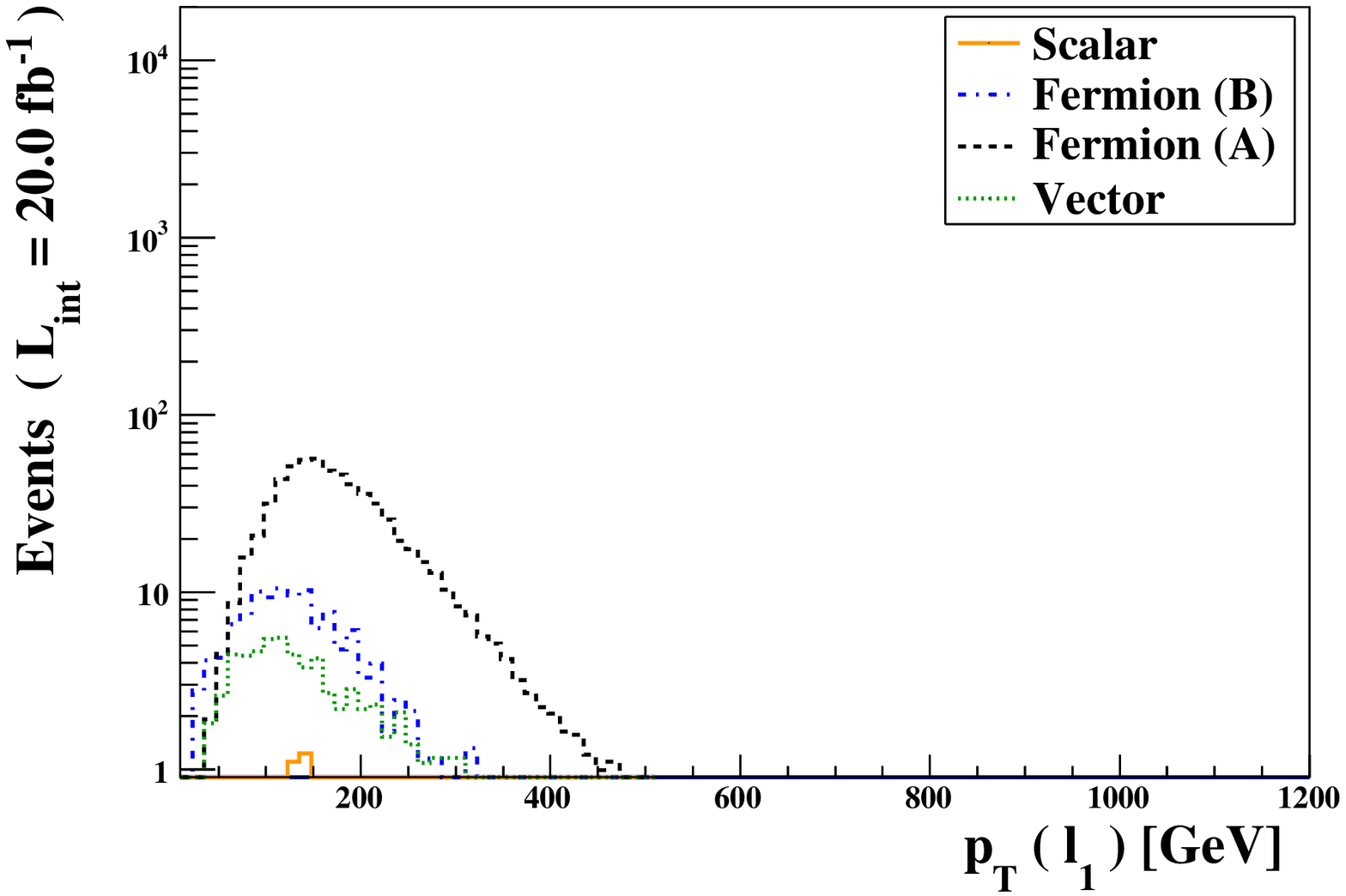}
  \caption{\label{fig:ptl1}Transverse-momentum spectrum of the leading
  lepton emerging from a possible doubly-charged particle signal with $N_\ell \geq 3$ charged
  leptons in the final states. Event generation has been performed in the context of
  the LHC and for 20~fb$^{-1}$
  of collisions at a center-of-mass energy of 8 TeV. In the top, middle and lower series
  of graphs,
  we require that the new states lie in the trivial, fundamental and adjoint representation
  of $SU(2)_L$, respectively, while their mass is set to 100~GeV, 250~GeV
  and 350~GeV in the left, central and right columns of the figure. In each subfigure,
	  we show distributions for scalar fields (plain orange curve), vector fields
  (dashed green curve) and fermionic fields whose singly-charged component is
  allowed to mix with the Standard Model $\tau$ lepton (dashed blue curve, dubbed scenario B) or not
  (dot-dashed black curve, dubbed scenario A).}
\end{figure}

\begin{figure}[!t]
  \centering
  \begin{picture}(400,2)
\put(80,2){$p p \to N_\ell$ charged leptons at the LHC (8 TeV), with $N_\ell \geq 3$.}
  \end{picture}
   \hspace*{-0.6cm}
  \includegraphics[width=.33\columnwidth]{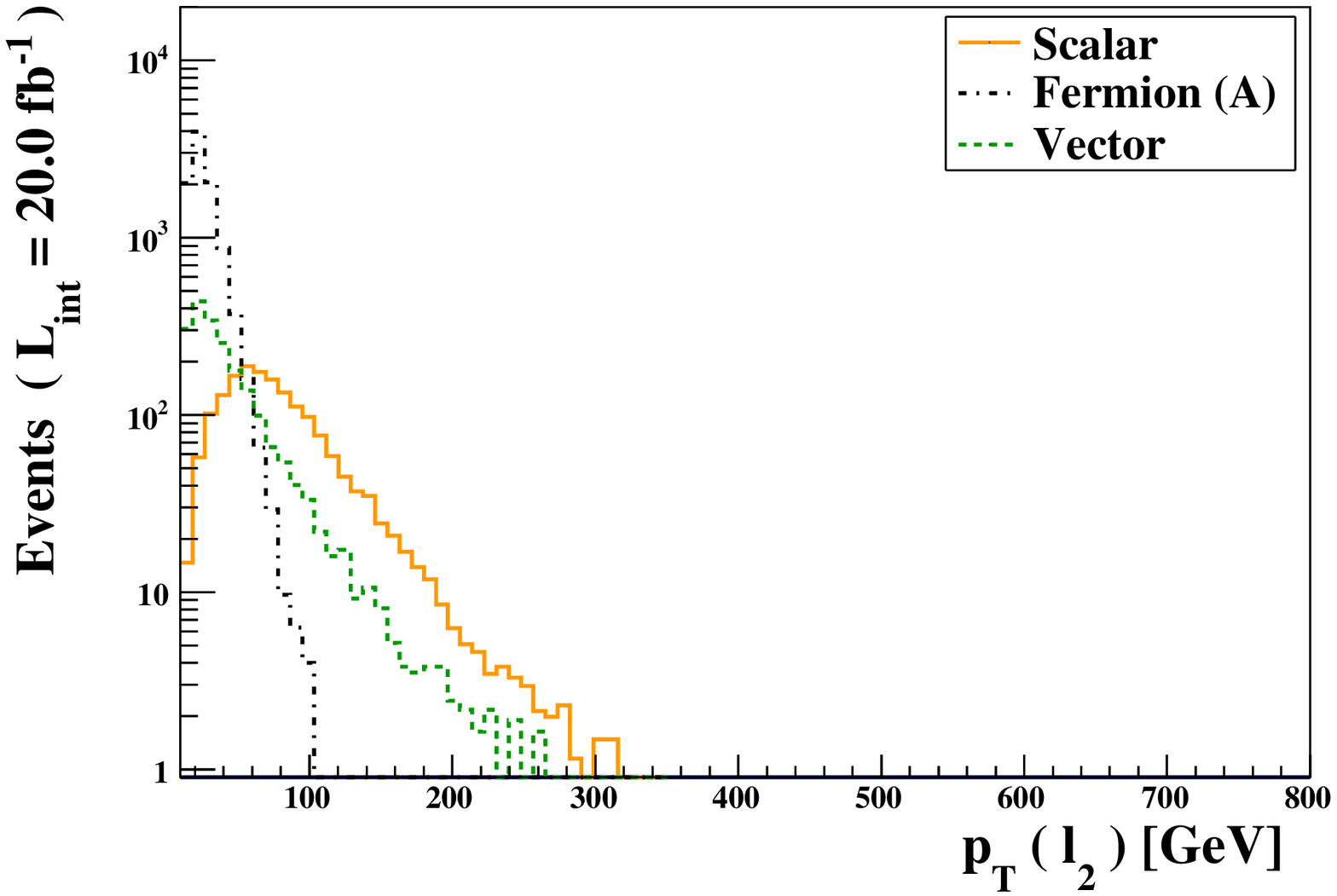}
  \includegraphics[width=.33\columnwidth]{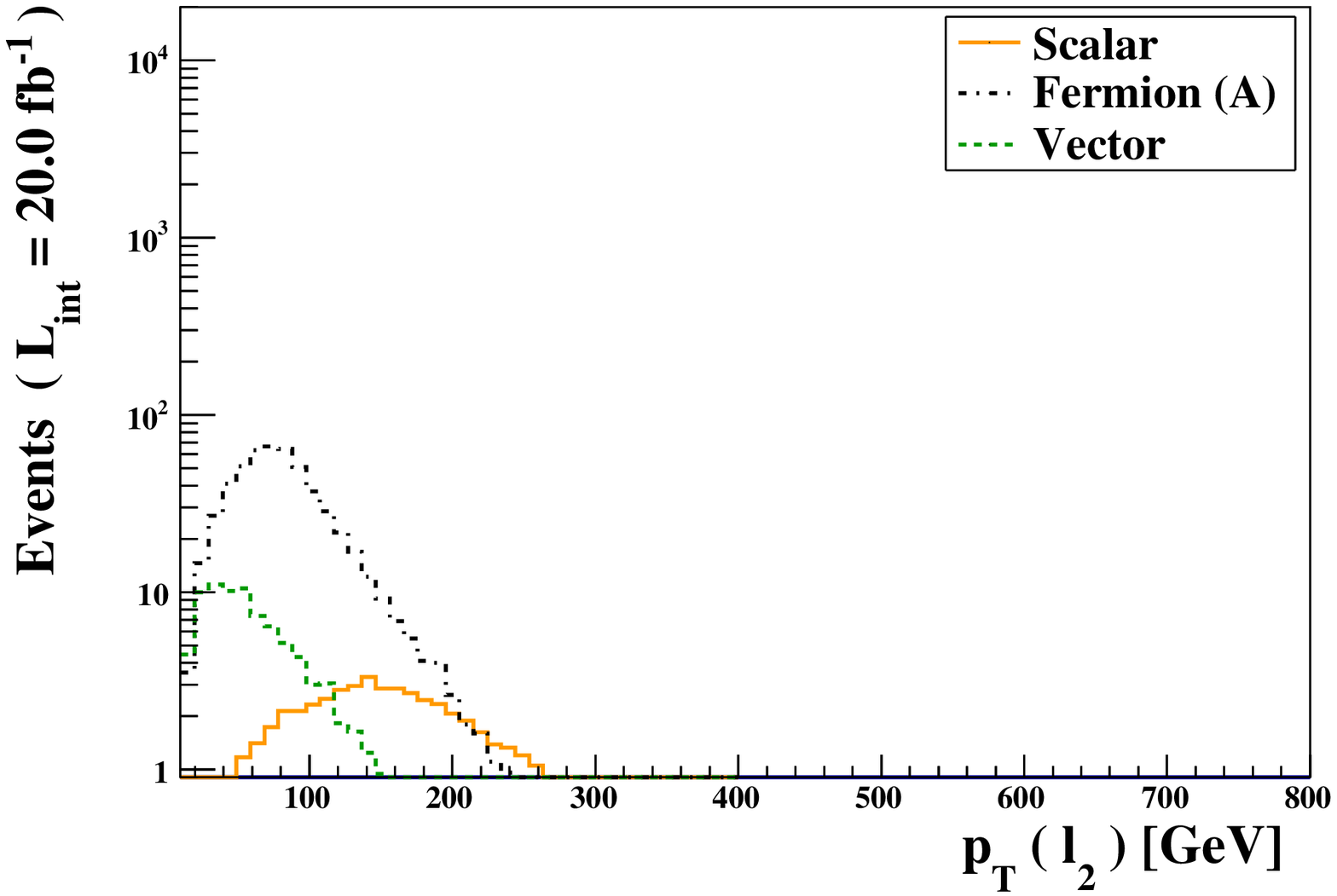}
  \includegraphics[width=.33\columnwidth]{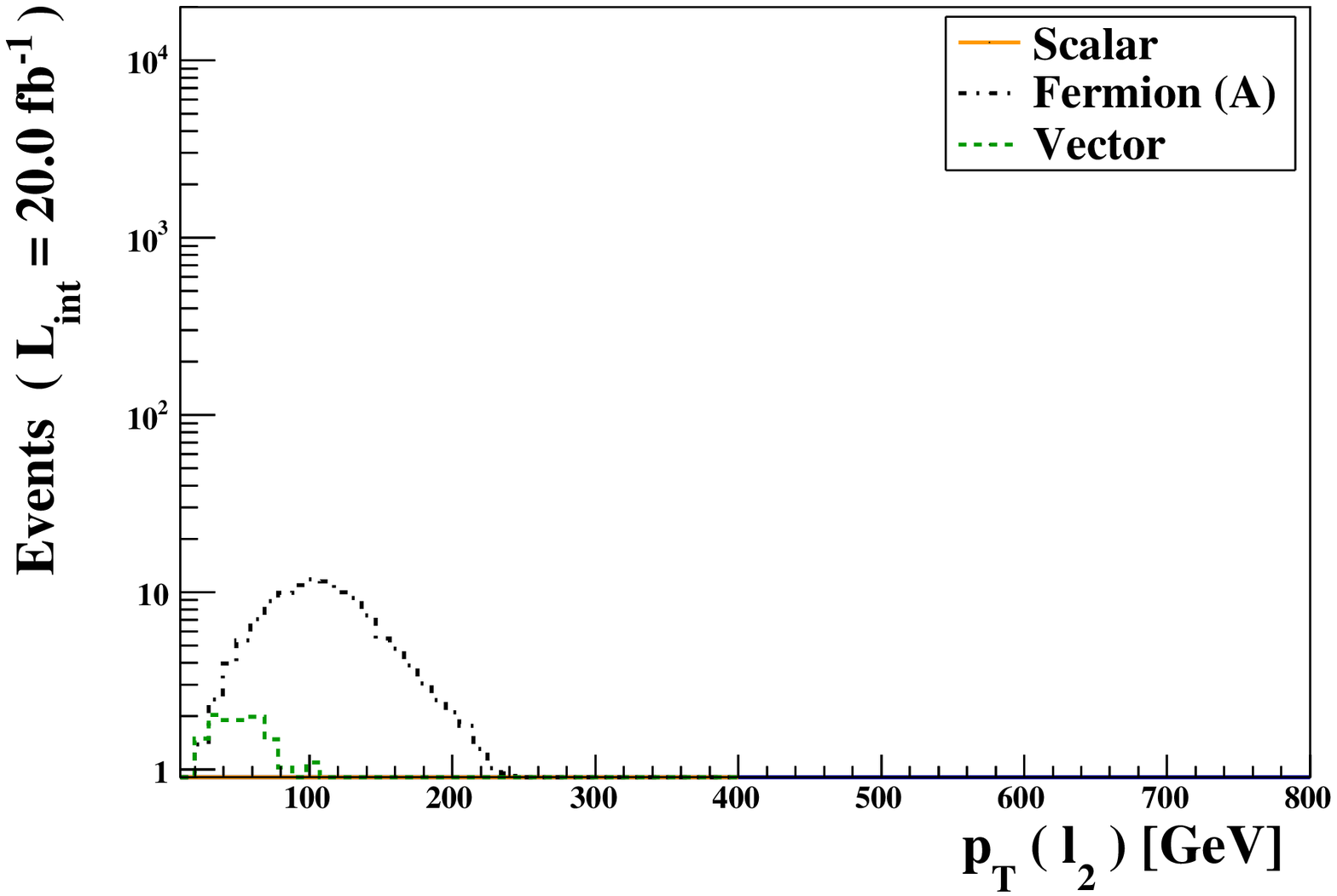}\\
   \hspace*{-0.6cm}
  \includegraphics[width=.33\columnwidth]{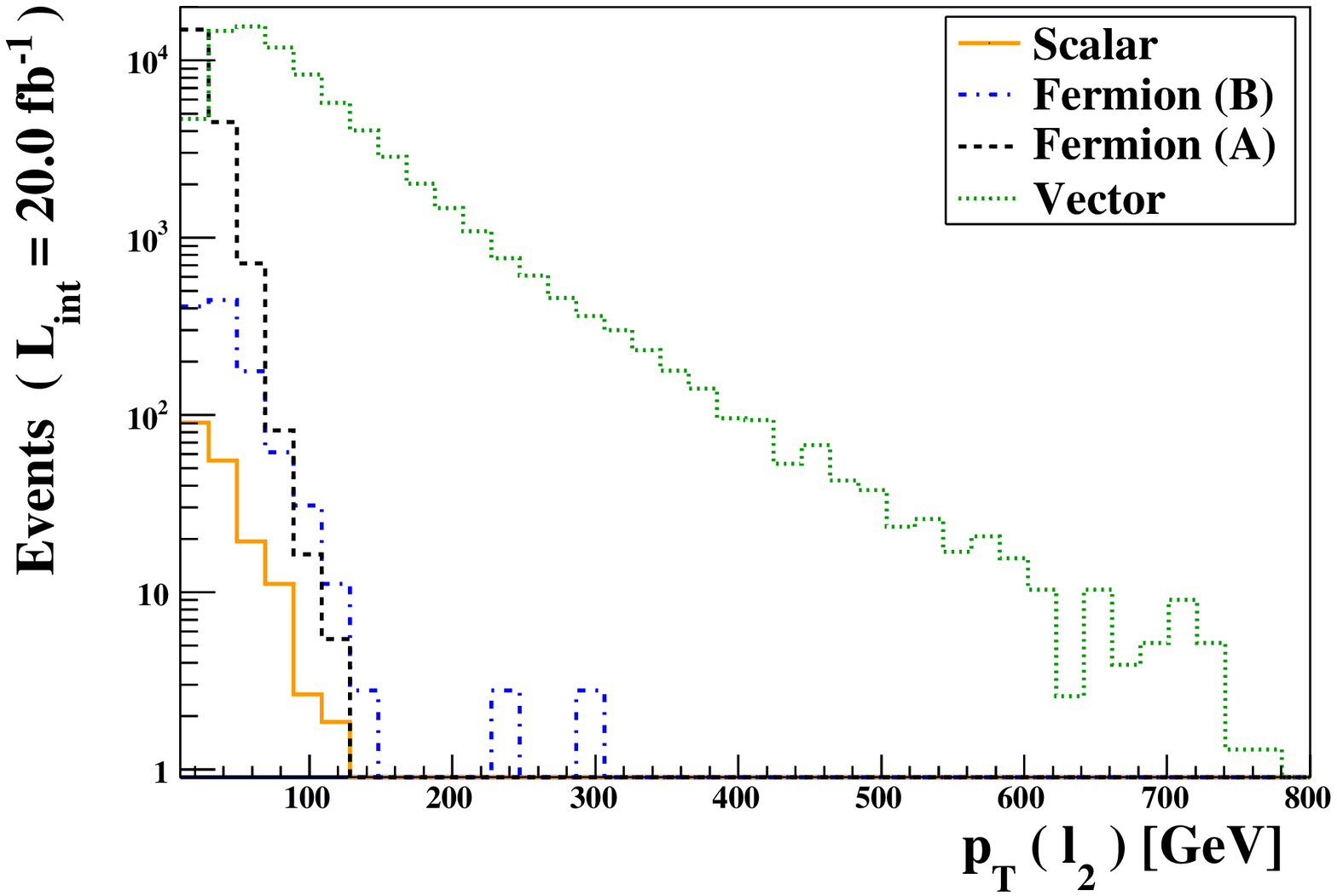}
  \includegraphics[width=.33\columnwidth]{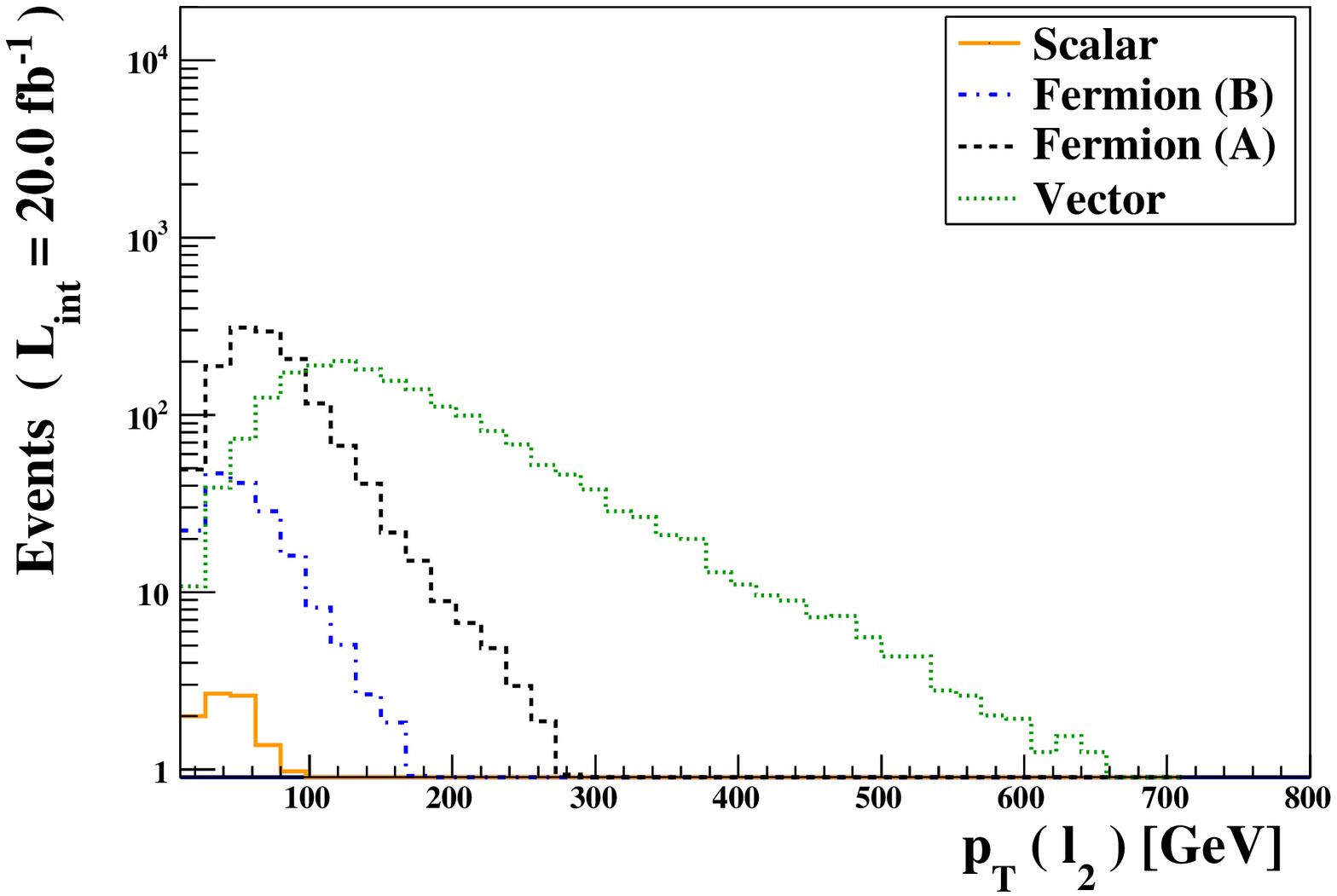}
  \includegraphics[width=.33\columnwidth]{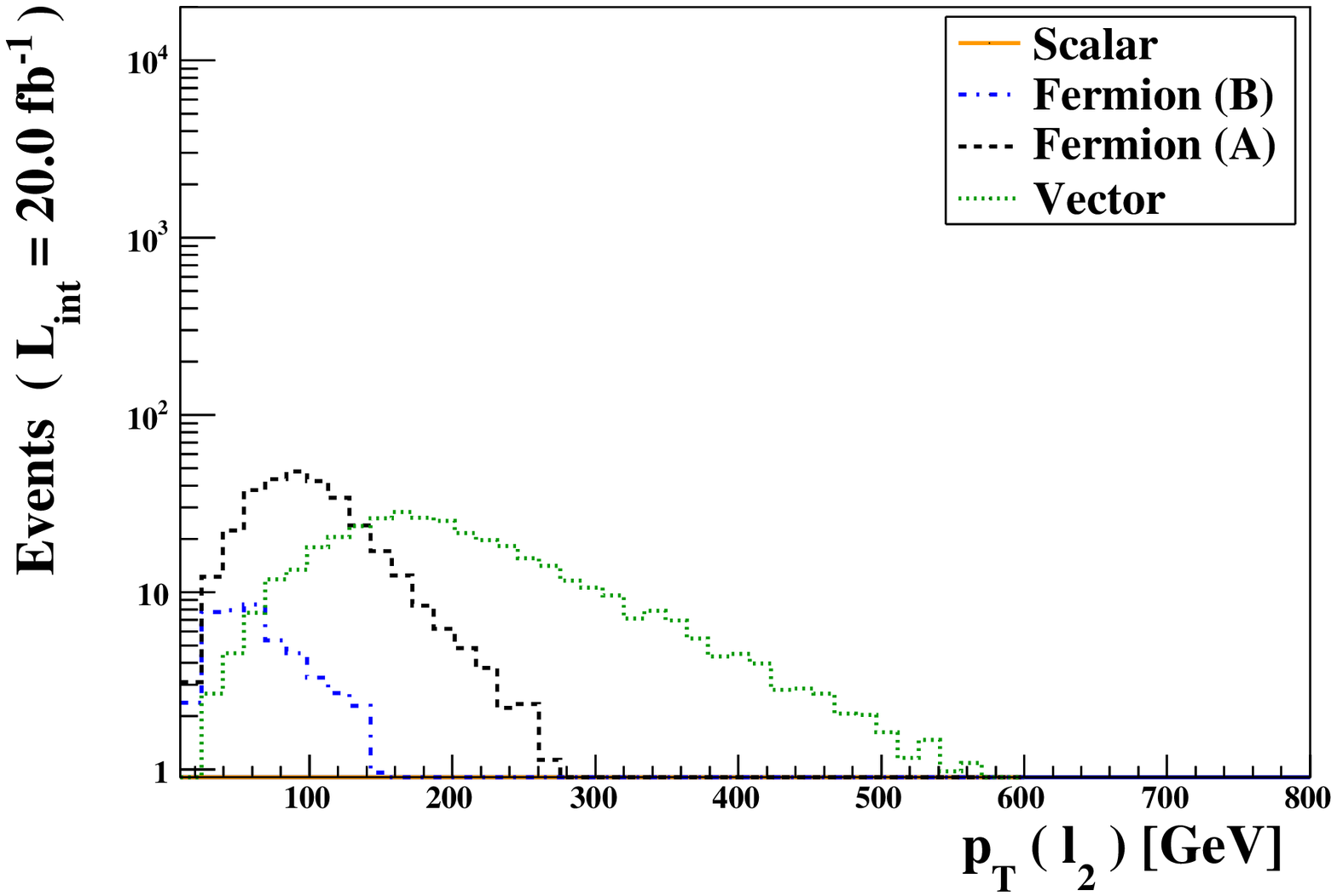}\\
   \hspace*{-0.6cm}
  \includegraphics[width=.33\columnwidth]{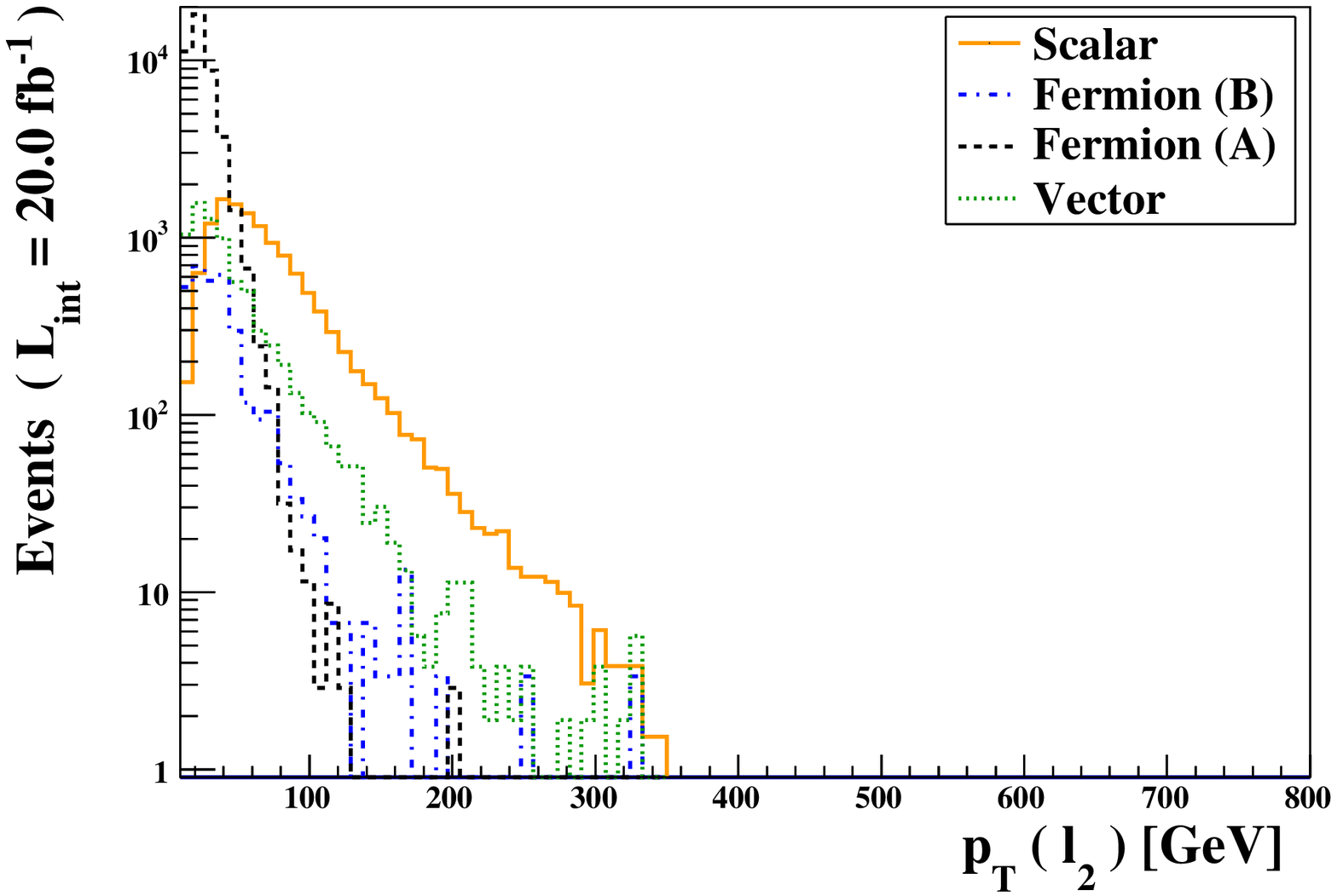}
  \includegraphics[width=.33\columnwidth]{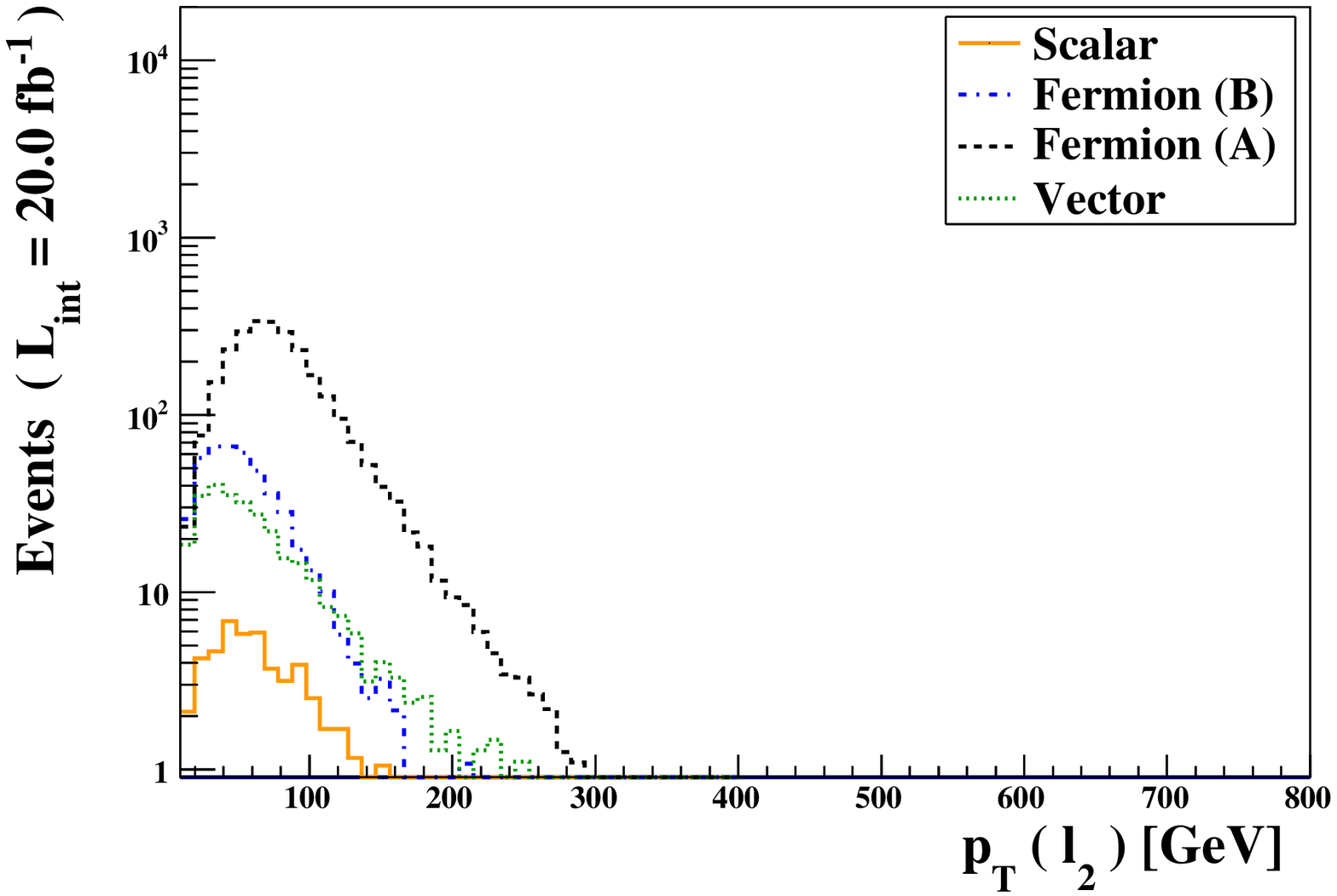}
  \includegraphics[width=.33\columnwidth]{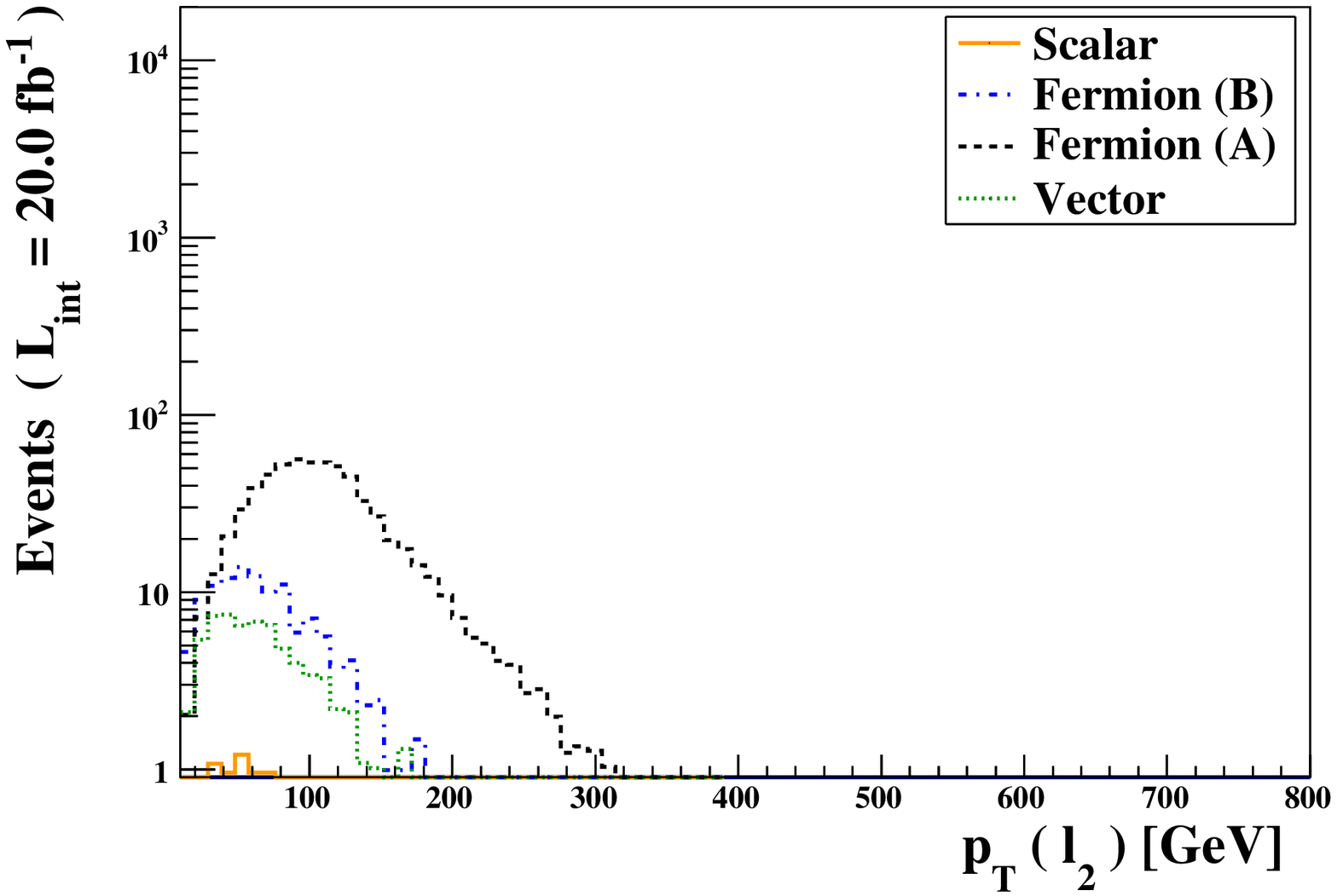}
  \caption{\label{fig:ptl2}Same as in Figure~\ref{fig:ptl1}, but for the 
  transverse-momentum spectrum of the next-to-leading leading charged
  lepton.}
\end{figure}

Since all experimental analyses focusing on
multileptonic signatures further select events by requiring specific thresholds
on the transverse-momentum of (at least) the two leading leptons $\ell_1$ and $\ell_2$,
we present
the related spectra in Figure~\ref{fig:ptl1} and Figure~\ref{fig:ptl2} in the context of
all the Standard Model extensions introduced in Section \ref{sec:themodel}.
In the top, middle and lower series of graphs shown on the
figures, respectively,  we focus on fields lying in the singlet, doublet and triplet
representations of $SU(2)_L$. In contrast, on the left, central and right
columns of the figure, we set the masses of the new states to
100~GeV, 250~GeV and 350~GeV, respectively. We recall that these choices have been adopted from
the three mass scenarios constructed in the previous
section. All the represented spectra exhibit a common global behavior.
The distributions start by steeply rising, then peak and are finally
extended by a tail up to (in general) moderate $p_T$ values smaller than 400~GeV.
It is therefore rather complicated to define a feature to a given spin and/or $SU(2)_L$
representation when one accounts for the possible different new particle masses.
Two exceptions are however allowed. First, events containing very hard leptons with
a transverse momentum larger than 500~GeV are expected to be copiously produced
in scenarios where the Standard Model is extended by a vectorial field $\cal V$
lying in the doublet representation of $SU(2)_L$, for any mass value. Next,
models with additional doubly-charged scalar fields that are singlet under
the $SU(2)_L$ gauge group lead to the production of multileptonic events
where the $p_T$ spectra of the two leading leptons are depleted in the low
and intermediate transverse-momentum regions. This feature nevertheless competes
with the low cross sections associated with heavy scalar masses larger than 250-300~GeV
(as illustrated in Figure~\ref{fig:scalar}).

From those considerations,
one concludes that the $p_T$ spectra of the two leading leptons $\ell_1$ and $\ell_2$
offer possible means to distinguish the spin and/or $SU(2)_L$ representations of the
doubly-charged particles in very specific cases, but not in general.
Similar features are found when analyzing
the transverse-momentum distribution of the next-to-next-to-leading lepton $\ell_3$, as well as
the transverse-mass and the invariant-mass spectra
of any pair of leptons present in the event. The corresponding figures
have therefore been omitted for brevity.

\begin{figure}[!t]
  \centering
  \begin{picture}(400,2)
\put(80,2){$p p \to N_\ell$ charged leptons at the LHC (8 TeV), with $N_\ell \geq 3$.}
  \end{picture}
   \hspace*{-0.6cm}
  \includegraphics[width=.33\columnwidth]{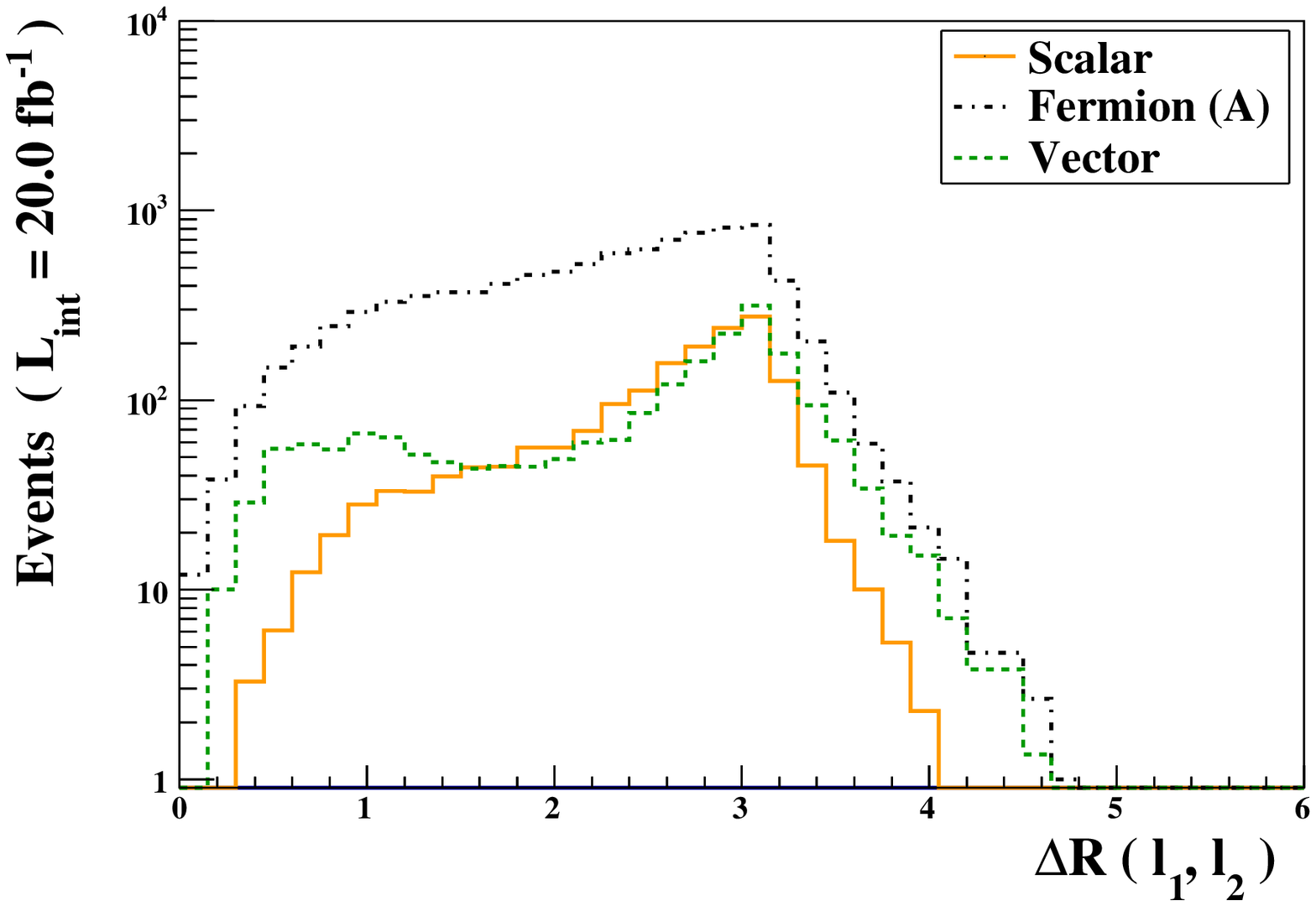}
  \includegraphics[width=.33\columnwidth]{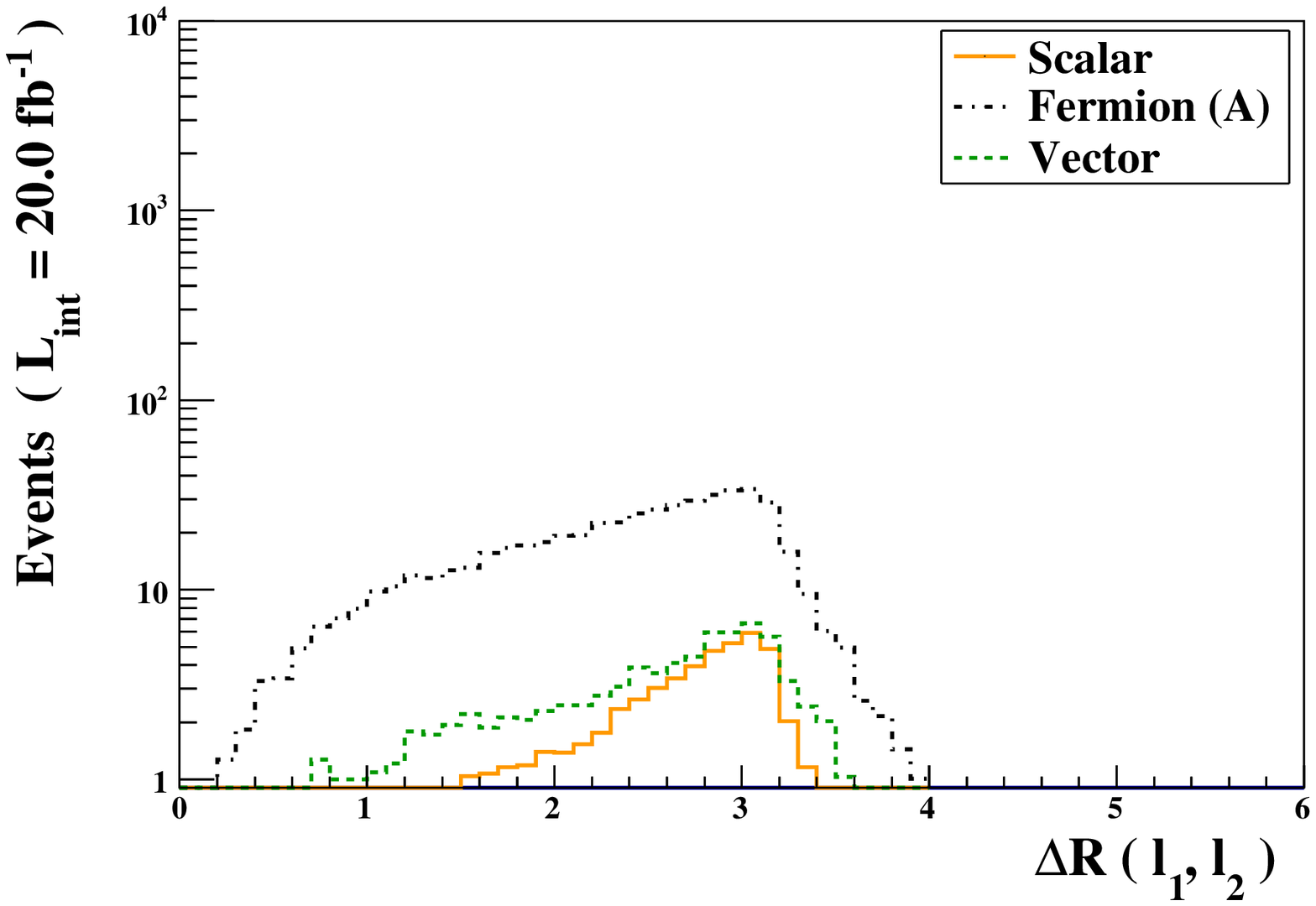}
  \includegraphics[width=.33\columnwidth]{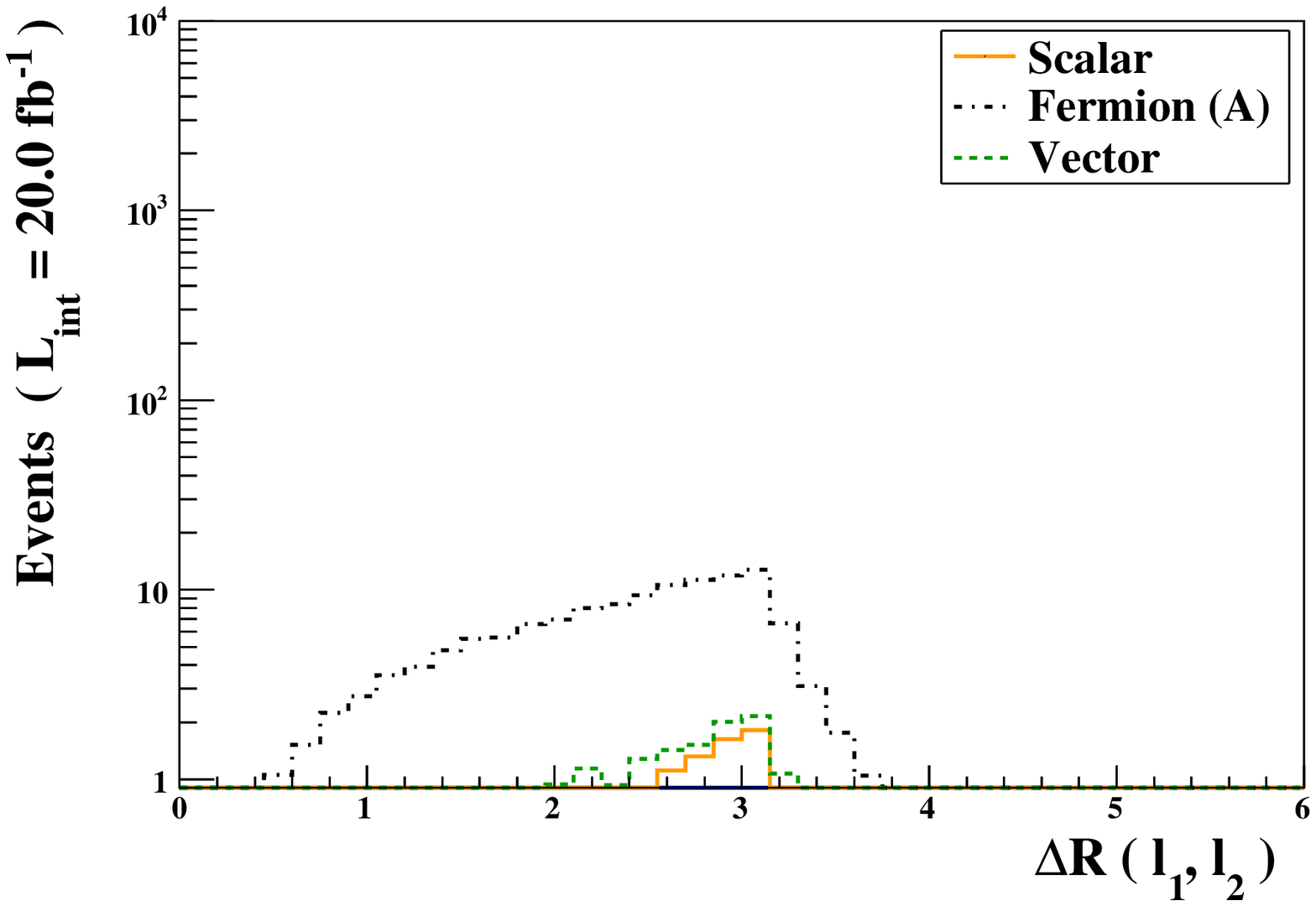}\\
   \hspace*{-0.6cm}
  \includegraphics[width=.33\columnwidth]{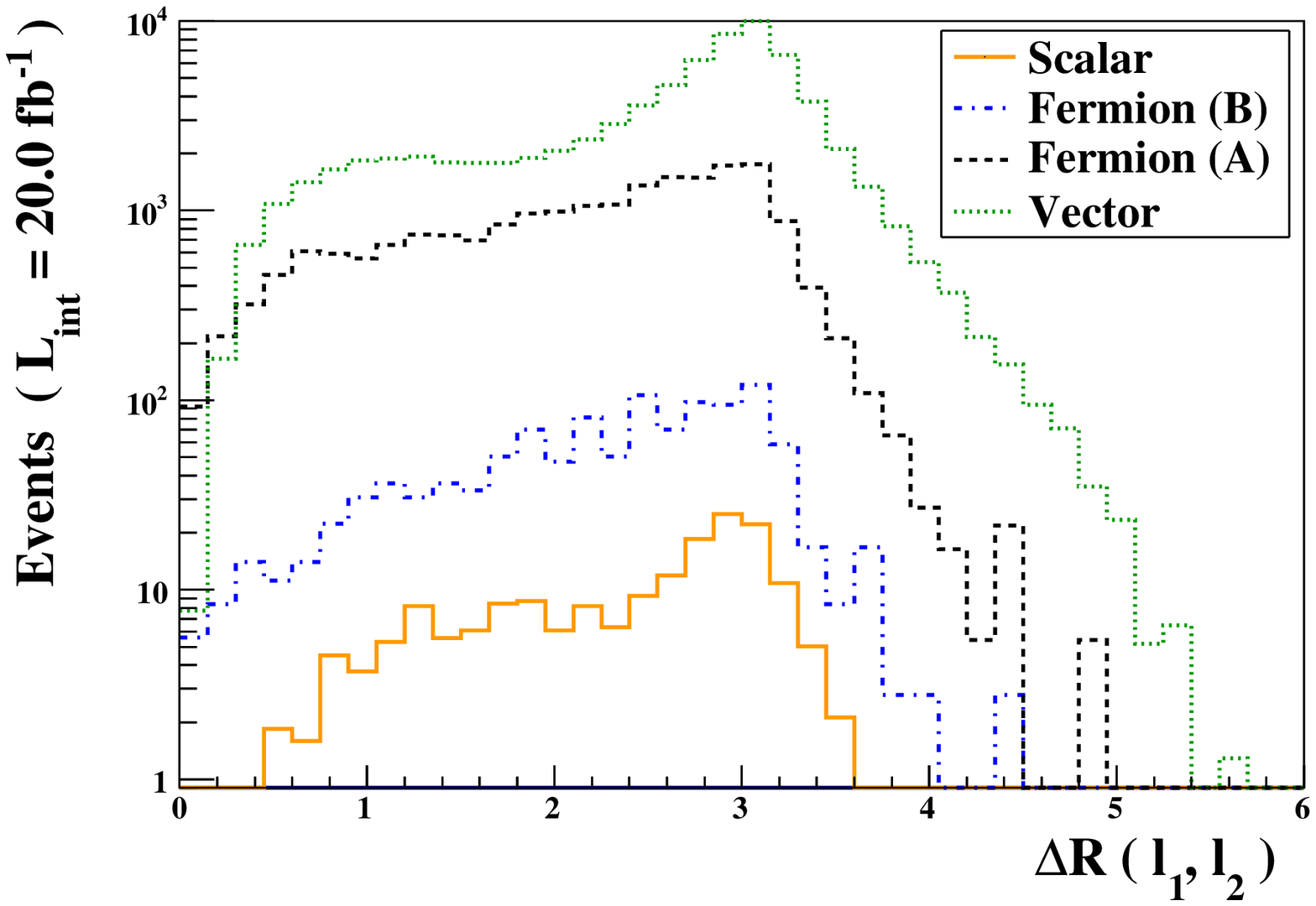}
  \includegraphics[width=.33\columnwidth]{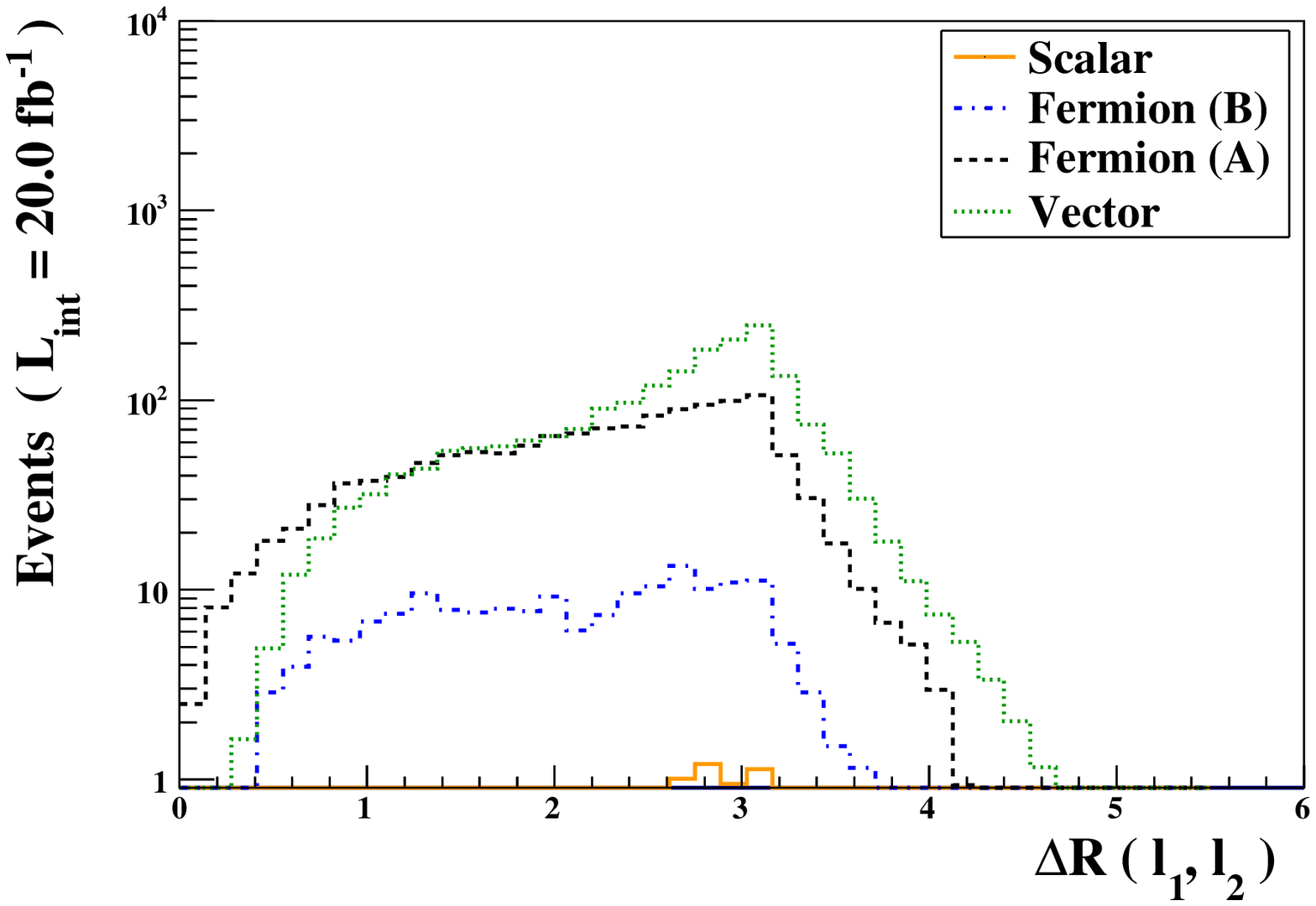}
  \includegraphics[width=.33\columnwidth]{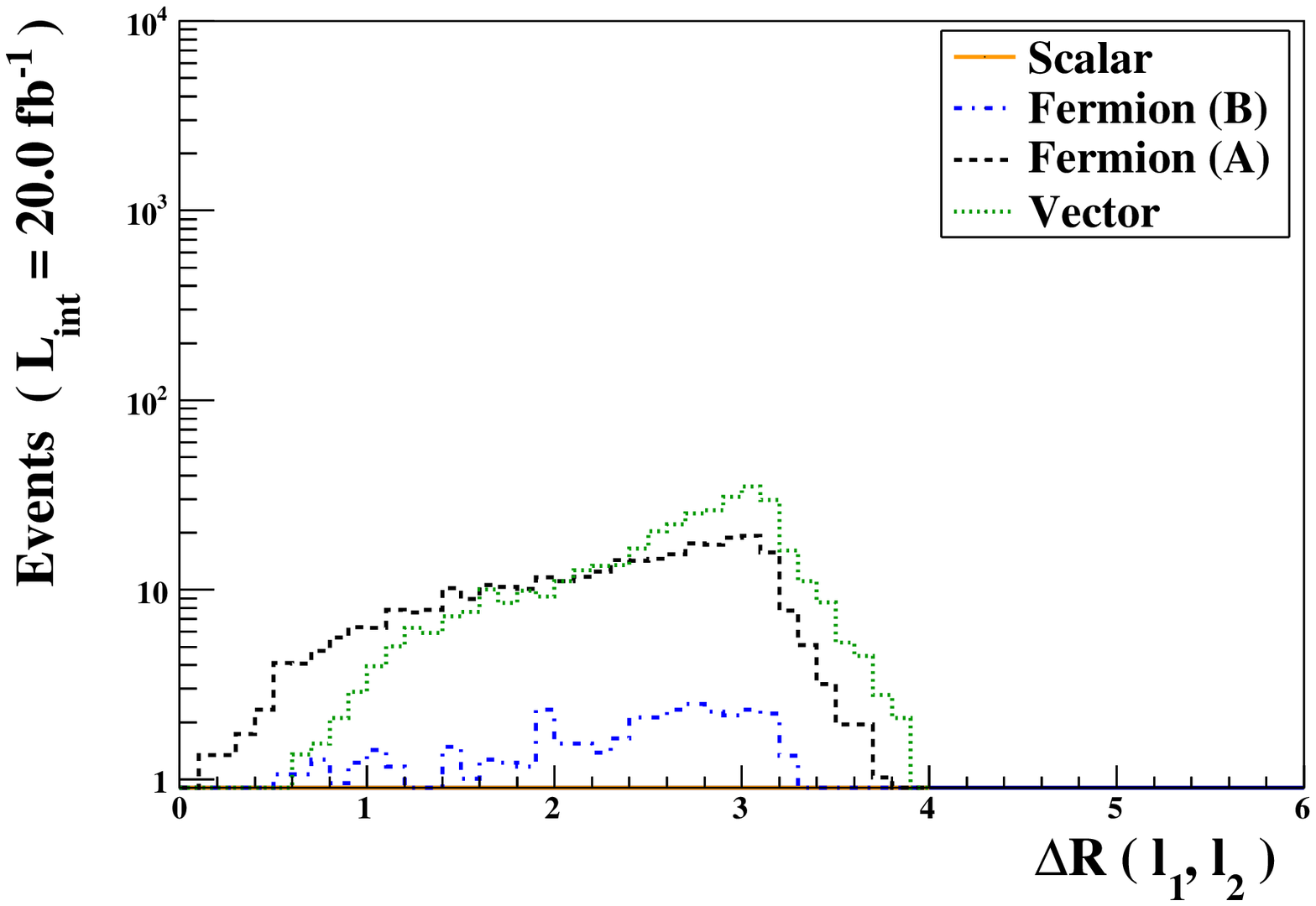}\\
   \hspace*{-0.6cm}
  \includegraphics[width=.33\columnwidth]{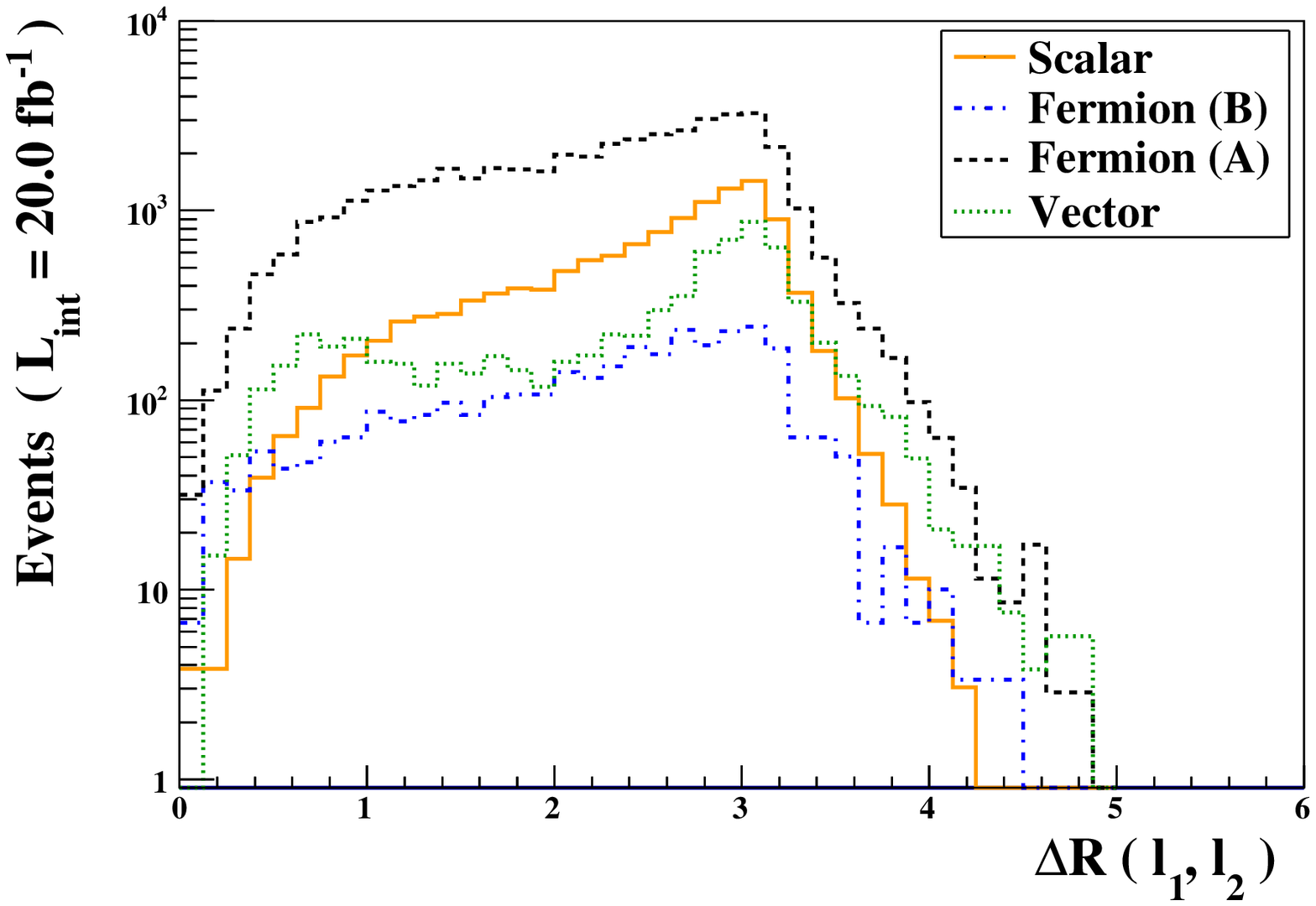}
  \includegraphics[width=.33\columnwidth]{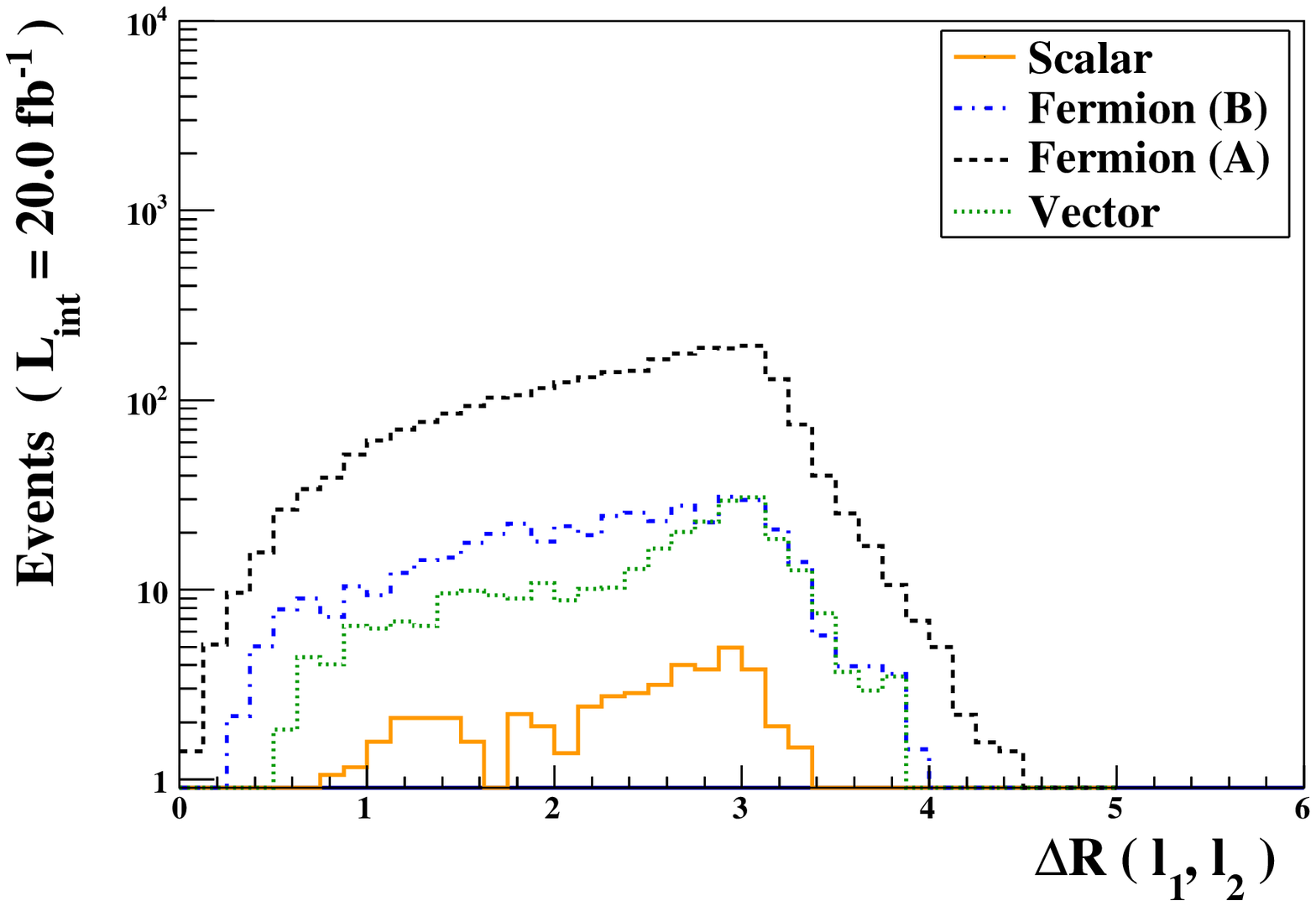}
  \includegraphics[width=.33\columnwidth]{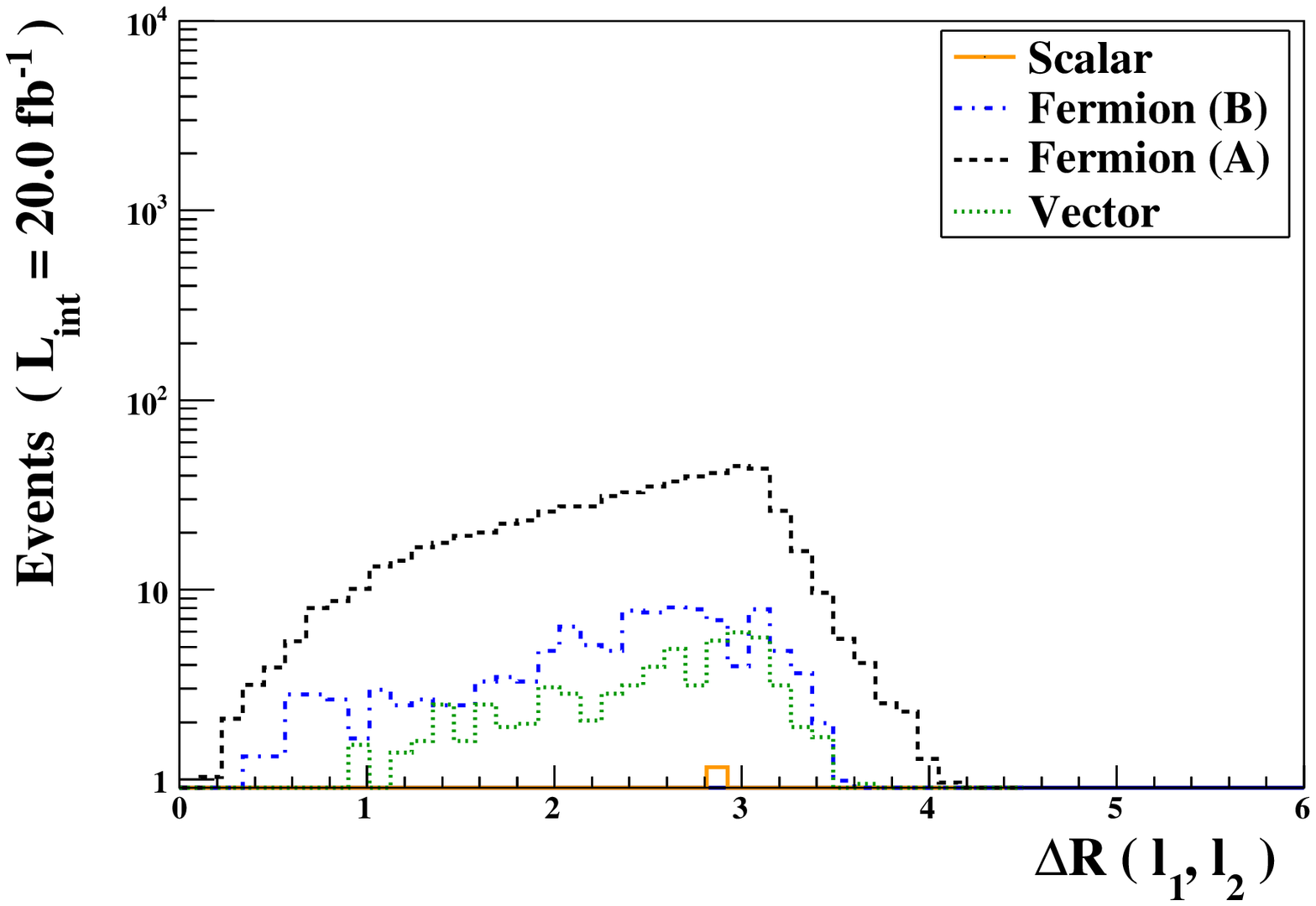}
  \caption{\label{fig:dr12}Same as in Figure~\ref{fig:ptl1}, but for the 
  angular-distance spectrum of a lepton pair comprised of the two leading leptons
  $\ell_1$ and lepton $\ell_2$.}
\end{figure}

Spin representations are highly correlated to angular distributions. In this way,
scalar, fermionic and vectorial doubly-charged particles are expected to give rise to
signals with largely different features, when investigating kinematical variables such as
	angular distances between final state particles. As an example, in
Figure~\ref{fig:dr12} we show the distribution in the angular distance between the
two leading leptons $\Delta R(\ell_1 \ell_2)  = \sqrt{\Delta\phi^2_{12}
+ \Delta\eta_{12}^2}$. In this expression,
$\Delta \phi_{12}$ stands for the azimuthal angular separation
of the two leptons with respect to the beam direction and $\Delta\eta_{12}$ for
their pseudorapidity difference. Investigating the shapes of the spectra, we observe that
they not only depends on the Lorentz representation of the new field, but also
on their $SU(2)_L$ one. Therefore, these variables offer an important
discriminating features among the different scenarios and deserve to be studied in the context
	of a more realistic phenomenological analysis, including detector effects that could alter
the spectrum. As we wish to keep our considerations as general as possible, this however is beyond the scope of this prospective work.


\section{Conclusions}
\label{sec:conclusion}
In this paper we have investigated LHC signals related to the presence of doubly-charged
particles. We have considered different scenarios for such particles,  varying their Lorentz
and $SU(2)_L$ representations and constructing associated simplified models allowing for their
pair production, followed by their decays into Standard Model particles. We have studied the
contributions of doubly-charged states to the
production cross sections of final states containing three or more charged leptons,
a signature known to suffer from a reduced Standard Model background. Using analytical and numerical computations
we have hence deduced that
masses ranging up to 700~GeV are possibly accessible at the LHC for specific models.
We have then employed Monte Carlo simulations to probe
several kinematical distributions allowing to possibly distinguish the spin and $SU(2)_L$
representations of a doubly-charged state. The
results found have been encouraging, in particular in the case of variables such as the
angular distance $\Delta R$ between two of the final state leptons.
This motivates an extension of this
work, including an investigation of the detector effects which could spoil the shapes
of the angular variable spectra, possibly in the context of
ATLAS or CMS analyses.

\acknowledgments

The authors are grateful to Claude Duhr and Olivier Mattelaer for discussing
fermion number violating interactions in {\sc MadGraph}. We acknowledge NSERC of Canada for partial
financial support under grant number SAP105354 and the Theory-LHC-France initiative of the CNRS/IN2P3.


\end{document}